\documentclass{article}

\usepackage{microtype}

\usepackage{hyperref}

\usepackage{amsmath,amsfonts,bm}

\def\1{\bm{1}}

\def\vz{{\bm{z}}}

\def\vone{{\bm{1}}}

\def\vtheta{{\bm{\theta}}}

\DeclareMathAlphabet{\mathsfit}{\encodingdefault}{\sfdefault}{m}{sl}
\SetMathAlphabet{\mathsfit}{bold}{\encodingdefault}{\sfdefault}{bx}{n}

\def\bP{{\mathbb{P}}}

\def\bR{{\mathbb{R}}}

\def\sD{{\mathcal{D}}}

\def\sN{{\mathcal{N}}}

\def\sP{{\mathcal{P}}}

\def\sY{{\mathcal{Y}}}
\def\sZ{{\mathcal{Z}}}

\newcommand{\E}{\mathbb{E}}

\newcommand{\R}{\mathbb{R}}

\newcommand{\KL}{D_{\mathrm{KL}}}

\DeclareMathOperator*{\argmin}{arg\,min}

\usepackage[accepted]{icml2026}

\usepackage{amsmath}
\usepackage{amssymb}
\usepackage{mathtools}
\usepackage{amsthm}

\usepackage{url}
\usepackage{booktabs}           
\usepackage{subcaption}
\usepackage{longtable}          
\usepackage{caption}            
\usepackage{colortbl}           
\usepackage{array}              
\usepackage{multirow}           
\usepackage{natbib}
\usepackage{tabularx}
\usepackage[normalem]{ulem}
\usepackage{graphicx}    
\usepackage{makecell}
\usepackage{algorithm}
\usepackage{algorithmic}
\usepackage{dsfont}

\newcommand{\blue}[1]{#1} 

\newcommand{\zobs}{\vz_{1:n}}
\newcommand{\TV}{\operatorname{TV}}

\usepackage{tikz}
\usetikzlibrary{arrows.meta, positioning}
\usepackage{xcolor} 
\usepackage{enumitem}

\renewcommand{\vz}{z}
\renewcommand{\vtheta}{\theta}
\newcommand{\TabPFN}{\mathrm{TabPFN}}

\graphicspath{{images/}}

\icmltitlerunning{TabMGP: Martingale Posterior with TabPFN}

\begin{document}

\twocolumn[
  \icmltitle{TabMGP: Martingale Posterior with TabPFN}



  \icmlsetsymbol{equal}{*}

  \begin{icmlauthorlist}
    \icmlauthor{Kenyon Ng}{monash}
    \icmlauthor{Edwin Fong}{hku}
    \icmlauthor{David T. Frazier}{monash}
    \icmlauthor{Jeremias Knoblauch}{ucl}
    \icmlauthor{Susan Wei}{monash}
  \end{icmlauthorlist}

  \icmlaffiliation{monash}{Department of Econometrics and Business Statistics, Monash University, Melbourne.}
  \icmlaffiliation{ucl}{Department of Statistical Science, University College London.}
  \icmlaffiliation{hku}{Department of Statistics and Actuarial Science, University of Hong Kong.}

  \icmlcorrespondingauthor{Susan Wei}{susan.wei@monash.edu}

  \icmlkeywords{Machine Learning, ICML}

  \vskip 0.3in
]



\printAffiliationsAndNotice{}  

\begin{abstract}
  Bayesian inference provides principled uncertainty quantification but is often
  limited by the challenges of prior and likelihood elicitation. The martingale
  posterior (MGP) \citep{fong23martingale} offers an alternative by replacing
  these requirements with a predictive rule. In addition, the MGP focuses inference
  on parameters defined through a loss function. This framework is especially
  resonant in the era of foundation transformers; practitioners increasingly
  leverage models like TabPFN for their state-of-the-art capabilities, yet often
  require epistemic uncertainty for a scientific estimand $\theta$ that need not
  parameterise the implicit latent model. The MGP provides a mechanism
  to recover these posterior distributions. We introduce TabMGP, an MGP built on
  TabPFN for tabular data. TabMGP produces credible sets with near-nominal
  coverage and often outperforms both handcrafted MGP constructions and standard
  Bayesian baselines.
\end{abstract}

\section{Introduction}

\begin{figure}[tb]
\centering
\resizebox{\columnwidth}{!}{
\begin{tikzpicture}[
  xscale=1.0, yscale=1.1,
  rollout/.style={font=\small}, 
  shared/.style={text=black, fill=gray!20, rounded corners, inner sep=3pt, minimum width=0.7cm, align=center},
  suffix/.style={text=black, fill=blue!15, rounded corners, inner sep=3pt, minimum width=0.7cm, align=center},
  input/.style={text=black, font=\scriptsize\sffamily, dashed, draw, rounded corners, inner sep=2pt},
  arrow/.style={->, thick, >=Stealth},
  header/.style={align=center, font=\scriptsize\sffamily, anchor=south}
]

\node[header] (h1) at (0.9, 1.1) {Observed\\ $z_{1:n}$};
\node[header] (h2) at (3.8, 1.1) {Forward Sampling via TabPFN \\ $y_{t} \sim \text{TabPFN}(\cdot \mid x_t, z_{1:t-1})$};
\node[header] (h3) at (6.8, 1.1) {Approximate\\ Posterior Sample};

\node[font=\sffamily\scriptsize, anchor=east] at (-0.5, 0) {Rollout 1};
\node[rollout, shared] (z1_1) at (0, 0) {$z_1$};
\node (z1_d) at (0.9, 0) {$\dots$};
\node[rollout, shared] (z1_n) at (1.8, 0) {$z_n$};

\node[input] (x1_1) at (2.9, 0.6) {$x_{n+1}^{(1)}$};
\node[input] (x1_N) at (4.7, 0.6) {$x_{N}^{(1)}$};

\node[rollout, suffix] (s1_1) at (2.9, 0) {$y_{n+1}^{(1)}$};
\node (s1_d) at (3.8, 0) {$\dots$};
\node[rollout, suffix] (s1_N) at (4.7, 0) {$y_{N}^{(1)}$};

\draw[arrow, dashed, thin] (x1_1) -- (s1_1);
\draw[arrow, dashed, thin] (x1_N) -- (s1_N);

\node[font=\small] (theta1) at (6.8, 0) {$\theta(F_N^{(1)})$};
\draw[arrow] (s1_N) -- (theta1) node[midway, above=0.01cm, font=\tiny\sffamily] {solve};

\node[font=\sffamily\scriptsize, anchor=east] at (-0.5, -1.2) {Rollout 2};
\node[rollout, shared] (z2_1) at (0, -1.2) {$z_1$};
\node (z2_d) at (0.9, -1.2) {$\dots$};
\node[rollout, shared] (z2_n) at (1.8, -1.2) {$z_n$};

\node[input] (x2_1) at (2.9, -0.6) {$x_{n+1}^{(2)}$};
\node[input] (x2_N) at (4.7, -0.6) {$x_{N}^{(2)}$};

\node[rollout, suffix] (s2_1) at (2.9, -1.2) {$y_{n+1}^{(2)}$};
\node (s2_d) at (3.8, -1.2) {$\dots$};
\node[rollout, suffix] (s2_N) at (4.7, -1.2) {$y_{N}^{(2)}$};

\draw[arrow, dashed, thin] (x2_1) -- (s2_1);
\draw[arrow, dashed, thin] (x2_N) -- (s2_N);

\node[font=\small] (theta2) at (6.8, -1.2) {$\theta(F_N^{(2)})$};
\draw[arrow] (s2_N) -- (theta2) node[midway, above=0.01cm, font=\tiny\sffamily] {solve};

\node at (0.9, -1.9) {$\vdots$};
\node at (3.8, -1.9) {$\vdots$};
\node at (6.8, -1.9) {$\vdots$};

\node[font=\sffamily\scriptsize, anchor=east] at (-0.5, -2.7) {Rollout $L$};
\node[rollout, shared] (zL_1) at (0, -2.7) {$z_1$};
\node (zL_d) at (0.9, -2.7) {$\dots$};
\node[rollout, shared] (zL_n) at (1.8, -2.7) {$z_n$};

\node[input] (xL_1) at (2.9, -2.1) {$x_{n+1}^{(L)}$};
\node[input] (xL_N) at (4.7, -2.1) {$x_{N}^{(L)}$};

\node[rollout, suffix] (sL_1) at (2.9, -2.7) {$y_{n+1}^{(L)}$};
\node (sL_d) at (3.8, -2.7) {$\dots$};
\node[rollout, suffix] (sL_N) at (4.7, -2.7) {$y_{N}^{(L)}$};

\draw[arrow, dashed, thin] (xL_1) -- (sL_1);
\draw[arrow, dashed, thin] (xL_N) -- (sL_N);

\node[font=\small] (thetaL) at (6.8, -2.7) {$\theta(F_N^{(L)})$};
\draw[arrow] (sL_N) -- (thetaL) node[midway, above=0.01cm, font=\tiny\sffamily] {solve};

\end{tikzpicture}
}
\caption{TabMGP for obtaining posterior samples of
  $\theta(F_\infty) \mid z_{1:n}$, where each $z = (x, y)$ represents a
  covariate-response pair. We perform \textbf{forward sampling} with TabPFN to
  generate $L$ independent continuations of the observed dataset $z_{1:n}$.
  Since TabPFN does not model covariates, the forward sampling of $x_i$ is
  performed via a separate process, while TabPFN provides the conditional
  response $y_i \sim \TabPFN(\cdot \mid x_i, z_{1:i-1})$. At the end of
  each rollout $l$, we form the empirical measure
  $F_N^{(l)} = \frac{1}{N} \left(\sum_{i=1}^{n} \delta_{z_i} + \sum_{i=n+1}^{N} \delta_{z_{i}^{(l)}}\right)$
  and collect $\theta(F_N^{(l)})$ as one \textbf{approximate posterior sample}.}
\label{fig:tabmgp-sampling}
\end{figure}

Classical Bayesian inference provides a principled framework for uncertainty
quantification, but its formulation hinges on the specification of a
prior–likelihood pair. This requirement can be burdensome in practice. Priors
are often placed on uninterpretable parameters, and ``uninformative'' defaults
can dominate inference in data-scarce or high-dimensional regimes
\citep{zhang22bayesian}. The problem is further compounded by the frequent
misspecification of the likelihood in modern applications. Even when a model is
settled upon, approximating the resulting posterior in high
dimensions remains a formidable computational challenge.

These limitations have catalysed a wave of generalisations, including
Bayesian predictive inference (BPI) \citep{fortini25exchangeability},
generalised Bayes (GB) \citep{bissiri16general}, and optimisation-centric Bayes
\citep[see, e.g.,][]{knoblauch22optimizationcentric, wild23rigorous,
  shen25predictioncentric}, collectively marking the emergence of a
\textbf{post-Bayesian paradigm.} Within this paradigm, BPI shifts the focus from
prior and likelihood elicitation to the sequence of predictions themselves.
Specifically, the analyst provides a \textbf{predictive rule}, defined as a
sequence of one-step-ahead predictive distributions $(P_i)_{i \geq 0}$. BPI is
post-Bayesian in the sense that the predictive rule need not correspond to a
standard Bayesian posterior predictive distribution (PPD). Instead, the
framework requires only that the sequence $(P_i)_{i \geq 0}$ converges weakly to
a random limiting measure $F_\infty$ almost surely.

We can then ask for the posterior distribution of $F_\infty$ given data
$z_{1:n}$. In this role, $F_\infty$ represents the
\textbf{latent Bayesian model} internalised by the predictive rule, a state of
knowledge that remains implicit but can be queried through simulation.
Specifically, we may obtain posterior samples of $F_\infty \mid z_{1:n}$ through
a process of \textbf{forward sampling}: starting from the observed dataset
$z_{1:n}$, one repeatedly draws the next observation from the predictive rule
conditional on the data so far, appends it to the dataset, and continues. In
practice, the forward sampling process is terminated at some large $N$ such that
$F_N$ serves as an approximate posterior sample of $F_\infty \mid z_{1:n}$.

Complementary to BPI is GB, a framework that argues we can update a prior to a
posterior distribution over parameters that are connected to observations
through a loss function rather than a likelihood. This allows for significant
flexibility as the object of interest $\theta$ is generally defined as the
minimiser of an expected loss and need not index the underlying Bayesian model.
Despite this flexibility, GB still requires the specification of an explicit
prior.

The martingale posterior (MGP) emerges as a powerful synthesis of these
directions, one that is especially resonant in the era of foundation
transformers \citep{bommasani22opportunities}, which have recently begun to
dominate tabular data tasks---a domain long considered a stronghold of
traditional, non-deep-learning methods. Practitioners confronted with a new
tabular dataset will increasingly turn to pretrained models like TabPFN
\citep{hollmann22tabpfn} rather than design a bespoke likelihood--prior pair and
navigate the complexities of Markov chain Monte Carlo or variational inference.
For these users, the parameters of scientific interest are effectively decoupled
from the internals of the transformer, necessitating a framework that can
recover the posterior distribution of a functional $\theta(F_\infty)$ directly
from the transformer’s predictive outputs. The MGP provides this mechanism
by treating the transformer as a predictive engine for the latent model
$F_\infty$, while the GB loss function specifies the functional $\theta$ to be
extracted.

The forward sampling mechanism above for $F_\infty \mid z_{1:n}$ provides the
computational means for this recovery. The procedure remains the same except
that, at termination, we report $\theta(F_N)$ as an approximate posterior sample
from $\theta(F_\infty) \mid z_{1:n}$. This sampling scheme is well suited to
transformers: \textbf{next-token prediction} \textit{is} the predictive rule,
and forward sampling mirrors \textit{autoregressive generation}, where each
token is sampled from the conditional distribution given the context, appended
to the context, and the process is repeated. This procedure is visualised in
Figure~\ref{fig:tabmgp-sampling}.

Yet, despite this alignment, the martingale posterior literature has focused
almost exclusively on handcrafted predictive rules that satisfy exact martingale
properties
\citep[e.g.,][]{fong23martingale,ghalebikesabi23quasibayesian,huk24quasibayes};
see Section~\ref{sec:related-works} for a review. These approaches are
theoretically rigorous, but they require problem-specific design choices to tune
the predictive rule for the dataset at hand. Moreover, the emphasis on exact
martingale structure overstates this condition's importance. Although the
martingale property is a sufficient condition to guarantee the existence of a
limiting law $F_\infty$, it is \textbf{not necessary} for delivering epistemic
uncertainty for $\theta(F_\infty)$. We argue that the reliance on this
restrictive sufficient condition has impeded the broader adoption of MGPs and
prevents their application to high-capacity models.

This motivates our central question: can modern foundation transformers be used as
predictive rules in martingale posteriors? We answer in the affirmative by
introducing \textbf{TabMGP}, the first martingale posterior powered by a large
pretrained foundation transformer. Specifically, we instantiate the predictive
rule with \textbf{TabPFN} \citep{hollmann22tabpfn,hollmann25accurate}, a
transformer trained on a broad range of synthetic tabular data that now achieves
state-of-the-art results for tabular classification and regression tasks. 

Our results show that TabMGP consistently provides reliable uncertainty
quantification, often matching or surpassing both handcrafted martingale
posteriors and classical Bayesian inference. These findings demonstrate that the
MGP framework is considerably more robust than initially theorised. Even when
high-capacity predictive rules like TabPFN do not strictly satisfy formal
sufficient conditions such as the martingale property, the resulting posteriors
remain empirically stable, providing superior frequentist coverage and tighter
credible sets than standard baselines. By establishing a new method for the
modern Bayesian toolkit, this work opens new avenues for broadening the MGP
framework to the era of foundation models.

\section{Martingale Posterior}
\label{sec:mgp}
The martingale posterior \citep{fong23martingale} is a recently proposed
statistical inference method within the BPI framework \citep[see][for a
review]{fortini25exchangeability}. We follow the notation of
\citet{fortini25exchangeability} in our discussion.

\paragraph{Notation.} Let $(Z_i)_{i \ge 1}$ be random variables taking values in
a space $\sZ$.  For
any probability law $\bP$ for the sequence $(Z_i)_{i \ge 1} \in \sZ^{\infty}$,
define the \emph{predictive rule} as the sequence of predictive distributions
$P_0(\cdot) \coloneq \bP(Z_1 \in \cdot)$ and, for $i \ge 1$,
\begin{equation*}
  P_i(\cdot) \coloneq \bP(Z_{i+1} \in \cdot \mid Z_{1:i}).
\end{equation*}
We write $P_i(\cdot \mid z_{1:i})$ for the realisation given $z_{1:i}$. We also
write $\bP$-a.s.~for ``with $\bP$-probability one''. We use $p_{i}$ to denote
the density or mass function of $P_{i}$.

The martingale posterior requires two ingredients:
\begin{enumerate}[noitemsep, topsep=0pt, leftmargin=*]
  \item A predictive rule $(P_{i} )_{i \ge 0}$; and
  \item A functional of interest
        $\vtheta: \sP(\sZ) \to \Theta \subseteq \R^{p}$, where $\sP(\sZ)$
        denotes the set of all distributions $F$ on $\sZ$.
\end{enumerate}
In this work, we focus on a risk minimiser that minimises a given loss function
$\ell : \sZ \times \Theta \to \R$:
\begin{equation}
  \label{eq:risk-minimiser}
  \vtheta(F) \coloneq \argmin_{\vartheta\in\Theta} \int \ell(\vz, \vartheta) \, \mathrm{d}F(z) \, .
\end{equation}
For the martingale posterior distribution to be well-defined, the sequence of
predictive distributions $(P_i)_{i \ge 0}$ must converge $\bP$-a.s.~to a random
probability measure $F_\infty$, which represents the limiting empirical
distribution of $(Z_{i})_{i \ge 1}$. Here, $F_{\infty}$ is analogous to the
statistical model in classical Bayes, and the probability laws of $F_\infty$ and
$F_\infty \mid z_{1:n}$ are the prior and posterior, respectively. This posterior
law of $F_\infty \mid z_{1:n}$ then induces a posterior distribution on
functionals of $F_\infty$, which is called the martingale posterior.

\paragraph{Definition.} Let $\zobs \in \sZ^n$ denote the observed data and
$\Pi(F_{\infty} \mid \zobs)$ the posterior law of $F_{\infty}$ given $\zobs$.
The \emph{martingale posterior} distribution is defined as the posterior law of
$\vtheta(F_{\infty})$:
\begin{equation}
  \label{eq:mgp}
  \Pi_\infty (\vtheta(F_{\infty}) \in A \mid \zobs)
  = \int \vone(\vtheta(F_{\infty}) \in A) \, \mathrm{d} \Pi(F_{\infty} \mid \zobs) ,
\end{equation}
for all measurable sets $A \subseteq \Theta$. The classical Bayesian posterior
can be recovered from \eqref{eq:mgp} as a special case by using a Bayesian PPD
as the predictive rule and the corresponding negative log-likelihood as the loss
function (Appendix~\ref{sec:mgp-bayes-equiv}).

\paragraph{Sufficient conditions.} A sufficient, but not necessary, condition for
the existence of $F_\infty$ and hence for the martingale posterior distribution
to be well-defined, is that the sequence $( P_i )_{i \ge 0}$ is a martingale, or
equivalently, that the sequence $(Z_i)_{i \ge 1}$ is conditionally identically
distributed (c.i.d.) \citep{berti04limit}. The martingale property stipulates
that $( P_i )_{i \ge 0}$ satisfies, for every $i \ge 0$ and every measurable set
$A$:
\begin{equation}
  \label{eq:martingale-cond}
  \E[ P_{i+1}(A) \mid Z_{1:i} ] = P_i(A).
\end{equation}
Relaxations of this sufficient condition remain an active area of research
\citep[see, e.g.,][]{battiston25bayesian}. As noted in the introduction, the
term ``martingale posterior'' is somewhat of a misnomer. Any predictive rule
with a well-defined $F_\infty$ limit induces a posterior distribution as
in~\eqref{eq:mgp}. In this paper, we use the term broadly, so even
predictive rules that violate the martingale property may still induce valid
``martingale'' posteriors.

\paragraph{Computation.} The martingale posterior \eqref{eq:mgp} generally does not
admit a closed-form expression and is approximated by the \emph{finite
  martingale posterior}. This is the posterior law of $\theta(F_{N})$, where
$F_{N}$ is the empirical distribution of $Z_{1:N}$. A draw from $\theta(F_{N})$
is obtained by \emph{forward sampling} a simulation rollout $Z_{n+1:N}$ from the
predictive rule:
\begin{equation*}
  Z_{i+1} \sim P_i, \quad i = n, \dots, N-1,
\end{equation*}
then forming $F_N$ and computing $\theta(F_N)$ \citep{fong23martingale}.
Repeating this produces $L$ samples $\{\theta^{(l)}\}_{l=1}^{L}$. These rollouts
are independent and thus embarrassingly parallel.

\section{TabMGP}
\label{sec:tabmgp}
Given the strong predictive performance of modern foundation models
\citep{bommasani22opportunities}, it is natural to ask whether such models can
serve as a predictive rule in the martingale posterior framework. We describe
our proposed methodology, which employs TabPFN
\citep{hollmann22tabpfn,hollmann25accurate} as the predictive rule.

Given that TabPFN is designed for supervised prediction, we specialise the
martingale posterior framework to the supervised setting, where each observation
is a feature–response pair $z = (x,y)$. In this setting, a predictive rule
generates future pairs $Z_{n+1}, \ldots, Z_N$. \citet{fong23martingale} describe
this as the ``joint method,'' which requires modelling both the distribution of
$X$ and the conditional distribution $Y \mid X$. Because modelling $X$ can be
challenging even in moderate dimensions, we follow \citet{fong23martingale} and
use the Bayesian bootstrap (see Section~\ref{sec:related-works}) to forward
sample features, drawing $X_{i+1}$ at random from the empirical distribution of
$x_{1:i}$. In our work, we adopt this strategy for $X$, while TabPFN supplies
the conditional $Y \mid X$.

\subsection{Proposed Methodology}
We propose instantiating the martingale posterior with the predictive rule
induced by TabPFN through in-context learning. TabPFN is a
\emph{prior-data-fitted network} \citep{muller22transformers}, explicitly trained to
approximate a Bayesian PPD by sampling training data from exchangeable sequences
defined by likelihood--prior pairs; see Appendix~\ref{sec:tabpfn-review} for a
review. We term the resulting martingale posterior \textbf{TabMGP}.

Several distinctive features of TabPFN facilitate its integration as a
predictive rule within the martingale posterior framework. Through in-context
learning, TabPFN consumes the context alongside a new query to return a
one-step-ahead predictive distribution without the need for retraining or
fine-tuning. This allows the model to adapt to novel datasets instantaneously,
handling classification via deterministic probability vectors and regression
through a multiclass classification approach that bins the response variable.

The architecture further benefits from native row-permutation invariance,
ensuring that predictions depend solely on the multiset rather than the sequence
order. This built-in property provides a significant computational advantage
over many existing martingale posteriors, such as copula-based constructions
\citep{fong23martingale}, which must manually enforce invariance by averaging
over permutations. Finally, as a prior-data-fitted network, TabPFN is designed
to be approximately Bayesian, having been trained to approximate a
  Bayesian PPD. In the martingale posterior framework, the Bayesian PPD serves
as the ideal predictive rule; when the two coincide, the martingale
posterior recovers the classical Bayesian posterior. By training on synthetic
data from a vast array of likelihood--prior pairs, TabPFN is encouraged to
approximate this ideal behaviour across a highly diverse range of tasks.

\paragraph{Why not other transformers?}
Large language models and other pretrained transformers often exhibit
in-context learning: given a context $z_{1:i}$ and query $x_{i+1}$, they can
produce a prediction for $y_{i+1}$. In this sense, they share the same basic
mechanism as TabPFN. However, they lack the two further properties that make
TabPFN especially well-suited as a predictive rule for martingale posteriors.

First, they are not row-permutation invariant. Since such models rely on
positional embeddings, their predictions change with the ordering of training
examples, unless explicit averaging is performed across permutations. Second,
they are not trained to approximate a Bayesian PPD: their training
objectives are tailored to next-token text prediction rather than calibrated
probabilistic prediction for supervised tabular tasks. These differences mean
that, although generic transformers can perform in-context learning, their
predictive distributions are not naturally aligned with the martingale posterior
framework. By contrast, TabPFN combines in-context learning with built-in
permutation invariance and Bayesian alignment, making it a more principled
choice.

\begin{algorithm}[tb]
  \caption{TabMGP}
  \label{alg:tabpfn-credible}
  \begin{algorithmic}[1]
    \REQUIRE Observed data $\vz_{1:n}$, loss function $\ell(\vz, \vtheta)$,
    forward sampling depth $N$, predictive rule
    $\TabPFN(\cdot \mid x_{i+1}, \vz_{1:i})$, number of posterior samples $L$

    \FOR{$l = 1$ to $L$}
    \STATE Set $\vz_{1:n}^{(l)} \gets \vz_{1:n}$.
    \FOR{$i = n$ to $N-1$}
    \STATE Sample $x_{i+1}^{(l)} \gets \operatorname{Empirical}(x_{1:i}^{(l)})$.
    \STATE Sample $y_{i+1}^{(l)} \gets \TabPFN(\cdot \mid x_{i+1}^{(l)}, \vz_{1:i}^{(l)})$.
    \STATE Form $\vz^{(l)}_{i+1} \gets (x^{(l)}_{i+1}, y^{(l)}_{i+1})$.
    \ENDFOR
    \STATE Compute $\vtheta^{(l)} \gets \argmin_{\vtheta} \sum_{i=1}^{N} \ell(\vz_{i}^{(l)}, \vtheta)$.
    \ENDFOR

    \STATE Return $\{\vtheta^{(1)}, \dots, \vtheta^{(L)}\}$ as samples from the martingale posterior.
  \end{algorithmic}
\end{algorithm}

\subsection{Validity of TabMGP}
A central question for any MGP construction is whether the chosen predictive
rule leads to a well-defined limiting law $F_\infty$. While the framework
traditionally points to conditions like the martingale property
\citep{fong23martingale} or the \textit{almost conditionally identically
  distributed} (a.c.i.d.) condition \citep{battiston25bayesian} as sufficient
guarantees, verifying these conditions for high-capacity models like TabPFN is currently
impossible with existing tools in deep learning theory. Preliminary visual
inspections of the predictive rule associated with TabPFN suggest there may be
subtle deviations from these sufficient conditions
(see Appendix~\ref{sec:tabmgp-validity}). However, we argue that such properties are
sufficient rather than necessary for delivering useful epistemic uncertainty.

\begin{figure*}
  \centering
  \includegraphics[width=\linewidth]{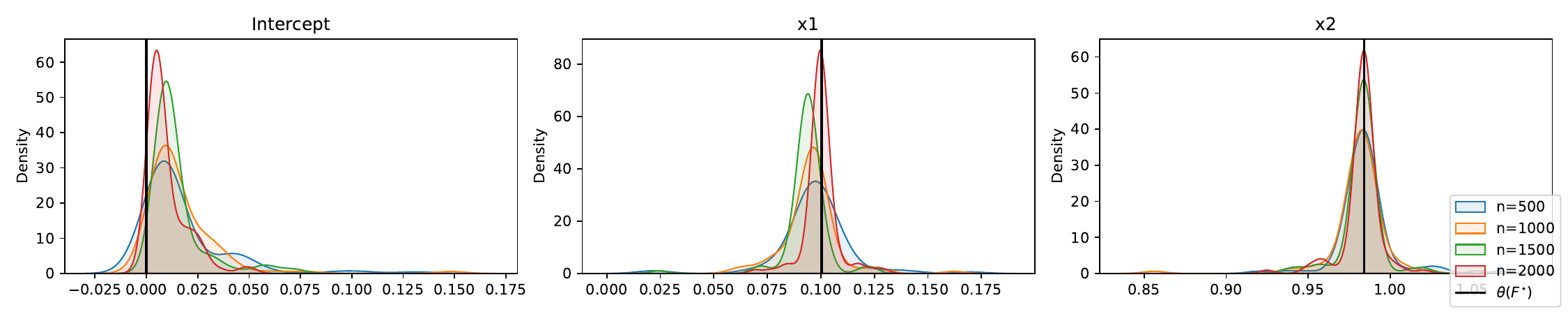}
  \caption{Concentration of TabMGP. The black vertical line indicates the
    population risk minimiser
    $\theta(F^{\star}) = (0, {\beta^{\star}}^{\top})$. See Appendix~\ref{sec:tabmgp-validity} for experimental details.}
  \label{fig:tabmgp-concentration}
\end{figure*}

In the absence of formal theoretical guarantees, we rely on empirical
diagnostics to assess the behaviour of TabMGP:

\begin{itemize}[nosep,leftmargin=*,labelsep=0.5em]
  \item \textbf{Path Stability (Scaled $L_1$ Convergence):} The stability of the
        functional $\theta(F_N)$ as the rollout length $N$ increases is the
        primary diagnostic for the existence of a stable limit. We monitor this
        convergence by tracking the expected $L_1$-norm between the initial
        empirical risk minimiser $\theta(F_n)$ and the terminal estimate
        $\theta(F_N)$, scaled by the parameter dimension $p$:
        $\mathbb{E}_{F_N} [ \frac{1}{p} \lVert \theta(F_n) - \theta(F_N) \rVert_1 ]$.
        If the transformer is internalising a consistent law, these trajectories
        should stabilise at a non-zero constant as $N$ grows. Systematic drift
        \blue{or divergent $L_1$ paths would indicate that
        $\theta(F_{\infty})$ is ill-defined and the resulting finite-$N$
        posterior should be treated cautiously. In practice, we also use this to
        determine the minimum $N$ needed for convergence.}

  \item \textbf{Frequentist Coverage Assessment:} We use frequentist coverage as
        an empirical check on the relationship between the transformer's
        predictive rule and the population~$F^\star$. A $(1-\alpha)$ credible
        set should contain the population risk minimiser $\theta(F^\star)$ at a
        rate of at least $1-\alpha$.

  \item \textbf{Posterior Contraction:} A valid epistemic uncertainty measure
        must represent uncertainty that vanishes as the amount of data
        increases. One can verify this behaviour by observing the width of the
        posterior of $\theta(F_N)$ as the observed sample size $n$ increases. A
        sensible predictive rule should result in a posterior narrowing towards
        the population risk minimiser $\theta(F^\star)$ as evidence accumulates,
        reflecting a reduction in epistemic uncertainty
        (Figure~\ref{fig:tabmgp-concentration}).
\end{itemize}

By focusing on these observed behaviours, practitioners can assess the outputs
of TabMGP based on its empirical consistency, bypassing the currently
insurmountable challenge of proving the transformer's limiting properties.

\paragraph{Implications of non-convergence of $F_N$.} \blue{If $F_N$ fails to
  converge, $\theta(F_N)$ may oscillate or diverge. This implies that the
  martingale posterior, which is defined as the distribution of
  $\theta(F_\infty)$, is ill-defined. Even when $\theta(F_\infty)$ does exist,
  practical implementation requires stopping at a finite $N$, meaning the
  quality of the approximation hinges on the proximity of $\theta(F_N)$ to
  $\theta(F_\infty)$. Our path stability diagnostic
  (Appendix~\ref{sec:diagnostics}) is designed to detect this. If the $L_1$
  trajectory has not plateaued, practitioners should increase $N$.}


\section{Related Work}
\label{sec:related-works}


The martingale posterior framework admits many possible predictive rules, and a
growing literature has explored different constructions. We review these
constructions below, since they form the competing methods for TabMGP in Section
\ref{sec:experiments}. In parallel, a separate line of work has asked whether
the predictive behaviour induced by transformer in-context learning satisfies
martingale properties. This is especially relevant here as TabPFN is a
transformer foundation model whose in-context learning defines our predictive
rule.

Recent analyses \citep{falck24incontext,ye24exchangeable,nagler25uncertainty}
show that the predictive rules induced by foundation models, such as TabPFN,
need not satisfy the classical martingale property. Building on this line of
work, we take the further step of assessing the more flexible almost
conditionally identically distributed (a.c.i.d.) condition
\citep{battiston25bayesian} and observe that TabPFN departs from this
requirement as well. Nevertheless, we find that the resulting TabMGP exhibits
consistent path stability and near-nominal coverage in practice. This suggests
that the usual sufficient conditions are stricter than necessary and that
validity can be achieved under weaker assumptions than previously
theorised.

\blue{
Employing TabPFN as the predictive rule in the martingale posterior framework
has also been considered in \citet{nagler25uncertainty} for conditional
inference (whereas we focus on unconditional inference for loss-defined functionals).
However, their finding that TabPFN is not a martingale leads them to abandon it
as a predictive rule after the initial step, instead using its output to
initialise a copula-based update. This approach preserves martingale validity
but also inherits the drawbacks of copula-based updates (e.g., fragile bandwidth
tuning). By contrast, we retain TabPFN as the predictive rule throughout and
interpret its departure from both the martingale and a.c.i.d.\ conditions as
evidence that validity can be achieved under weaker assumptions.
}

A concurrent effort by \citet{fortini26principled} focuses on posterior
inference for $F_{\infty} \mid z_{1:n}$ using a predictive central limit theorem
(CLT). However, their framework does not incorporate the generalised Bayes
flavour of inference for a parameter $\theta$ defined by a user-specified loss
function. The predictive CLT has distinct advantages in supervised settings, and
extending the predictive CLT to functionals $\theta(F_{\infty})$ would be an
interesting direction for future work.

We now give a brief summary of the main predictive rules developed in the
martingale posterior literature.

\paragraph{Bayesian bootstrap.}
The \emph{Bayesian bootstrap}
\citep{rubin81bayesian} is a predictive rule where future observations
$Z_{i+1}$ are uniformly drawn from the observations up to $i$:
\begin{equation*}
  Z_{i+1} \sim F_{i}, \quad i \ge n,
\end{equation*}
with $F_{i}$ the empirical distribution of $z_{1:i}$. As this rule is discrete,
its use is recommended only when the functional of interest arises from a
parametric likelihood loss, for which the smoothness of $F_{\infty}$ is
irrelevant \citep{fong23martingale, dellaporta22robust}.

\paragraph{Newton's algorithm.}
Rooted in Bayesian nonparametrics, \emph{Newton's algorithm}
\citep{newton98nonparametric} was initially developed for unsupervised
sequential learning in mixture models. It was subsequently re-interpreted as a
predictive rule that generates a c.i.d.~sequence of observations
\citep{fortini20quasibayes}. This approach, however, is not suitable for
high-dimensional $\theta$, as it requires numerical integration of an
intractable normalising constant.

\paragraph{Copula-based updates.} For applications requiring smoothness of
$F_{\infty}$, the mainstream implementation of the martingale posterior is
dominated by \emph{copula-based updates}. Introduced by \citet{hahn18recursive}
and adopted as the default predictive rule in \citet{fong23martingale}, these
updates smooth $F_{N}$ by attaching a bivariate copula to each new observation
while preserving the martingale property of the predictive rule exactly.
Subsequent work has extended this framework with autoregressive Gaussian-process
likelihoods for higher-dimensional features
\citep{ghalebikesabi23quasibayesian}, direct quantile updates
\citep{fong25bayesian}, log-concave shape constraints \citep{cui26martingale},
and vine copulas for very high-dimensional features \citep{huk24quasibayes}. All
these variants preserve the exact martingale guarantee of the copula framework,
but each introduces at least one smoothing or dependence hyperparameter, whose
value controls the spread of the resulting posterior. These hyperparameters are
generally difficult to tune and must be retuned for each new dataset.


\paragraph{Parametric plug-in predictives.} For parametric inference,
\citet{walker22chapter} and \citet{holmes23statistical} proposed a predictive
rule of the form $Z_{i+1} \sim p(\cdot \mid \widehat{\theta}_{i})$, where
$p(\cdot \mid \theta)$ is an assumed parametric model and $\widehat{\theta}_{i}$
is a point estimate (e.g., maximum likelihood estimate) computed from all
previous observations $z_{1:i}$. The asymptotic properties of this predictive
rule are studied in \citet{fong26asymptotics}. However, it requires that the
assumed parametric model provides a reasonable approximation to the data
distribution in order to yield sensible uncertainty quantification.

\paragraph{Permutation-equivariant neural network.}
Martingale posteriors have also been explored in the context of neural processes
\citep{lee23martingale}, where a permutation-equivariant neural network is used
to construct a martingale posterior over latent variables. While this approach
demonstrates the potential of using neural networks within a martingale
posterior framework, the focus is still on prediction rather than inference, and
it remains unclear how to generalise to broader statistical inference tasks.

\section{Experiments}
\label{sec:experiments}

We compare TabMGP with other martingale posterior constructions using synthetic
and real-world data. Our experiments focus on small- to moderate-$n$ settings
where the prior specification of the classical Bayesian posterior is important,
making the MGP particularly well suited.

To evaluate a given posterior $\Pi(\cdot \mid \vz_{1:n})$, we compute the
\emph{coverage} and ``size'' of its joint $(1-\alpha)$ credible set
$C_{1-\alpha}(\vz_{1:n}) \subset \Theta$:
\begin{equation*}
  \Pr_{\vtheta \sim \Pi(\cdot \mid \vz_{1:n})}\bigl( \vtheta \in C_{1-\alpha}(\vz_{1:n}) \bigr) = 1-\alpha.
\end{equation*}
Although the set can be constructed in different ways, an ellipsoidal
approximation is often used for computational efficiency; see
Appendix~\ref{sec:hpd-credible-set} for a discussion.

The coverage of $C_{1-\alpha}(\vz_{1:n})$ is defined as
\begin{equation}
  \label{eq:coverage}
  \Pr_{\vz_{1:n} \sim F^{\star} } \bigl( \vtheta(F^{\star}) \in C_{1-\alpha}(\vz_{1:n}) \bigr),
\end{equation}
namely, the probability with which the credible set contains the population risk
minimiser,
$\vtheta(F^{\star}) = \arg\min_{\vtheta} \int \ell(\vz, \vtheta)\, \mathrm{d}F^{\star}(\vz)$,
under repeated draws of $z_{1:n}$ from the true data-generating distribution
$F^{\star}$. Although frequentist in nature, this metric is commonly used for
comparing Bayesian posteriors.


\paragraph{Posterior constructions.} We compare TabMGP against the martingale
posteriors suggested in \citet{fong23martingale}: Bayesian bootstrap
(\textit{BB}) and the bivariate copula update (\textit{Copula}). We also include
two non-martingale baselines built from the Gaussian centred at the loss
minimiser $\widehat{\vtheta}_{n}$ and scaled by the inverse-Hessian of the loss:
\textit{Asymptotic} uses this Gaussian directly as the credible set, while
\textit{Bayes} uses it as the prior in a posterior under a parametric likelihood
matched to the loss. The model choice is discussed later in this section when
introducing the loss functions. The exact computation details of the posteriors
are provided in Appendix~\ref{sec:posterior-const}. \blue{Additional coverage
  experiments using alternative posterior constructions are reported in
  Appendix~\ref{sec:coverage-extra}. These include a standard Bayesian posterior
  with a diffuse $\sN(0,10^2)$ prior and a TabPFN-initialised copula baseline
  adapted from \citet{nagler25uncertainty}.}

\paragraph{Loss functions.} We focus exclusively on interpretable linear
  models. We use negative log-likelihood as the loss in
\eqref{eq:risk-minimiser} and use the corresponding model for the Bayes posterior.
For continuous responses (linear regression), we use
$\ell(x, y, \vtheta) = (y - [1 \ x^{\top}] \vtheta)^{2}$, which is the negative
log-likelihood of $\sN([1 \ x^{\top}] \vtheta, 1)$. For $K$-class categorical
responses (logistic regression), we use
$\ell(x, y, \vtheta) = - \log \Pr(y = k)$, where $\Pr(y = k)$ is the $k$-th
entry of the softmax applied to the logits
$([1 \ x^{\top}] \vtheta_{1}, \ldots, [1 \ x^{\top}] \vtheta_{K})$, and
$\vtheta_{1}, \ldots, \vtheta_{K}$ are the class-specific coefficients. We also
set appropriate constraints on the coefficients to ensure identifiability in the
loss function; see Appendix~\ref{sec:identifiable-loss} for details.

\paragraph{Data-generating setups.} We describe the true data-generating
distribution $F^{\star}$ used here. For each $F^{\star}$, we draw 100
independent datasets $\{z_{1:n}^{(r)}\}_{r=1}^{100}$ from $F^{\star}$. We refer to
$F^{\star}$ together with the procedure used to generate $z_{1:n}$ as a
\emph{setup}. The sample size $n$ is chosen to be as small as possible without
rendering the design matrix ill-conditioned. In total, we have 30 setups (11
synthetic, 19 real-data).

Let $\beta^{\star}$ be a vector of regression coefficients and let $x_{i}$ be
sampled i.i.d.~from the hypercube $[-1, 1]^{10}$. For the synthetic setups, we generate
$y \mid x$ with:
\begin{enumerate}[noitemsep, topsep=0pt, leftmargin=*]
\item \label{item:dgp-linreg} $y_{i} = x^{\top}_{i} \beta^{\star} + \epsilon_{i}$, where
        $\epsilon_{i} \overset{iid}{\sim} \sN(0,1)$;
  \item $y_{i} = x^{\top}_{i} \beta^{\star} + \epsilon_{i}$, where $\epsilon_{i}$ are
        i.i.d.~draws from Student-$t$ distributions with 5, 4, or 3 degrees of freedom;
  \item $y_{i} = x^{\top}_{i} \beta^{\star} + \epsilon_{i}$, where $\epsilon_{i}$ has
        heteroscedastic variance that depends on $x_{i}$. We use three strengths
        of heteroscedasticity: $s_{1}$, $s_{2}$, $s_{3}$, from weak to strong;
  \item \label{item:dgp-logreg}
        $y_{i} \mid x_{i} \overset{\mathrm{iid}}{\sim} \mathrm{Bernoulli}(L(x^{\top}_{i} \beta^{\star}))$,
        with a logistic link function $L$;
  \item
        $y_{i} \mid x_{i} \overset{\mathrm{iid}}{\sim} \mathrm{Bernoulli}(L(x^{\top}_{i} \beta^{\star}))$,
        where $L$ is the distribution function of a Gaussian mixture model
        (GMM). We consider three choices of location parameters: GMM$(0)$,
        GMM$(-1)$, and GMM$(-2)$, where the argument denotes the mean of one
        mixture component.
\end{enumerate}
For the 19 real-data setups, we use datasets from the OpenML and UCI
repositories (9 continuous response, 8 binary response, and 2 multiclass
categorical response). As $F^{\star}$ is unknown in the real-data setups, we
take the empirical distribution of the entire dataset as the truth
$\widetilde{F}^{\star}$. The details of all data-generating setups are given in
Appendix~\ref{sec:data-setups}. \blue{We also include experiments that combine
  real covariates with known nonlinear response mechanisms, providing a
  complementary check between the fully synthetic and real-data regimes; see
  Appendix~\ref{sec:coverage-semi-synthetic-mlp}.}

\paragraph{Metrics for comparing posteriors.}
For each posterior, we construct joint credible sets
$\widehat{C}_{1-\alpha}(z_{1:n})$ and compare their coverage~\eqref{eq:coverage}
and ``size''. Our approach to constructing $\widehat{C}_{1-\alpha}(z_{1:n})$ is
described in Appendix~\ref{sec:empirical-credible-set}. We consider a posterior
more reliable and meaningful when its credible set is small while achieving
coverage at least $1 - \alpha$.

In the synthetic setups, the coverage is given by
$\frac{1}{100} \sum^{100}_{r=1} \vone \bigl[ \theta(F^{\star}) \in \widehat{C}_{1-\alpha}(z^{(r)}_{1:n}) \bigr]$,
where $\{z_{1:n}^{(r)}\}_{r=1}^{100}$ are 100 replicates of $z_{1:n}$ drawn from
$F^{\star}$. In the real-data setups, as $F^{\star}$ is unknown, we take the
empirical distribution of the entire dataset $\sD$ as the truth
$\widetilde{F}^{\star}$, and compute the coverage on
$\theta(\widetilde{F}^{\star}) = \arg\min_{\theta} \sum_{z \in \sD} \ell(z, \theta)$.
That is, the coverage is
$\frac{1}{100} \sum^{100}_{r=1} \vone \bigl[ \theta(\widetilde{F}^{\star}) \in \widehat{C}_{1-\alpha}(z^{(r)}_{1:n}) \bigr]$,
where $z^{(r)}_{1:n}$ denotes the $r$-th random draw (without replacement) from
$\sD$; see Appendix~\ref{sec:data-setups} for details.

We use the trace of the posterior covariance matrix as a proxy for the
``size'' of a credible set.

\subsection{Results}

\paragraph{Convergence of $\theta(F_{N})$.} To conduct statistical inference, we
require $\theta(F_{N})$ to converge $\bP$-a.s.\ as $N \to \infty$. In practice,
for a fixed observed dataset $z_{1:n}$, we approximate this limit by drawing $T$
additional forward-sampling steps, so that the rollout length is $N=n+T$. This
convergence is commonly assessed using trace plots of each component of
$\theta(F_{N})$, but this approach quickly becomes impractical in high
dimensions. Instead, we track the expected $L_{1}$-norm of $\theta(F_{N})$
relative to the empirical risk minimiser $\theta(F_{n})$ as $T$ increases; see
Appendix~\ref{sec:diagnostics} for the precise definition. For a convergent
$\theta(F_{N})$, this diagnostic should stabilise at a non-zero constant as $T$
increases.

\begin{figure}[tb]
  \centering
  \includegraphics[width=\linewidth,draft=false]{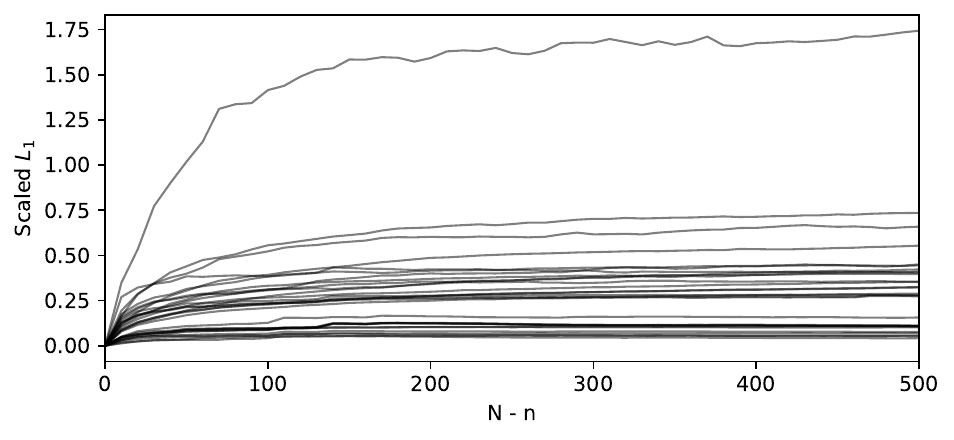}
  \caption{Expected $L_{1}$-norm between $\theta(F_{N})$ from TabMGP and
    $\theta(F_{n})$ as $N$ increases. Each of the 30 trajectories corresponds to
    a realisation of $z_{1:n}$ from one setup.}
  \label{fig:convergence-l1-all-data-mean}
\end{figure}

\blue{We show these trajectories in
Figure~\ref{fig:convergence-l1-all-data-mean}, where each trajectory corresponds
to one of the 30 setups. For completeness, we also report the $L_{1}$-norm
between each realisation of $\theta(F_{N})$ and $\theta(F_{n})$ in
Figure~\ref{fig:convergence-l1-all-data-full}. In both
Figure~\ref{fig:convergence-l1-all-data-mean}
and~\ref{fig:convergence-l1-all-data-full}, the diagnostic is computed from a
single realisation of $z_{1:n}$, reflecting the practical setting in which an
analyst has access to only one observed dataset. We also report the same
diagnostic over 20 independent realisations of $z_{1:n}$ for each setup in
Figure~\ref{fig:convergence-l1-20reps}.}

\blue{In general, we observe that $T=500$ forward-sampling steps is sufficient
  for convergence in most setups. For slower-converging setups, we use longer
  forward-sampling runs and verify convergence by $T=1000$; see
  Figure~\ref{fig:convergence-l1-extra-long-full}. Unless stated otherwise, all
  subsequent results use $T=500$.}

\begin{figure}[tb]
  \centering
  \includegraphics[width=\linewidth,draft=false]{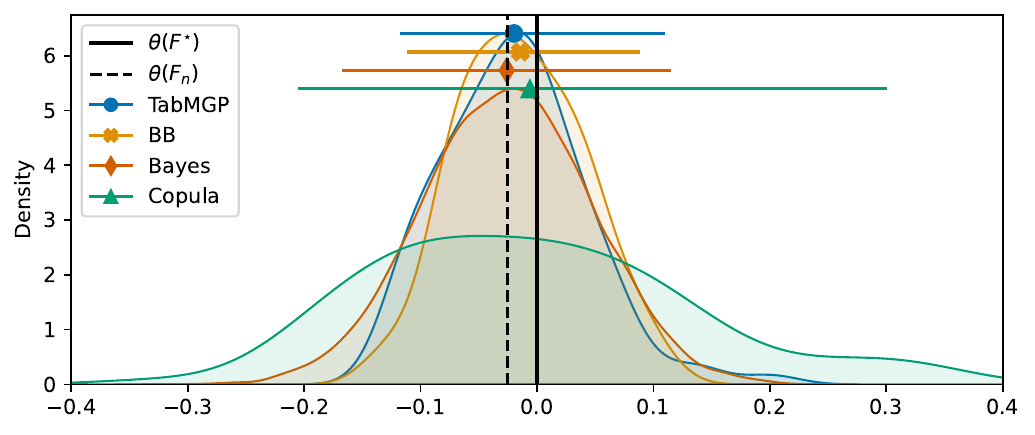}
  \caption{Posterior densities of the intercept in the `concrete' setup. For
    each density, the 95\% marginal credible interval is shown as a horizontal
    bar, and the posterior mean is marked. The solid and dashed vertical lines
    correspond to $\theta(F^{\star})$ and $\theta(F_{n})$, respectively. }
  \label{fig:concrete-intercept-kde}
\end{figure}

\paragraph{Shape of the posteriors.} We present the marginal densities of the
posteriors for a realisation of $z_{1:n}$ in
Figure~\ref{fig:concrete-intercept-kde} (intercept of the `concrete' setup),
with the extra dimensions and setups provided in
Appendix~\ref{sec:marginal-density}. In general, the posterior means are similar
across methods, but TabMGP often exhibits skewness and multimodal structure
compared with the other posteriors, which are typically Gaussian-like.

\paragraph{Joint credible sets for linear regression.}
The results for synthetic data (top and middle) and real-data (bottom) setups
are shown in Table~\ref{tab:coverage-linear-reg}. In the real-data setups, the
rows are ordered by decreasing $n/p$, with $p \coloneq \dim(\theta)$.

TabMGP is competitive in both coverage and credible set size, performing
particularly well on real data. Notably, it can attain nearly 100\% coverage in
many setups while maintaining small credible sets. Similarly, Copula provides
good coverage with small credible sets, although its performance is less
consistent across real-data setups (e.g., severe undercoverage in
\texttt{kin8nm} and overcoverage in \texttt{quake}). This is unsurprising, as
bespoke Gaussian-copula-based predictive rules have an advantage in Gaussian (or
near-Gaussian) setups but can be sensitive to departures from that structure.
Bayes and Asymptotic produce excessively large credible sets in all synthetic
setups, as both rely on asymptotic approximations that are less accurate in
low-$n$ settings. They are also more susceptible to model misspecification and
tend to underperform relative to TabMGP. In contrast, BB tends to produce overly
small sets with poor coverage in the synthetic setups (top and middle) and in
low $n/p$ settings (last few rows of Table~\ref{tab:coverage-linear-reg}). This
reflects limited variability in forward samples when bootstrapping from a small
$z_{1:n}$ (e.g., $n=20$ in the synthetic data setups).

\begin{table}
  \centering
  \scriptsize
  \setlength{\tabcolsep}{3pt}
  \caption{Experiments on linear regression: i.i.d.~errors (top), dependent
    errors (middle), and real data (bottom). We compare TabMGP (ours), BB,
    Copula, Bayes and Asymptotic. The rows corresponding to real data are sorted
    by $n/p$ ratio, i.e., bottom rows are harder. Reported values are coverage
    (Rate) and the median trace of the posterior covariance (Size), computed
    over 100 repetitions. Target coverage is 0.95.}
  \label{tab:coverage-linear-reg}
\begin{tabular}{lcccccccccc}
\toprule
\multicolumn{1}{c}{ } & \multicolumn{2}{c}{TabMGP} & \multicolumn{2}{c}{BB} & \multicolumn{2}{c}{Copula} & \multicolumn{2}{c}{Bayes} & \multicolumn{2}{c}{Asymptotic} \\
\cmidrule(l{3pt}r{3pt}){2-3} \cmidrule(l{3pt}r{3pt}){4-5} \cmidrule(l{3pt}r{3pt}){6-7} \cmidrule(l{3pt}r{3pt}){8-9} \cmidrule(l{3pt}r{3pt}){10-11}
Setup & Rate & Size & Rate & Size & Rate & Size & Rate & Size & Rate & Size\\
\midrule
$\sN(0, 1)$ & 1.00 & 0.45 & 0.55 & 0.09 & \textbf{0.99} & \textbf{0.35} & 1.00 & 0.65 & 1.00 & 1.31\\
$t_5$ & 1.00 & 0.49 & 0.65 & 0.11 & \textbf{0.99} & \textbf{0.37} & 1.00 & 0.65 & 1.00 & 1.31\\
$t_4$ & 0.99 & 0.51 & 0.64 & 0.12 & \textbf{0.98} & \textbf{0.37} & 0.99 & 0.65 & 1.00 & 1.31\\
$t_3$ & 1.00 & 0.48 & 0.66 & 0.14 & \textbf{0.97} & \textbf{0.35} & 0.98 & 0.65 & 0.98 & 1.31\\
\midrule
$s_1$ & 1.00 & 0.37 & 0.50 & 0.03 & \textbf{1.00} & \textbf{0.36} & 1.00 & 0.65 & 1.00 & 1.31\\
$s_2$ & \textbf{1.00} & \textbf{0.36} & 0.46 & 0.02 & 1.00 & 0.36 & 1.00 & 0.65 & 1.00 & 1.31\\
$s_3$ & \textbf{1.00} & \textbf{0.33} & 0.53 & 0.02 & 1.00 & 0.37 & 1.00 & 0.65 & 1.00 & 1.31\\
\midrule
concrete & 0.91 & 0.06 & 0.80 & 0.05 & 1.00 & 0.12 & 0.87 & 0.05 & \textbf{1.00} & \textbf{0.10}\\
quake & \textbf{1.00} & \textbf{0.03} & 0.95 & 0.07 & 1.00 & 0.18 & 0.91 & 0.04 & 0.98 & 0.09\\
airfoil & 0.96 & 0.08 & 0.93 & 0.05 & 0.97 & 0.11 & \textbf{0.96} & \textbf{0.06} & 1.00 & 0.12\\
energy & \textbf{1.00} & \textbf{0.04} & 0.80 & 0.01 & 1.00 & 0.06 & 1.00 & 0.07 & 1.00 & 0.14\\
fish & 0.90 & 0.10 & 0.79 & 0.07 & \textbf{0.95} & \textbf{0.10} & 0.91 & 0.08 & 1.00 & 0.16\\
kin8nm & \textbf{0.95} & \textbf{0.13} & 0.77 & 0.09 & 0.63 & 0.08 & 0.84 & 0.11 & 0.99 & 0.22\\
auction & \textbf{0.99} & \textbf{0.66} & 0.86 & 0.35 & 0.97 & 1.44 & 1.00 & 0.81 & 1.00 & 1.65\\
grid & \textbf{0.97} & \textbf{0.13} & 0.72 & 0.08 & 0.88 & 0.10 & 0.97 & 0.16 & 1.00 & 0.31\\
abalone & \textbf{0.97} & \textbf{0.95} & 0.66 & 0.69 & 0.94 & 0.98 & 0.84 & 0.81 & 0.98 & 1.64\\
\bottomrule
\end{tabular}
\end{table}

\paragraph{Joint credible sets for logistic regression.}
The results for synthetic (top) and real-data (middle and bottom) setups are
shown in Table~\ref{tab:coverage-logistic-reg}. Copula was not applied to
multinomial logistic regression (bottom) as the implementation of
\citet{fong23martingale} is not applicable. In the real-data setups, the rows
are ordered by decreasing $n/p$ within each section.

In the synthetic setups, the Asymptotic credible sets achieve the best coverage
and size, as expected, since the asymptotic approximation is reasonably accurate
at the larger sample size ($n=100$). TabMGP, however, shows the most consistent
performance in the real-data setups, often achieving good coverage with the
smallest credible sets. TabMGP performs particularly well relative to competing
methods in the `harder' setups, i.e., the bottom rows with low $n/p$ ratios.
Asymptotic also provides reasonable coverage in the real-data setups, although
its credible sets are typically much wider than those of TabMGP. This is likely
due to model misspecification, which results in miscalibrated credible sets. BB
performs reasonably well in binary logistic regression (top and middle), but its
credible sets are larger than those of TabMGP. However, BB severely undercovers
in multinomial logistic regression (bottom), where $p$ is much larger than for
the other logistic regression setups. Copula performs poorly in logistic
regression, producing overly large credible sets. This suggests that Copula is
sensitive to the choice of bandwidth, as the same value yields good performance
in the linear regression experiment.

\begin{table}
  \centering
  \scriptsize
  \setlength{\tabcolsep}{3pt}
  \caption{Experiments on logistic regression: synthetic data with various link
    functions (top), real data with binary logistic regression (middle), and
    real data with multinomial logistic regression (bottom). The rows
    corresponding to real data are sorted by $n/p$ ratio, i.e., bottom rows are
    harder. We compare TabMGP (ours), BB, Copula, Bayes and Asymptotic. Reported
    values are coverage (Rate) and the median trace of the posterior
    covariance (Size), computed over 100 repetitions. Target coverage is 0.95.}
  \label{tab:coverage-logistic-reg}
\begin{tabular}{lcccccccccc}
\toprule
\multicolumn{1}{c}{ } & \multicolumn{2}{c}{TabMGP} & \multicolumn{2}{c}{BB} & \multicolumn{2}{c}{Copula} & \multicolumn{2}{c}{Bayes} & \multicolumn{2}{c}{Asymptotic} \\
\cmidrule(l{3pt}r{3pt}){2-3} \cmidrule(l{3pt}r{3pt}){4-5} \cmidrule(l{3pt}r{3pt}){6-7} \cmidrule(l{3pt}r{3pt}){8-9} \cmidrule(l{3pt}r{3pt}){10-11}
Setup & Rate & Size & Rate & Size & Rate & Size & Rate & Size & Rate & Size\\
\midrule
Logistic & 0.84 & 1.30 & 0.83 & 1.96 & 1.00 & 2.72 & 0.53 & 0.78 & \textbf{0.96} & \textbf{1.53}\\
GMM(0) & 0.91 & 2.44 & 0.91 & 4.14 & 1.00 & 4.01 & 0.60 & 1.34 & \textbf{0.98} & \textbf{2.63}\\
GMM(-1) & 0.83 & 1.27 & 0.86 & 1.95 & 1.00 & 2.64 & 0.58 & 0.79 & \textbf{0.99} & \textbf{1.54}\\
GMM(-2) & 0.78 & 0.88 & 0.78 & 1.36 & 1.00 & 2.08 & 0.52 & 0.57 & \textbf{1.00} & \textbf{1.13}\\
\midrule
rice & \textbf{1.00} & \textbf{ 3.74} & 0.88 & 7.35 & 1.00 & 28.41 & 0.91 & 2.87 & 1.00 & 5.92\\
sepsis & \textbf{0.95} & \textbf{ 0.69} & 1.00 & 2.42 & 1.00 & 10.00 & 0.98 & 1.18 & 1.00 & 2.42\\
banknote & 0.99 & 1.16 & 1.00 & 1.48 & 1.00 & 1.53 & \textbf{0.99} & \textbf{0.64} & 1.00 & 1.28\\
mozilla & \textbf{0.96} & \textbf{1.42} & 0.97 & 2.04 & 1.00 & 2.68 & 0.87 & 0.65 & 0.93 & 1.31\\
skin & 1.00 & 1.39 & 1.00 & 1.84 & 1.00 & 1.80 & \textbf{1.00} & \textbf{0.57} & 1.00 & 1.16\\
blood & 0.87 & 1.07 & 0.95 & 2.51 & 1.00 & 5.40 & 0.84 & 0.86 & \textbf{0.99} & \textbf{1.71}\\
phoneme & \textbf{1.00} & \textbf{1.32} & 0.95 & 1.92 & 1.00 & 5.62 & 0.73 & 0.83 & 1.00 & 1.66\\
telescope & \textbf{0.99} & \textbf{ 5.96} & 0.99 & 10.66 & 1.00 & 50.69 & 0.67 & 3.50 & 0.96 & 7.10\\
\midrule
yeast & \textbf{0.97} & \textbf{11.10} & 0.68 & 25.13 & -- & -- & 0.33 & 10.82 & 0.88 & 19.22\\
wine & \textbf{0.93} & \textbf{11.04} & 0.79 & 39.75 & -- & -- & 0.09 & 12.51 & 0.91 & 26.25\\
\bottomrule
\end{tabular}
\end{table}

\paragraph{Sensitivity to the rollout length.}
\blue{TabMGP inherits a single tuning parameter from the martingale posterior
  framework: the rollout length $N$. Ideally, $N$ should be large enough that
  the resulting credible sets are insensitive to additional forward sampling,
  although computational constraints can limit this in practice.
  Appendix~\ref{sec:coverage-sensitivity} assesses the sensitivity of these
  credible sets by varying $N$ in the slower-converging setups. The results show
  that both coverage and credible-set size stabilise once $N$ is sufficiently
  large.}

\paragraph{Marginal credible intervals.} In addition to the results of joint
credible sets presented above, we also \blue{report marginal credible-interval
  results in Appendix~\ref{sec:coverage-marginal-ci}. These one-dimensional
  intervals provide an alternative assessment of the coverage that does not rely
  on the empirical-cutoff ellipsoids used for the joint credible sets. We report
  both the coverage and the median width of the credible intervals. The results
  are broadly consistent with the results from the joint credible sets above.}


\paragraph{Computation time.} The main computational cost of martingale posteriors
is forward sampling, which is readily parallelisable. Per sample, BB and
Asymptotic are the fastest (near-instantaneous); Copula takes 10--20s; and
TabMGP takes about 70s (binary/categorical response) or 200s (continuous
response). However, despite being the most computationally expensive method,
TabMGP requires minimal manual intervention compared to the others. \blue{The
  No-U-Turn algorithm \citep{hoffman14nouturn} used for Bayes runs in under
  60s.} BB, Copula, Bayes, and Asymptotic are implemented in compiled JAX,
whereas TabMGP currently runs in uncompiled PyTorch. We expect this gap to
narrow with further code optimisation and improvements in the inference speed of
tabular foundation models. All experiments were conducted on an NVIDIA L40S.

\section{Conclusion}
While bespoke Bayesian modelling remains the gold standard when a well-specified
likelihood--prior pair is available, tabular foundation models like TabPFN are
increasingly used in practice. In scientific workflows, prediction is usually a
means to an end: one ultimately wants uncertainty about a scientific estimand
$\theta$. However, TabPFN does not by itself return posterior uncertainty for
$\theta$. In TabPFN, the carrier of epistemic uncertainty is the
\emph{predictive rule itself} and the associated random limiting measure
$F_\infty$, so the natural inference target is the functional
$\theta(F_\infty)$. This motivates a framework that returns posterior
uncertainty for $\theta(F_\infty)$ using only queries to TabPFN's predictive
outputs.

For this purpose, we introduce TabMGP. The procedure requires neither
likelihood--prior specification, as in classical Bayesian inference, nor bespoke
predictive rules or hyperparameter tuning, as in existing martingale posterior
frameworks, relying instead solely on the predictive rule induced by TabPFN. We
evaluated TabMGP across a large collection of synthetic and real-world setups
and compared it extensively with existing martingale posterior
constructions as well as standard Bayesian baselines. Across these experiments,
TabMGP consistently performs well from a frequentist perspective and delivers
stable posteriors with near-nominal coverage. Our empirical evaluation shows
that TabMGP provides reliable uncertainty quantification.

The experiments also reveal concrete failure modes. TabMGP can exhibit
  undercoverage in classification problems with small $n/p$ and can exhibit
slow convergence of $\theta(F_N)$ in some setups. \blue{Accordingly,
  practitioners should treat TabMGP as a diagnostic-driven procedure. Under
  finite computational budgets, one should inspect the path-stability
  diagnostics described in Appendix~\ref{sec:diagnostics} and interpret the
  posterior with non-stabilising paths cautiously.}

Our findings also expose limitations in the current theoretical understanding of
martingale posteriors. Existing analyses emphasise sufficient conditions such as
the martingale or a.c.i.d.\ properties; in our experiments, finite-horizon
diagnostics do not \emph{certify} these sufficient conditions, yet the resulting
posteriors maintain empirical stability and near-nominal coverage. This suggests
that the present theory may be overly restrictive and does not yet capture the
range of predictive rules that work well empirically. Developing weaker and more
realistic conditions is an important direction for future research, particularly
as martingale posteriors are likely to be combined with an increasing variety of
black-box predictive models. More broadly, deep learning theory continues to lag
far behind practice. For predictors like TabPFN, we should not expect to be able
to verify theoretically that any proposed condition holds, so progress
will also require a battery of theory-guided empirical diagnostics to test these
conditions.




\section*{Acknowledgements}
The authors would like to thank the reviewers and area chair for their helpful
comments and suggestions. We also thank Abdelhamid Ezzerg, Joshua Bon, Christian
Robert, Lorenzo Cappello, Jack Jewson, Rubén Loaiza-Maya and Anastasios
Panagiotelis for helpful comments in the development of this work. KN was
supported by the Australian Government Research Training Program and the
Statistical Society of Australia PhD Top-up Scholarship. EF was supported by the
Research Grants Council of Hong Kong through the General Research Fund
(17307321) and the Early Career Scheme (27304424). DTF acknowledges funding from
the Australian Research Council (DE200101070, DP200101414). JK was supported
through the UK's Engineering and Physical Sciences Research Council (EPSRC) via
EP/W005859/1. SW was supported by the Australian Research Council (DE200101253).
Computational resources were provided by the ARDC Nectar Research Cloud.

\section*{Impact Statement}
This paper presents work whose goal is to advance the field of Machine Learning.
There are many potential societal consequences of our work, none of which we feel
must be specifically highlighted here.

\bibliography{main}
\bibliographystyle{icml2026}

\newpage
\appendix
\onecolumn
\section{Review of TabPFN}
\label{sec:tabpfn-review}
A precursor to TabPFN appears in \citet{muller22transformers}, where the
transformer is referred to as a prior-data-fitted network (PFN). The underlying
premise is that the Bayesian posterior predictive distribution (PPD) is optimal
in an \textit{average} sense. We present this result in the unsupervised setting
to keep the notation lightweight, drawing on a result from
\citet{aitchison75goodness}\footnote{The work of \citet{muller22transformers}
  does not appear to be aware of this result.}.

Suppose we have an exchangeable distribution
\begin{equation}
  \label{eq:exchangeable}
  p(z_1, \ldots, z_n) = \int \prod_{i=1}^n p(z_i \mid \nu) \phi(\nu) \, \mathrm{d}\nu \, .
\end{equation}
Consider the Kullback-Leibler (KL) divergence between $p(z \mid \nu)$ and some
predictive density $q(z \mid z_{1:n})$ where $z_{1:n}$ is drawn from
\eqref{eq:exchangeable}, i.e., first draw $\nu$ from the prior $\phi(\nu)$ and
then draw $z_i$ i.i.d.~from $p(z \mid \nu)$. Then,
\begin{equation*}
  \KL(p(z \mid \nu) \, \Vert \, q(z \mid z_{1:n}))
\end{equation*}
is random because (a) $\nu$ is random and (b) the data $z_{1:n}$ are random. By
integrating out these sources of randomness, we get an average KL divergence, a
deterministic quantity
\begin{equation}
  \E_{\nu, z_{1:n}} \KL(p(z \mid \nu) \, \Vert \, q(z \mid z_{1:n}))
  = \int \phi(\nu) \, \mathrm{d} \nu \int p(z_{1:n} \mid \nu) \, \mathrm{d} z_{1:n} \int  p(z \mid \nu) \log \frac{p(z \mid \nu)}{q(z \mid z_{1:n})} \, \mathrm{d} z \, .
  \label{eq:averageKL}
\end{equation}
\citet{aitchison75goodness} then shows that the predictive density $q$ that
minimises this average KL divergence in \eqref{eq:averageKL} is none other than
the Bayesian PPD corresponding to the likelihood--prior pair in
\eqref{eq:exchangeable}:
\begin{equation}
  p(z \mid z_{1:n}) = \int p(z \mid \nu) \pi(\nu \mid z_{1:n}) \, \mathrm{d} \nu
\end{equation}
where $\pi(\nu \mid z_{1:n}) \propto \prod_{i=1}^n p(z_i \mid \nu) \phi(\nu)$,
the standard Bayesian posterior construction.

This optimality result for the Bayesian PPD immediately suggests a meta-learning
training objective. Since minimising the average KL divergence with respect to
$q$ is indifferent to the entropy term of $p(z \mid \nu)$ in the KL divergence,
we may rewrite the objective as
\begin{equation*}
  \argmin_{q(z \mid z_{1:n})} \E_{\nu, z_{1:n}} \operatorname{KL}(p(z \mid \nu) \, \Vert \,  q(z \mid z_{1:n}))
  = \argmin_{q(z \mid z_{1:n})} \E_{\nu, z_{1:n}} \E_{p(z \mid \nu)} -\log q(z \mid z_{1:n})
\end{equation*}
If we use a model $q_w$, the empirical risk counterpart is
\begin{equation*}
  \argmin_{w} - \sum_{m=1}^M \sum_{k=1}^K \log q_w(z_{n+k}^m  \mid z_{1:n}^m).
\end{equation*}
Here, $q_w$ can be any model that can accept a variable-length set $z_{1:n}$ and
returns a distribution over future $z$, and the samples $z_{1:n+K}$ are drawn
from an exchangeable distribution $p(z_1,\ldots,z_n, z_{n+1}, \ldots, z_{n+K})$
as in \eqref{eq:exchangeable}. \citet{muller22transformers} choose to use a
transformer for $q_w$. This is most natural when $z$ is categorical. In the case
of real-valued $z$, \citet{muller22transformers} perform an adaptive binning
operation that essentially converts regression into a classification task.

Later developments led to TabPFN \citep{hollmann22tabpfn,hollmann25accurate},
which is trained on a vast amount of synthetic data generated from a very large
array of likelihood--prior pairs (e.g., structural causal models and Bayesian
neural network models), rather than a single pair as in the original PFN work.
This gives TabPFN its \textit{foundation model} qualifier.

\section{Recovering the Classical Bayesian Posterior from the Martingale Posterior}
\label{sec:mgp-bayes-equiv}
A sample from the classical Bayesian posterior
$\pi(\theta \mid z_{1:n}) \propto \pi(\theta) \prod_{i=1}^{n} p(z_{i} \mid \theta)$
can be obtained equivalently from the martingale posterior. This is achieved by
forward sampling with the Bayesian PPD:
\begin{equation*}
  p(\vz \mid \vz_{1:i}) = \int p(z \mid \theta) \, \pi(\theta \mid z_{1:i}) \, \mathrm{d}\theta,
\end{equation*}
then computing the posterior mean estimator $\E[\vtheta \mid \vz_{1:\infty}]$.
This follows from Doob's martingale convergence theorem
\citep{doob49application}. Alternatively, we can use the maximum likelihood
estimator as our functional of interest $\theta(F)$, i.e., by setting the loss in
\eqref{eq:risk-minimiser} as $\ell(z, \theta) = - \log p(z \mid \theta)$, and
the distribution of $\theta(F_\infty)$ is $\pi(\theta \mid z_{1:n})$. This
is due to the asymptotic equivalence between a posterior mean estimator and the
maximum likelihood estimator.

\section{Validity of TabMGP as a Martingale Posterior}
\label{sec:tabmgp-validity}

As discussed in Section~\ref{sec:mgp}, a sufficient condition for the existence
of $F_\infty$ and hence for the martingale posterior distribution to be
well-defined is that the sequence $( P_i )_{i \ge 0}$ is a martingale, or
equivalently, that the sequence $(Z_i)_{i \ge 1}$ is conditionally identically
distributed \citep[c.i.d.][]{berti04limit}. The martingale property stipulates
that $( P_i )_{i \ge 0}$ satisfies, for every $i \ge 0$ and every measurable set
$A$:
\begin{equation}
  \label{eq:cid}
  \E(P_{i+1}(A) \mid Z_{1:i}) = P_i(A).
\end{equation}
A relaxation of this sufficient condition is the \emph{almost conditionally
  identically distributed} (a.c.i.d.) condition \citep{battiston25bayesian}. Let
$(\xi_{i})_{i \ge 1}$ be a sequence of non-negative random variables. The
sequence $(Z_i)_{i \ge 1}$ is a.c.i.d.~if the corresponding predictive rule
$(P_{i})_{i \ge 0}$ satisfies
\begin{equation}
  \label{eq:acid}
  \lvert \E_{Z_{i+1}}(P_{i+1}(A) \mid Z_{1:i}) - P_{i}(A) \rvert \leq \xi_{i}, \quad \text{a.s.,}
\end{equation}
for all $i \ge 1$ and all measurable sets $A$. \citet{battiston25bayesian} show that
$F_{\infty}$ exists if $\sum_{i=0}^{\infty} \xi_{i} < \infty$, $\bP$-a.s., and
that the sequence $(Z_i)_{i \ge 1}$ is $F_{\infty}$-asymptotically exchangeable.
We will refer to these conditions collectively as the \emph{a.c.i.d.~condition}.
Clearly, \eqref{eq:acid} generalises the stronger martingale condition
\eqref{eq:cid}, i.e., $\xi_{i} = 0$ for all $i \ge 1$, and we will use
\eqref{eq:acid} as an empirical diagnostic for TabMGP instead
of~\eqref{eq:cid}.

Verifying~\eqref{eq:acid} empirically is generally challenging, as it must hold
for all measurable sets $A$. Instead, we adopt an equivalent formulation in terms
of the total variation distance \citep{battiston25bayesian}:
\begin{equation*}
  \sum_{i=1}^{\infty} \TV \left( \E_{Z_{i+1}}(P_{i+1}(\cdot) \mid Z_{1:i}), P_{i}(\cdot) \right) \leq \xi_{i}, \quad \text{a.s.,}
\end{equation*}
for all $i \ge 0$. Since the predictive distributions of both TabPFN and the
Bayesian bootstrap are probability mass functions $p_i$, the condition can be
further simplified in terms of the $L_{1}$ distance:
\begin{equation}
  \label{eq:acid-l1}
  \sum_{i=1}^{\infty} \sum_{z \in \sZ} \lvert \E_{Z_{i+1}}[p_{i+1}(z) \mid Z_{1:i}] - p_i(z)
  \rvert < \infty, \quad \text{a.s.}.
\end{equation}
There are several practical considerations when estimating \eqref{eq:acid-l1}.
First, we start the outer summation from $i = n$, as TabPFN requires a small
number of initial observations to produce reliable predictions. Second, in our
supervised learning setting where $z = (x, y)$, the expectation
\begin{equation*}
  \E_{Z_{i+1}}[p_{i+1}(z) \mid z_{1:i}] =
  \E_{Y_{i+1}, X_{i+1}}[p_{i+1}(y \mid x_{i+2}) \, p_{i+1}(x) \mid z_{1:i}] ,
\end{equation*}
requires a large number of Monte Carlo samples and is computationally demanding.
To mitigate this, we fix all future covariates $x_{n+1:N+2}$ to a constant value
$x^{\star}$ and consider the weaker condition:
\begin{equation}
  \label{eq:acid-conditional}
  \sum_{i=n}^{N} \sum_{y \in \sY} \lvert
  \E_{Y_{i+1}}[p_{i+1}(y \mid x^{\star}) \mid z_{1:i}] - p_i(y \mid x^{\star})
  \rvert < \infty,
\end{equation}
for a sufficiently large $N$. We then compute \eqref{eq:acid-conditional}
repeatedly over many realisations of $Y_{n+1:N}$.

We evaluate this condition on an initial dataset of size $n = 100$ generated as
follows: $x_i$ are drawn independently and uniformly from the two-dimensional
cube $[-1,1]^2$, and
$y_i \overset{\mathrm{i.i.d.}}{\sim} \operatorname{Bernoulli}(\operatorname{logistic}(x_i^\top \beta^\star))$,
for $i = 1, \ldots, n$, with a fixed $\beta^\star \in \R^2$. We consider 10
different values of $x^\star$, randomly selected from the initial sequence
$x_1, \ldots, x_n$. Due to computational constraints, we evaluate
\eqref{eq:acid-conditional} only on a sparse grid
$i \in \{100, 120, 140, \ldots, 1100\}$ and use the trapezoidal rule to
approximate the partial sum.

\begin{figure}
  \centering
  \includegraphics[width=\linewidth]{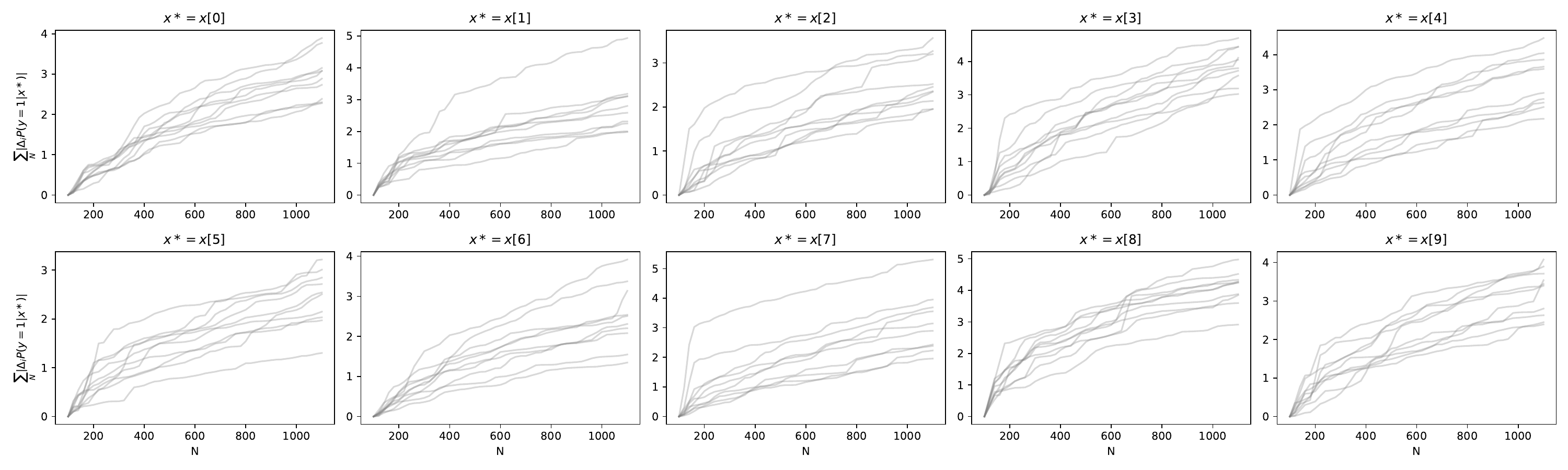}
  \caption{Cumulative sum (over $N$) of the $L_{1}$-distance between
    $\E_{y_{i+1}}[p_{i+1}(y \mid x^{\star}) \mid z_{1:i}]$ and
    $p_i(y \mid x^{\star})$ across ten choices of~$x^{\star}$. Ideally, this
    cumulative sum converges to satisfy the a.c.i.d.~condition.}
  \label{fig:tabmgp-l1-sum}
\end{figure}

Our empirical results (Figure~\ref{fig:tabmgp-l1-sum}) suggest that the
predictive rule induced by TabPFN does not strictly adhere to the a.c.i.d.
condition as $N$ increases. However, at these scales ($N > 1000$), any
measurable drift falls below the numerical resolution of the model, making it
impossible to distinguish genuine theoretical departures from the cumulative
numerical instability inherent to long-horizon autoregressive inference. This
empirical ambiguity, however, does not imply non-convergence to $F_{\infty}$ for
TabMGP, as the a.c.i.d.~condition is sufficient but not necessary. Moreover, the
observed divergence may be attributable to the finite numerical precision of
TabPFN. In particular, both
$\E_{Y_{i+1}}[p_{i+1}(y \mid x^{\star}) \mid z_{1:i}]$ and
$p_i(y \mid x^{\star})$ are expected to be accurate only up to machine
precision, so that their difference behaves as random numerical noise for
sufficiently large $N$, and the partial sum is infeasible to compute in this
regime.

As an alternative, we track the sample path of $\theta(F_{N})$ as $N$ increases,
as in \citet{fong23martingale}. After all, we need $\theta(F_{N})$ to converge
$\bP$-a.s. to conduct meaningful statistical inference. We refer to these
diagnostic plots as the ``trace plot'' and inspect them in
Appendix~\ref{sec:diagnostics}. We observe that the sample path of
$\theta(F_{N})$ does converge in most of our experiments, thus suggesting that
the a.c.i.d.~condition is too restrictive for our purposes in establishing the
existence of $\theta(F_{\infty})$.

For completeness, we further demonstrate that TabMGP can concentrate close to
$\theta(F^{\star})$ as the sample size $n$ grows. To illustrate this, we use a
regression dataset, since the sample size needed for meaningful
concentration, together with the forward-sampling steps required for convergence in
classification, exceeds TabPFN’s 10,000-context length limit. The regression
dataset is defined as
$y_{i} \overset{i.i.d.}{\sim} \sN(x^{\top}_{i} \beta^{\star}, 0.1^{2}), \, i = 1, \dots, n$,
for some fixed $\beta^{\star} \in \R^{2}$, with the loss function
$\ell(x, y, \vtheta) = (y - [1 \ x^{\top}] \vtheta)^{2}$. Ideally, TabMGP should
concentrate around $\theta(F^{\star}) = [0, {\beta^{\star}}^{\top}]$ as $n$ increases.
Figure~\ref{fig:tabmgp-concentration} confirms that TabMGP indeed concentrates
around $\theta(F^{\star})$.

\clearpage
\section{Highest-Posterior Density Credible Sets}
\label{sec:hpd-credible-set}
The \emph{highest-posterior-density} (HPD) credible set is a common construction
of a Bayesian credible set. The HPD set at level $1-\alpha$ is defined as
\begin{equation*}
  C_{1-\alpha}^{\mathrm{HPD}} = \left\{ \theta \in \bR^p : \pi(\theta \mid z_{1:n}) \ge t_{1-\alpha} \right\},
\end{equation*}
where $t_{1-\alpha}$ is chosen such that
$\int_{C_{1-\alpha}^{\mathrm{HPD}}} \pi(\theta \mid z_{1:n}) \, \mathrm{d}\theta = 1-\alpha$.
By construction, $C_{1-\alpha}^{\mathrm{HPD}}$ has minimum volume among all sets
of posterior probability $1-\alpha$, making it the ``gold standard'' for
summarising joint uncertainty \citep{hyndman96computing}.

In many models --- particularly linear and logistic regressions with
well-behaved posteriors --- the posterior is approximately multivariate normal:
\begin{equation*}
  \theta \mid z_{1:n} \approx \sN_p(\hat{\mu}, \hat{\Sigma}).
\end{equation*}
In these cases, the HPD set reduces to the $\chi^2$-ellipsoid
\begin{equation}
  \label{eq:chi-square-ellipsoid}
  C_{1-\alpha}^{\mathrm{HPD}} \approx \Bigl\{ \theta : (\theta-\hat{\mu})^\top \hat{\Sigma}^{-1} (\theta-\hat{\mu}) \le \chi^2_{p,1-\alpha} \Bigr\},
\end{equation}
treating the Mahalanobis ellipsoid as the HPD set under the normal
approximation.

\section{Experimental Settings and Hyperparameters}
\label{sec:hyperparameters}
The code is released at \url{https://github.com/weiyaw/tabmgp}.

\subsection{Posterior Constructions}
\label{sec:posterior-const}

\paragraph{Asymptotic approximation of the Bayesian posterior.}
Let
$\widehat{\vtheta}_{n} = \argmin_{\vtheta} \sum^{n}_{i=1} \ell(z_{i}, \vtheta)$
and
$\widehat{H}_{n} = \nabla^{2}_{\vtheta} \sum^{n}_{i=1} \ell(z_{i}, \vtheta) \vert_{\theta = \widehat{\vtheta}_{n}}$.
We use $\sN(\widehat{\vtheta}_{n}, \widehat{H}_{n}^{-1})$ as the asymptotic
posterior approximation for constructing credible sets and as the prior
for the classical Bayes baseline.

\paragraph{Bayes settings.}
We approximate the posterior using a No-U-Turn sampler
\citep[NUTS,][]{hoffman14nouturn} with 1000 warm-up steps, followed by 2000
sampling steps across 4 chains with different initialisations, for a total of
8000 samples.

\paragraph{Settings common to all martingale posteriors.}
We run 2000 forward-sampling steps for both BB and Copula. For TabMGP, we find
that $N = n + 500$ is sufficient for $\theta(F_{N})$ to
converge~(Figure~\ref{fig:convergence-l1-all-data-mean} and
Appendix~\ref{sec:diagnostics}) and use only 500 forward-sampling steps. We
approximate all martingale posterior constructions with 100 samples.

\paragraph{TabMGP settings.}
The predictive outputs are obtained from
\texttt{TabPFNClassifier.predict\_proba} for logistic regression and
\texttt{TabPFNRegressor.predict(output\_type = "full")} for linear regression.
By default, the corresponding logits are scaled by a temperature of 0.9. For our experiments,
we use the distributions induced by the untempered logits as our predictive distributions, i.e., setting the
temperature to 1. By construction, TabPFN is not permutation invariant to the
ordering of the response and feature dimensions. The default
implementation of TabPFN uses an ensemble of predictions obtained by randomly permuting
these dimensions. The resulting probability mass functions
are then averaged to obtain the final predictive mass function. We use the default
number of ensemble members in version \texttt{2.0.6}, which is 4 for classifiers and 8
for regressors.

We use the following two checkpoints:
\begin{itemize}
  \item Classifier: \url{https://huggingface.co/Prior-Labs/TabPFN-v2-clf/blob/main/tabpfn-v2-classifier.ckpt}
  \item Regressor: \url{https://huggingface.co/Prior-Labs/TabPFN-v2-reg/blob/main/tabpfn-v2-regressor.ckpt}
\end{itemize}
These are the default checkpoints when we ran our experiments, and
the transformer was pre-trained exclusively on synthetic, exchangeable sequences
of data. Subsequent iterations of the TabPFN checkpoints, however, have been
fine-tuned on real-world datasets, and these checkpoints are not used in our
experiments.

\paragraph{Copula settings.}
We use the copula construction in \citet[\S 4.4.2 and \S
  4.4.3]{fong23martingale} with the bandwidth fixed to 0.8, since for our
experiments the marginal-likelihood-based tuning procedure of
\citet{fong23martingale} produces values that are too small and delivers poor
results. Instead, we estimate
\begin{equation*}
  \vtheta(P_{N}) \approx \argmin_{\vtheta} \sum_{i} \ell(\widetilde{x}_{i}, \widetilde{y}_{i}, \vtheta),
\end{equation*}
where $(\widetilde{x}_{i}, \widetilde{y}_{i})_{i \ge 0}$ are generated by
iterating $\widetilde{x}_{i}$ over $x_{1:n}$ and, for each $\widetilde{x}_{i}$,
drawing $\widetilde{y}_{i}$ from $P_{N}(\cdot \mid \widetilde{x}_{i})$. We
repeat this procedure five times, yielding $5n$ pairs of
$(\widetilde{x}_{i}, \widetilde{y}_{i})$.

\subsection{Specification of the Loss Function}
\label{sec:identifiable-loss}
We set the following coefficients to 0 to ensure identifiability in the loss
function:
\begin{enumerate}
  \item The coefficients for the first level of each categorical feature;
  \item The coefficients corresponding to all (almost) collinear features;
  \item In logistic regression, the coefficients corresponding to the first
        category, $\vtheta_{1}$.
\end{enumerate}

We identify the collinear features as those with the largest loading in the
least important principal component of the population data, removing them one at
a time until the computation of $\theta(F^{\star})$ is numerically stable. The
resulting parameter dimension, $p \coloneq \dim(\theta)$, and the features used
in the functional are listed in Table~\ref{tab:data-info-numbers} and
Table~\ref{tab:data-info-functional}, respectively.

We then standardise all continuous responses and features using the population
mean and variance, and one-hot encode all categorical features. Using population
moments for standardisation ensures that $\vtheta$ remains on a common scale
across repetitions.

\subsection{Data Setups}
\label{sec:data-setups}

\paragraph{Synthetic data.} The 11 synthetic setups are adapted from
\citet[Section~5]{wu23comparison}: the features $x_{i}$ are i.i.d.~and uniformly
sampled from the hypercube $[-1, 1]^{10}$. The true coefficients $\beta^{\star}$
are sampled once from $[-2, 3]^{10}$ and held fixed across the 100 repetitions of
$z_{1:n}$. The response $y \mid x$ is generated as follows:
\begin{itemize}
  \item Continuous i.i.d.~response (4 setups): We generate $n=20$ samples from
        $y_{i} = x^{\top}_{i} \beta^{\star} + \epsilon_{i}, \, i = 1, \ldots, n$.
        The errors $\epsilon_i$ are either i.i.d.~draws from $\sN(0,1)$ or
        from Student-$t$ distributions with 5, 4, or 3 degrees of freedom. Note that $\theta(F^{\star})$ coincides
        with $(0, {\beta^{\star}}^{\top})$ in the case of i.i.d.~$\sN(0,1)$ errors.
  \item Continuous response with heteroscedastic errors (3 setups): We generate
        $n=20$ samples from
        $y_{i} = x^{\top}_{i} \beta^{\star} + \epsilon_{i}, \, i = 1, \ldots, n$,
        where $\epsilon_{i} \sim \sN(0,\sigma_i^2)$. The standard deviation of
        the error $\sigma_i$ is determined by the first coordinate of $x_i$:
        \begin{equation*}
          \sigma_{i} =
          \begin{cases}
            s_\mathrm{left}, & \text{if } x_{i1} < \hat{\xi}_{0.25},                          \\
            s_\mathrm{mid},  & \text{if } \hat{\xi}_{0.25} \leq x_{i1} \leq \hat{\xi}_{0.75}, \\
            1,               & \text{if } x_{i1} > \hat{\xi}_{0.75},
          \end{cases}
        \end{equation*}
        where $\hat{\xi}_{0.25}$ and $\hat{\xi}_{0.75}$ are the empirical
        quartiles of $x_{11}, \ldots, x_{n1}$. The hyperparameters
        $(s_\text{left}, s_\text{mid})$ are ordered from weak to strong by
        their deviation from $\sigma = 1$: $s_{1} = (0.25, 0.5)$,
        $s_{2} = (0.05, 0.25)$, and $s_{3} = (0.01, 0.1)$.
  \item Binary response (4 setups): We generate $n=100$ samples from
        $y_{i} \mid x_{i} \overset{\mathrm{iid}}{\sim} \mathrm{Bernoulli}(L(x^{\top}_{i} \beta^{\star})), \, i = 1, \ldots, n$,
        with different link functions $L$. The link is either the logistic
        function:
        \begin{equation*}
          L(u) = (1 + \exp(-u))^{-1},
        \end{equation*}
        or the distribution function of a
        Gaussian-mixture model (GMM):
        \begin{equation*}
          L(u) = 0.7 \,\Phi(u \mid a, 1) + 0.3 \,\Phi(u \mid 2, 1),
        \end{equation*}
        where $\Phi(\cdot \mid m, 1)$ denotes the distribution function of a
        Gaussian with mean $m$ and unit variance, and $a \in \{0, -1, -2\}$.
        Note that $\theta(F^{\star})$ coincides with
        $(0, {\beta^{\star}}^{\top})$ in the case of the logistic link.
\end{itemize}

\paragraph{Real data.}
We apply \texttt{train\_test\_split} in \texttt{scikit-learn}
\citep{pedregosa11scikitlearn} to the whole dataset, and use the training split
as the dataset $z_{1:n}$. We repeat this procedure 100 times to generate
repetitions of $z_{1:n}$. For datasets with highly skewed responses or features,
we apply stratified sampling on these variables. For \texttt{yeast} and
\texttt{wine}, which have highly skewed multiclass categorical responses, we
also remove classes with fewer than 3\% of observations. Multinomial logistic
regression on a small dataset with an extremely skewed response remains
difficult with our method, as the design matrix tends to be ill-conditioned when
applying the Bayesian bootstrap on $x$.

The training sets are drawn from the population dataset with stratification on all
categorical and numerical variables with few unique values. In certain
datasets with heavily skewed numerical variables, we also stratify using
bins of these numerical variables. The bin boundaries are determined to
ensure that each bin contains approximately the same number of data points. The
number of bins depends on the dataset and is chosen such that the marginal
distribution of the training set roughly matches that of the population dataset.

The details of the 19 real-world datasets are provided in
Tables~\ref{tab:data-info-numbers} and~\ref{tab:data-info-functional}.


\begin{table}[ht]
  \centering
  \caption{Details of the real-world datasets.}
  \label{tab:data-info-numbers}
  \begin{tabular}{@{} llccccccc @{}}
    \toprule
    $z_{1:n}$ & Data ID      & $n$ & $p$ & $n/p$ & \makecell{Population              \\size} & \makecell{Num.\\classes} & \makecell{Num.~cont.\\features} & \makecell{Num.~cat.\\features} \\
    \midrule
    concrete  & OpenML 44959 & 100 & 7   & 14.3  & 1030                 & 0 & 8  & 0 \\
    quake     & OpenML 550   & 50  & 4   & 12.5  & 2178                 & 0 & 3  & 0 \\
    airfoil   & OpenML 44957 & 50  & 5   & 10.0  & 1503                 & 0 & 5  & 0 \\
    energy    & OpenML 44960 & 50  & 6   & 8.3   & 768                  & 0 & 8  & 0 \\
    fish      & OpenML 44970 & 50  & 6   & 8.3   & 908                  & 0 & 6  & 0 \\
    kin8nm    & OpenML 44980 & 50  & 9   & 5.6   & 8192                 & 0 & 8  & 0 \\
    auction   & OpenML 44958 & 50  & 11  & 4.5   & 2043                 & 0 & 5  & 2 \\
    grid      & OpenML 44973 & 50  & 12  & 4.2   & 10000                & 0 & 12 & 0 \\
    abalone   & OpenML 45042 & 20  & 5   & 4.0   & 4177                 & 0 & 7  & 1 \\
    \midrule
    rice      & UCI 545      & 100 & 4   & 25.0  & 3810                 & 2 & 7  & 0 \\
    sepsis    & OpenML 827   & 100 & 4   & 25.0  & 110341               & 2 & 2  & 1 \\
    banknote  & OpenML 1462  & 100 & 4   & 25.0  & 1372                 & 2 & 4  & 0 \\
    mozilla   & OpenML 1046  & 100 & 5   & 20.0  & 15545                & 2 & 5  & 0 \\
    skin      & OpenML 1502  & 50  & 3   & 16.7  & 245057               & 2 & 3  & 0 \\
    blood     & OpenML 1464  & 50  & 4   & 12.5  & 748                  & 2 & 4  & 0 \\
    phoneme   & OpenML 1489  & 50  & 6   & 8.3   & 5404                 & 2 & 5  & 0 \\
    telescope & UCI 159      & 50  & 9   & 5.6   & 19020                & 2 & 10 & 0 \\
    \midrule
    yeast     & OpenML 181   & 200 & 28  & 7.1   & 1350                 & 5 & 8  & 0 \\
    wine      & OpenML 40498 & 200 & 40  & 5.0   & 4873                 & 5 & 11 & 0 \\
    \bottomrule
  \end{tabular}
\end{table}

\begin{table}[ht]
  \centering
  \caption{Target name and the features included in the functional of interest.}
  \label{tab:data-info-functional}
  \begin{tabular}{lll}
    \toprule
    $z_{1:n}$      & Target Name                      & Dropped features in the functional                              \\
    \midrule
    concrete       & strength                         & blast\_furnace\_slag, water                                     \\
    quake          & col\_4                           &                                                                 \\
    airfoil        & sound\_pressure                  & angle\_of\_attack                                               \\
    energy         & heating\_load                    & relative\_compactness, surface\_area, roof\_area                \\
    fish           & LC50                             & MLOGP                                                           \\
    kin8nm         & y                                &                                                                 \\
    auction        & verification.time                & property.winner                                                 \\
    grid           & stab                             & p1                                                              \\
    abalone        & Classnumberofrings               & Length, Diameter, Whole\_weight, Viscera\_weight, Shell\_weight \\
    \midrule rice  & Class                            & Area, Perimeter, Major\_Axis\_Length, Convex\_Area              \\
    sepsis         & hospital\_outcome\_1alive\_0dead &                                                                 \\
    banknote       & Class                            & V3                                                              \\
    mozilla        & state                            & start                                                           \\
    skin           & Class                            & V2                                                              \\
    blood          & Class                            & V3                                                              \\
    phoneme        & Class                            &                                                                 \\
    telescope      & class                            & fSize, fConc                                                    \\
    \midrule yeast & class\_protein\_localization     & erl, pox                                                        \\
    wine           & Class                            & V7, V8                                                          \\
    \bottomrule
  \end{tabular}
\end{table}

\subsection{Constructions of Credible Sets}
\label{sec:empirical-credible-set}
\paragraph{Joint credible sets.}
Let $\widehat{\mu}$ and $\widehat{\Sigma}$ denote the empirical mean and
covariance estimated from the posterior draws $\{\vtheta^{(l)}\}_{l=1}^L$. For
martingale posteriors, we use a diagonal covariance structure
$\widehat{\Sigma} = \operatorname{diag}(\widehat{\sigma}^2_1, \ldots, \widehat{\sigma}^2_p)$
due to the limited number of posterior samples available for estimating a full
covariance matrix. Our credible set is then defined as
\begin{equation*}
  \widehat{C}_{1-\alpha} = \Bigl\{ \vtheta : (\vtheta-\widehat{\mu})^\top \widehat{\Sigma}^{-1} (\vtheta-\widehat{\mu}) \le \widehat{r}^2_{1-\alpha} \Bigr\},
\end{equation*}
where
\begin{equation*}
  \widehat{r}^2_{1-\alpha} = \operatorname{quantile}_{1-\alpha} \{  (\vtheta^{(l)} - \widehat{\mu})^\top \widehat{\Sigma}^{-1} (\vtheta^{(l)} - \widehat{\mu}) \}_{l=1}^L,
\end{equation*}
which is an empirical cutoff. This empirical-cutoff ellipsoid provides a data-driven
approximation of the highest-posterior-density set without relying on
$\chi^2_{p,1-\alpha}$, as stated in \eqref{eq:chi-square-ellipsoid}.

\paragraph{Marginal credible interval.}
Let $\theta_j$ be the $j$-th dimension of $\theta$. The $(1-\alpha)$ credible interval is
computed from the empirical quantiles of its posterior draws:
\begin{equation}
  \label{eq:marginal-ci}
  \left[ \operatorname{quantile}_{\alpha/2} \{ \theta_j^{(l)} \}_{l=1}^L, \;\; \operatorname{quantile}_{1-\alpha/2} \{ \theta_j^{(l)} \}_{l=1}^L \right].
\end{equation}


\section{Diagnostics}
\label{sec:diagnostics}
\begin{figure}[h!]
  \centering
  \begin{subfigure}[t]{\linewidth}
    \centering
    \includegraphics[width=\linewidth]{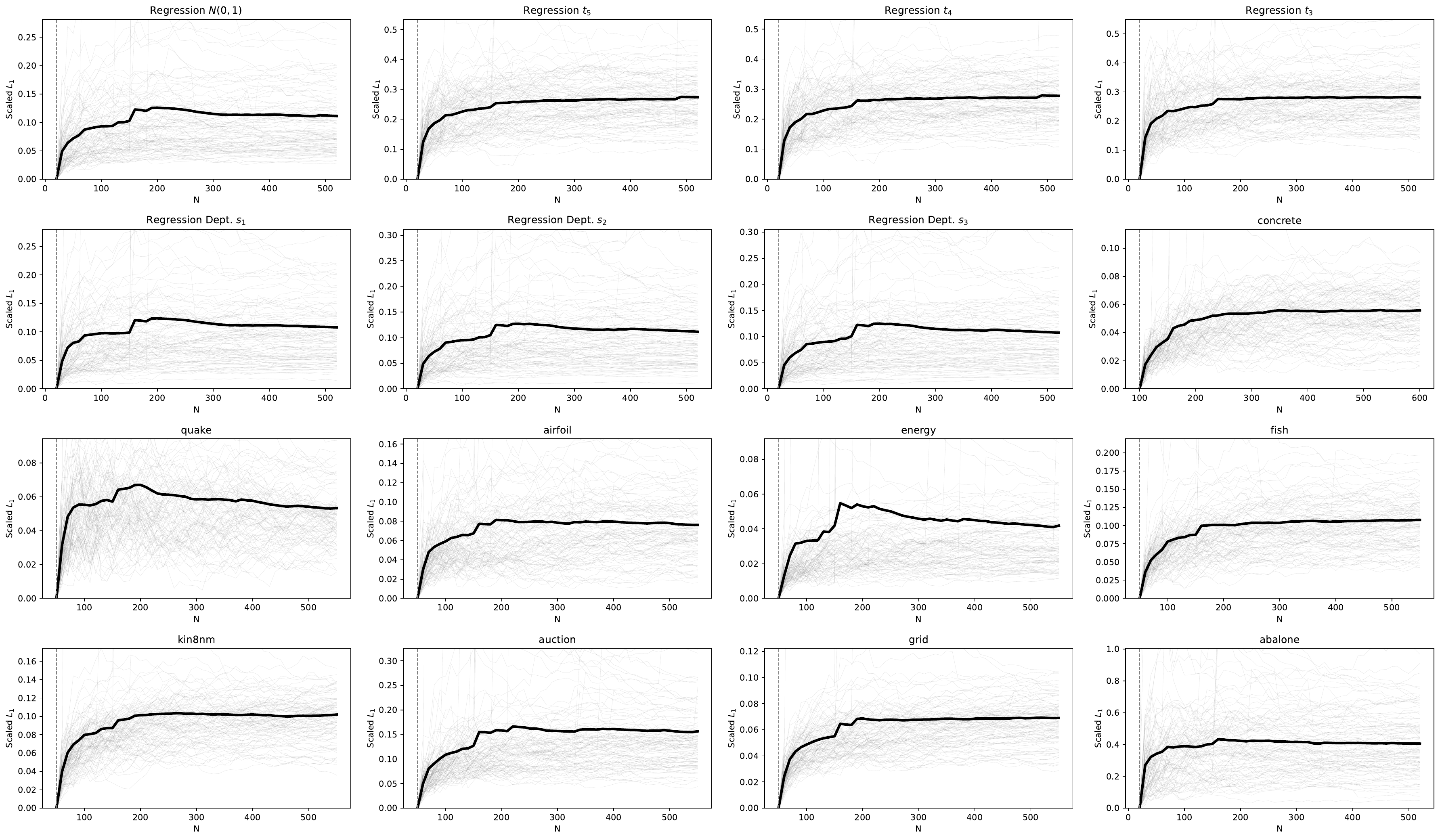}
    \caption{Linear regression}
  \end{subfigure}%
  \hfill
  \begin{subfigure}[t]{\linewidth}
    \centering
    \includegraphics[width=\linewidth]{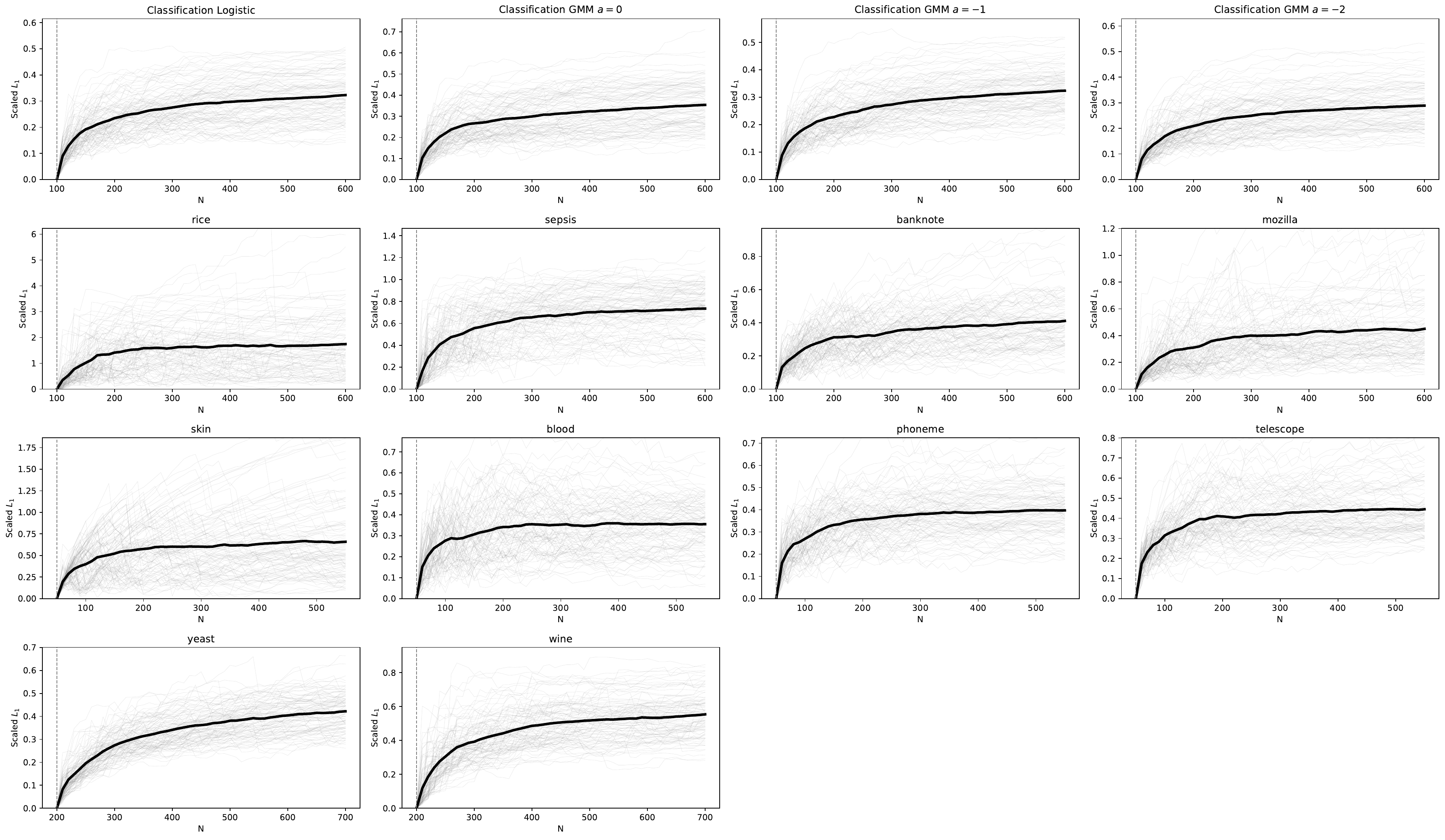}
    \caption{Logistic regression}
  \end{subfigure}
  \caption{$L_{1}$-norm between $\theta(F_{n})$ and $\theta(F_{N})$ as the
    rollout length $N$ increases. Each plot corresponds to a realisation
    $z_{1:n}$ of a particular setup. Each of the grey lines corresponds to a
    realisation of $F_{N}$, and the black solid lines are their average.}
  \label{fig:convergence-l1-all-data-full}
\end{figure}

Tracking each dimension of $\theta(F_{N})$ as $N$ grows is a common approach for
diagnosing the convergence of $\theta(F_{N})$. However, this becomes impractical
when $\theta$ is high-dimensional. Instead, we monitor convergence using the
(scaled) expected $L_{1}$-norm between $\theta(F_{n})$ and $\theta(F_{N})$:
\begin{equation*}
  \E_{F_{N}} \left[ \frac{1}{p} \lVert \theta(F_{n}) -
    \theta(F_{N}) \rVert_{1} \right],
\end{equation*}
where $p \coloneq \dim(\theta)$ and $\theta(F_{n})$ is the empirical risk
minimiser (or maximum likelihood estimate based on $z_{1:n}$ in our
experiments). In practice, we simulate many realisations of $F_{N}$ to
approximate this expectation. This expectation is shown in
Figure~\ref{fig:convergence-l1-all-data-mean} and corresponds to the black solid
trajectories in Figure~\ref{fig:convergence-l1-all-data-full}. In addition to
the expected value, we also present the $L_{1}$-norm for each realisation of
$F_{N}$ in Figure~\ref{fig:convergence-l1-all-data-full}. While the paths
corresponding to individual realisations of $F_{N}$ exhibit variability, they
generally stabilise as $N$ increases, providing further empirical evidence that
$\theta(F_N)$ converges.

\begin{figure}[h!]
  \centering
  \includegraphics[width=\linewidth]{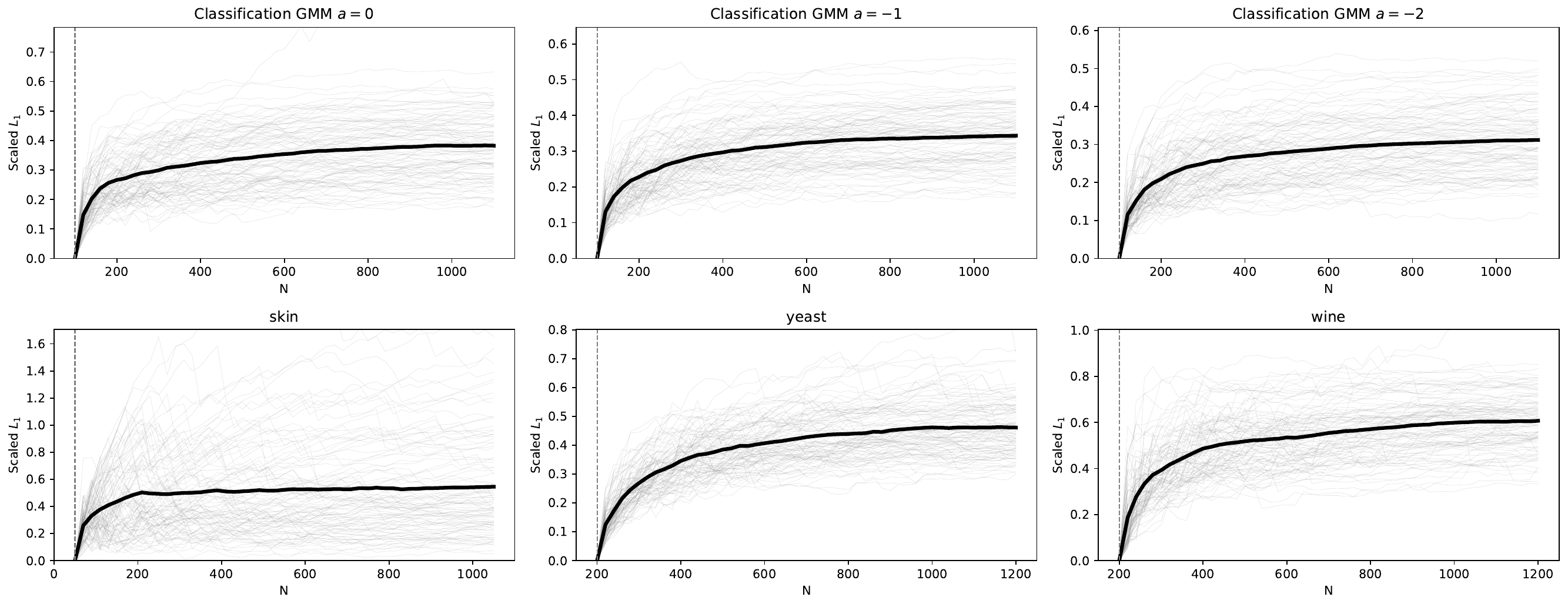}
  \caption{Scaled $L_1$-norm between $\theta(F_n)$ and $\theta(F_N)$ under
      longer TabMGP rollouts for the slower-converging setups. Each grey line
      corresponds to one rollout, and the black line is the average.}
  \label{fig:convergence-l1-extra-long-full}
\end{figure}

\blue{For the slower-converging setups, we extend the diagnostic to longer
  rollouts runs in Figure~\ref{fig:convergence-l1-extra-long-full}. The selected
  setups include the \texttt{skin}, \texttt{yeast}, \texttt{wine}, and GMM
  classification experiments.}

\begin{figure}[h!]
  \centering
  \begin{subfigure}[t]{\linewidth}
    \centering
    \includegraphics[width=\linewidth]{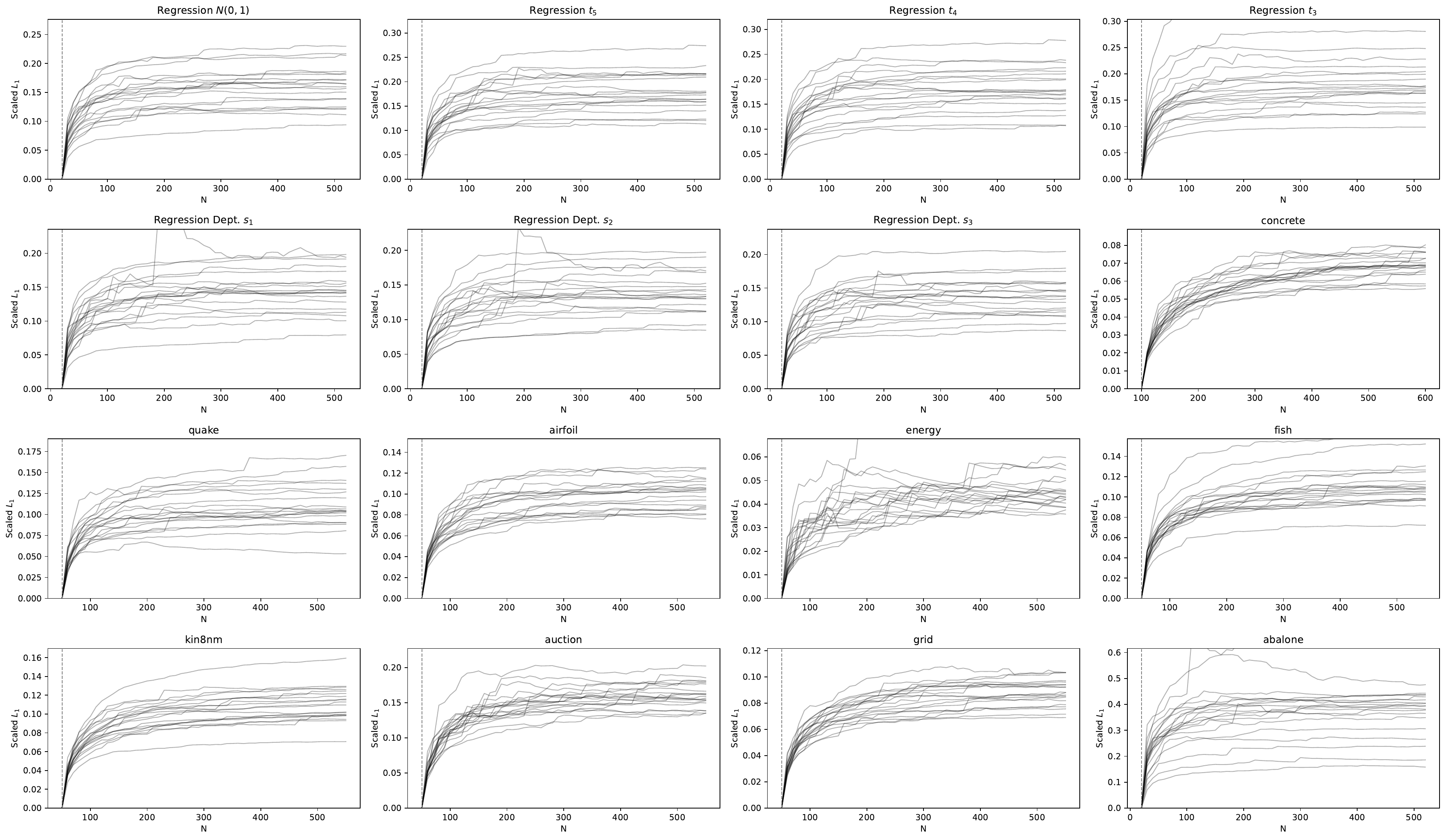}
    \caption{Linear regression}
  \end{subfigure}%
  \hfill
  \begin{subfigure}[t]{\linewidth}
    \centering
    \includegraphics[width=\linewidth]{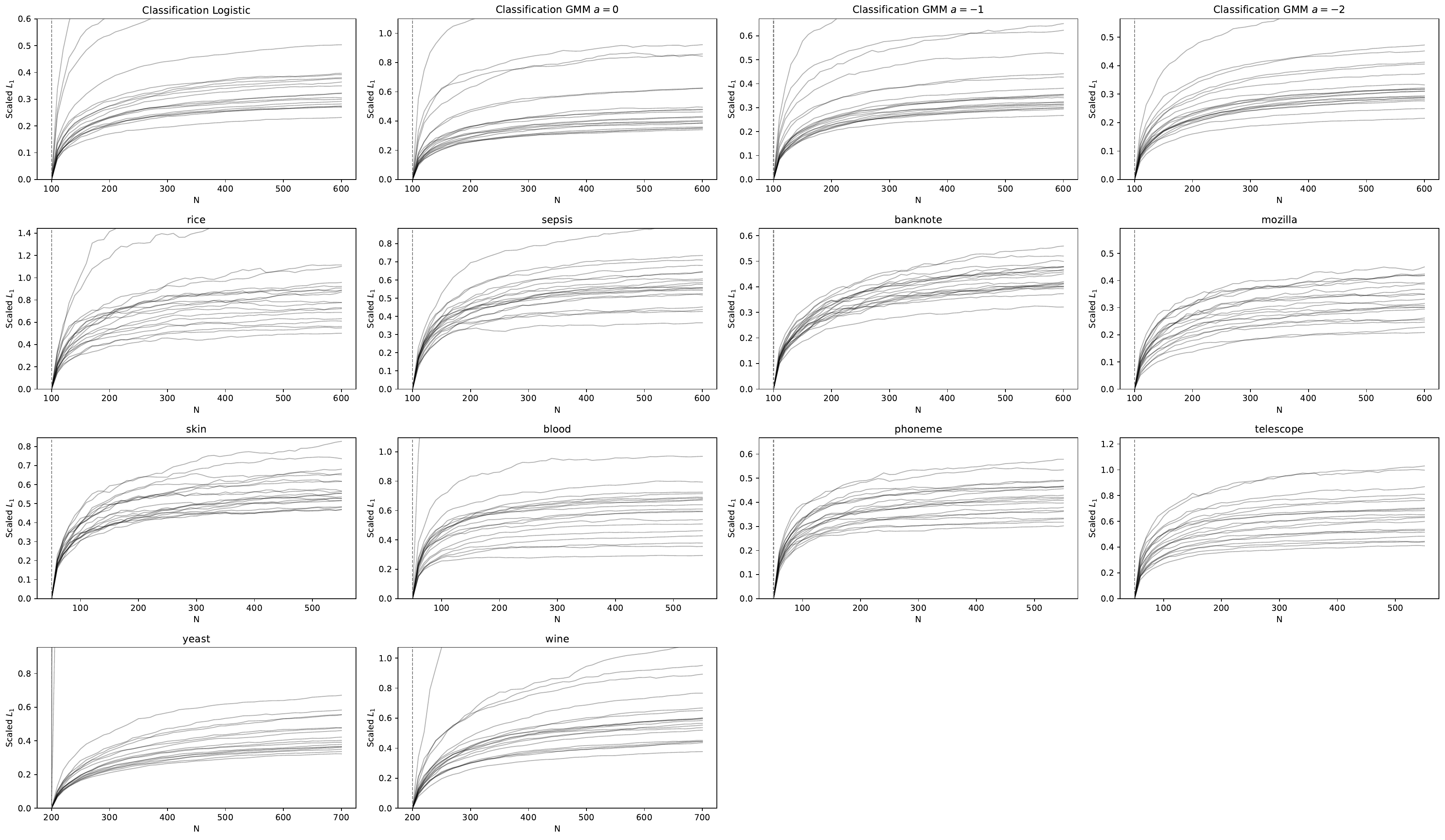}
    \caption{Logistic regression}
  \end{subfigure}
  \caption{\blue{Expected scaled $L_1$-norm between $\theta(F_n)$ and
      $\theta(F_N)$ for 20 independent realisations of $z_{1:n}$ in each setup.
      Most trajectories stabilise, while a few spiking trajectories indicate
      failure cases with low-diversity realisations of $z_{1:n}$.}}
  \label{fig:convergence-l1-20reps}
\end{figure}

\blue{We also evaluate whether the convergence diagnostic is specific to the
  particular observed dataset $z_{1:n}$ used in
  Figure~\ref{fig:convergence-l1-all-data-mean}.
  Figure~\ref{fig:convergence-l1-20reps} shows the expected scaled $L_1$
  trajectories over 20 independent realisations of $z_{1:n}$ for each setup. The
  trajectories generally stabilise, but a small number of trajectories spike or diverge. These
  edge cases typically correspond to realisations that lack diversity. As a
  result, TabPFN tends to produce predictions that lead to an ill-conditioned
  design matrix in the loss function (resulting in an unstable $\theta$). }


\section{Additional Coverage Experiments with Semi-Synthetic Setups}
\label{sec:coverage-semi-synthetic-mlp}

\blue{The real-data experiments preserve realistic covariate and response
  structure, but the population target must be approximated by the empirical
  distribution of the full dataset. Conversely, the synthetic experiments
  provide an exactly known data-generating process, but use simple covariate distributions. To
  combine these advantages, we include semi-synthetic setups that use real
  covariates together with a known nonlinear response mechanism. This allows us to
  evaluate coverage against a well-defined $\theta(F^\star)$ while retaining the
  covariate geometry of real tabular datasets.}

\blue{We construct two such setups, one for continuous-response regression and
  one for binary classification. For Concrete, we use the covariates from OpenML
  4353 and train a deterministic multi-layer perceptron (MLP) with hidden layers
  $(64,32)$ and ReLU activations on the original compressive-strength response.
  The fitted network is then treated as the response-generating mechanism,
  producing continuous synthetic responses from the observed covariates. For
  Phoneme, we use the covariates from OpenML 1489 and train a deterministic
  binary MLP with hidden layers $(128,64,32)$, batch normalisation, ReLU
  activations, and dropout during training. Synthetic binary responses are
  generated by thresholding the fitted network probabilities at $0.5$. In both
  cases, repeated datasets are obtained by sampling covariates from the
  empirical covariate distribution and applying the fitted MLP response rule.}

\blue{The corresponding coverage results are reported in
  Table~\ref{tab:coverage-semi-synthetic-mlp}. In these semi-synthetic setups,
  TabMGP remains competitive with BB while producing smaller credible sets, and
  Copula again tends to achieve coverage by using substantially larger sets.
  Bayes is well calibrated for Concrete but undercovers on Phoneme, whereas the
  asymptotic approximation overcovers in both setups.}

\begin{table}
  \centering
  \caption{\blue{Semi-synthetic MLP experiments. These setups combine real
    covariates with known nonlinear response mechanisms, allowing coverage to
    be evaluated against a known population target while preserving realistic
    tabular covariate structure. Reported values are coverage (Rate) and the
    median trace of the posterior covariance (Size), computed over 100
    repetitions. Target coverage is 0.95.}}
  \label{tab:coverage-semi-synthetic-mlp}
  \begin{tabular}{lcccccccccc}
    \toprule
    \multicolumn{1}{c}{ } & \multicolumn{2}{c}{TabMGP} &
    \multicolumn{2}{c}{BB} &
    \multicolumn{2}{c}{Copula} &
    \multicolumn{2}{c}{Bayes} &
    \multicolumn{2}{c}{Asymptotic} \\
    \cmidrule(l{3pt}r{3pt}){2-3}
    \cmidrule(l{3pt}r{3pt}){4-5}
    \cmidrule(l{3pt}r{3pt}){6-7}
    \cmidrule(l{3pt}r{3pt}){8-9}
    \cmidrule(l{3pt}r{3pt}){10-11}
    Setup & Rate & Size & Rate & Size & Rate & Size & Rate & Size & Rate & Size \\
    \midrule
    concrete-semi & 0.86 & 0.1465 & 0.84 & 0.1620 & 1.00 & 0.3580 & 0.94 & 0.2518 & 1.00 & 0.5189 \\
    phoneme-semi & 0.98 & 4.1763 & 0.94 & 5.1099 & 1.00 & 10.5623 & 0.76 & 1.5573 & 1.00 & 3.1998 \\
    \bottomrule
  \end{tabular}
\end{table}

\newpage
\section{Additional Coverage Experiments with Alternative Posterior Constructions}
\label{sec:coverage-extra}
\blue{
We reconstruct the standard Bayes posterior as detailed in
Appendix~\ref{sec:posterior-const} but replace the prior with a diffuse
$\sN(0, 10^{2})$. We find that the resulting credible sets are excessively wide
(Table~\ref{tab:coverage-linear-reg-extra} and
Table~\ref{tab:coverage-logistic-reg-extra}) and thus exclude it from the main
text to ensure a fair comparison with TabMGP.
}

\blue{
We also compare TabMGP against the copula method initialised with TabPFN, as
suggested in \citet{nagler25uncertainty}. However, the results are mixed and we
do not observe a substantial difference between the standard and
TabPFN-initialised copula methods (Table~\ref{tab:coverage-linear-reg-extra} and
Table~\ref{tab:coverage-logistic-reg-extra}).
}

\begin{table*}
  \centering
  \caption{Coverage for linear regression, comparing TabMGP with standard
    Bayesian posteriors under Gaussian priors (Hessian and Diffuse) and
    copula-based methods with either standard initialisation
    \citep{fong23martingale} or TabPFN initialisation
    \citep{nagler25uncertainty}. Reported values are coverage (Rate) and the
    median trace of the posterior covariance (Size), computed over 100
    repetitions. The values of TabMGP, Bayes (Hessian) and Copula (Standard) are
    copied from Table~\ref{tab:coverage-linear-reg} for ease of comparison.
    Target coverage is 0.95.}
  \label{tab:coverage-linear-reg-extra}
  \begin{tabular}{lcccccccccc}
    \toprule
    \multicolumn{1}{c}{ } & \multicolumn{2}{c}{TabMGP} &
    \multicolumn{2}{c}{Bayes (Hessian)} &
    \multicolumn{2}{c}{Bayes (Diffuse)} &
    \multicolumn{2}{c}{Copula (Standard)} &
    \multicolumn{2}{c}{Copula (TabPFN)} \\
    \cmidrule(l{3pt}r{3pt}){2-3}
    \cmidrule(l{3pt}r{3pt}){4-5}
    \cmidrule(l{3pt}r{3pt}){6-7}
    \cmidrule(l{3pt}r{3pt}){8-9}
    \cmidrule(l{3pt}r{3pt}){10-11}
    Setup & Rate & Size & Rate & Size &
    Rate & Size & Rate & Size & Rate &
    Size\\
    \midrule
    $\sN(0, 1)$ & 1.00 & 0.45 & 1.00 & 0.65 & 1.00 & 1.29 & \textbf{0.99} & \textbf{0.35} & 0.99 & 0.46\\
    $t_5$ & 1.00 & 0.49 & 1.00 & 0.65 & 1.00 & 1.29 & \textbf{0.99} & \textbf{0.37} & 0.99 & 0.52\\
    $t_4$ & 0.99 & 0.51 & 0.99 & 0.65 & 1.00 & 1.30 & \textbf{0.98} & \textbf{0.37} & 0.99 & 0.53\\
    $t_3$ & 1.00 & 0.48 & 0.98 & 0.65 & 0.98 & 1.30 & \textbf{0.97} & \textbf{0.35} & 1.00 & 0.55\\
    \midrule
    $s_1$ & 1.00 & 0.37 & 1.00 & 0.65 & 1.00 & 1.30 & 1.00 & 0.36 & \textbf{1.00} & \textbf{0.32}\\
    $s_2$ & 1.00 & 0.36 & 1.00 & 0.65 & 1.00 & 1.30 & 1.00 & 0.36 & \textbf{1.00} & \textbf{0.28}\\
    $s_3$ & 1.00 & 0.33 & 1.00 & 0.65 & 1.00 & 1.30 & 1.00 & 0.37 & \textbf{1.00} & \textbf{0.26}\\
    \midrule
    concrete & 0.91 & 0.06 & 0.87 & 0.05 & \textbf{1.00} & \textbf{0.10} & 1.00 & 0.12 & 0.67 & 0.04\\
    quake & \textbf{1.00} & \textbf{0.03} & 0.91 & 0.04 & 0.98 & 0.09 & 1.00 & 0.18 & 1.00 & 0.18\\
    airfoil & 0.96 & 0.08 & \textbf{0.96} & \textbf{0.06} & 1.00 & 0.12 & 0.97 & 0.11 & 0.88 & 0.07\\
    energy & \textbf{1.00} & \textbf{0.04} & 1.00 & 0.07 & 1.00 & 0.14 & 1.00 & 0.06 & 0.17 & 0.00\\
    fish & 0.90 & 0.10 & 0.91 & 0.08 & 1.00 & 0.16 & \textbf{0.95} & \textbf{0.10} & 0.94 & 0.14\\
    kin8nm & \textbf{0.95} & \textbf{0.13} & 0.84 & 0.11 & 0.99 & 0.22 & 0.63 & 0.08 & 0.99 & 0.17\\
    auction & \textbf{0.99} & \textbf{0.66} & 1.00 & 0.81 & 1.00 & 1.64 & 0.97 & 1.44 & 0.67 & 0.22\\
    grid & \textbf{0.97} & \textbf{0.13} & 0.97 & 0.16 & 1.00 & 0.31 & 0.88 & 0.10 & 0.98 & 0.14\\
    abalone & \textbf{0.97} & \textbf{0.95} & 0.84 & 0.81 & 0.98 & 1.62 & 0.94 & 0.98 & 0.87 & 0.78\\
    \bottomrule
  \end{tabular}
\end{table*}

\begin{table*}
  \centering
  \caption{Coverage for logistic regression, comparing TabMGP with standard
    Bayesian posteriors under Gaussian priors (Hessian and Diffuse) and
    copula-based methods with either standard initialisation
    \citep{fong23martingale} or TabPFN initialisation
    \citep{nagler25uncertainty}. Reported values are coverage (Rate) and the
    median trace of the posterior covariance (Size), computed over 100
    repetitions. The values of TabMGP, Bayes (Hessian) and Copula (Standard) are
    copied from Table~\ref{tab:coverage-logistic-reg} for ease of comparison.
    Target coverage is 0.95.}
  \label{tab:coverage-logistic-reg-extra}
  \begin{tabular}{lcccccccccc}
    \toprule
    \multicolumn{1}{c}{ } & \multicolumn{2}{c}{TabMGP} &
    \multicolumn{2}{c}{Bayes (Hessian)} &
    \multicolumn{2}{c}{Bayes (Diffuse)} &
    \multicolumn{2}{c}{Copula (Standard)} &
    \multicolumn{2}{c}{Copula (TabPFN)} \\
    \cmidrule(l{3pt}r{3pt}){2-3}
    \cmidrule(l{3pt}r{3pt}){4-5}
    \cmidrule(l{3pt}r{3pt}){6-7}
    \cmidrule(l{3pt}r{3pt}){8-9}
    \cmidrule(l{3pt}r{3pt}){10-11}
    Setup & Rate & Size & Rate & Size &
    Rate & Size & Rate & Size & Rate &
    Size\\
    \midrule
    Logistic & 0.84 & 1.30 & 0.53 & 0.78 & \textbf{0.99} & \textbf{1.94} & 1.00 & 2.72 & 1.00 & 3.44\\
    GMM(0) & 0.91 & 2.44 & 0.60 & 1.34 & \textbf{0.99} & \textbf{3.66} & 1.00 & 4.01 & 1.00 & 5.75\\
    GMM(-1) & 0.83 & 1.27 & 0.58 & 0.79 & \textbf{1.00} & \textbf{1.95} & 1.00 & 2.64 & 1.00 & 3.33\\
    GMM(-2) & 0.78 & 0.88 & 0.52 & 0.57 & \textbf{1.00} & \textbf{1.37} & 1.00 & 2.08 & 1.00 & 2.60\\
    \midrule
    rice & \textbf{1.00} & \textbf{3.74} & 0.91 & 2.87 & 1.00 & 6.66 & 1.00 & 28.41 & 1.00 & 31.80\\
    sepsis & \textbf{0.95} & \textbf{0.69} & 0.98 & 1.18 & 1.00 & 3.23 & 1.00 & 10.00 & 1.00 & 10.80\\
    banknote & 0.99 & 1.16 & \textbf{0.99} & \textbf{0.64} & 1.00 & 1.48 & 1.00 & 1.53 & 1.00 & 1.54\\
    mozilla & \textbf{0.96} & \textbf{1.42} & 0.87 & 0.65 & 0.96 & 1.42 & 1.00 & 2.68 & 1.00 & 2.36\\
    skin & 1.00 & 1.39 & \textbf{1.00} & \textbf{0.57} & 1.00 & 1.49 & 1.00 & 1.80 & 1.00 & 1.78\\
    blood & 0.87 & 1.07 & 0.84 & 0.86 & \textbf{1.00} & \textbf{2.06} & 1.00 & 5.40 & 1.00 & 6.16\\
    phoneme & \textbf{1.00} & \textbf{1.32} & 0.73 & 0.83 & 1.00 & 2.15 & 1.00 & 5.62 & 1.00 & 6.92\\
    telescope & \textbf{0.99} & \textbf{5.96} & 0.67 & 3.50 & 1.00 & 9.39 & 1.00 & 50.69 & 1.00 & 61.30\\
    \midrule
    yeast & \textbf{0.97} & \textbf{11.10} & 0.33 & 10.82 & 1.00 & 26.90 & -- & -- & -- & --\\
    wine & 0.93 & 11.04 & 0.09 & 12.51 & \textbf{0.98} & \textbf{33.10} & -- & -- & -- & --\\
    \bottomrule
  \end{tabular}
\end{table*}

\section{Sensitivity to the Rollout Length}
\label{sec:coverage-sensitivity}

\blue{We evaluate the sensitivity of the joint credible sets produced by TabMGP
  to the rollout length $N$. In our implementation, $N = n + T$, where $n$ is
  the number of observed data points and $T$ is the number of forward-sampling
  steps. We therefore vary $T$ while keeping the observed sample size fixed. We
  focus on the slower-converging settings: GMM, \texttt{skin}, \texttt{yeast},
  and \texttt{wine}.}

\begin{figure}[h!]
  \centering
  \caption{\blue{Coverage and credible-set size of TabMGP for selected
      slow-converging setups across several forward-sampling steps $T = N - n$.
      Coverage is evaluated over 100 repetitions of $z_{1:n}$. Size is the
      median trace of the posterior covariance. We expect the curves to
      stabilise as $T$ increases. The target coverage is 0.95.}}
  \label{fig:coverage-various-N}
  \includegraphics[width=\linewidth]{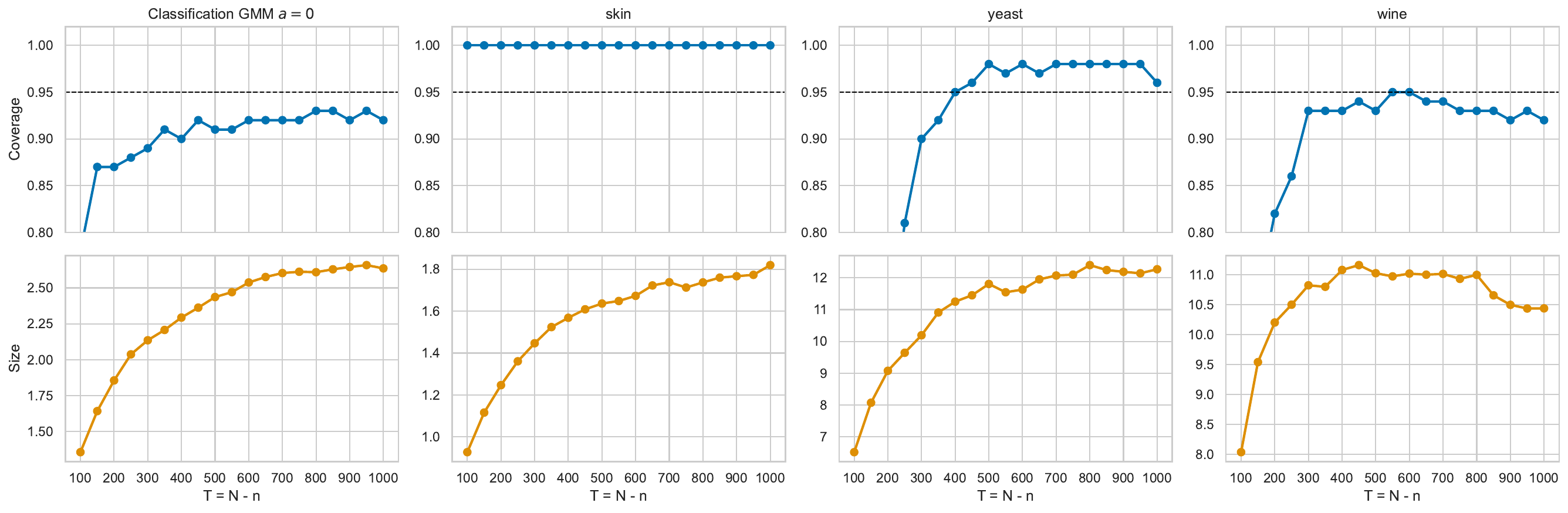}
\end{figure}

\blue{Figure~\ref{fig:coverage-various-N} reports the results for $T$ ranging
  from $100$ to $1000$ forward-sampling steps. As $T$ increases, the estimated
  coverage and credible-set size should stabilise once the distribution of
  $\theta(F_{N})$ is sufficiently close to its limiting behaviour. We find that
  most curves are largely stable by $T = 750$. The exception is a slight upward
  trend in credible-set size for \texttt{skin}. Together with
  Figure~\ref{fig:convergence-l1-20reps}, this suggests that the rollout
  trajectories for some repetitions of $z_{1:n}$ have not yet converged. We also
  observe a drop in credible-set size for \texttt{wine} between $T = 750$ and
  $T = 1000$, which we attribute to numerical instability in some repetitions.}

\newpage
\section{Marginal Density and Credible Intervals of the Posteriors}
\label{sec:marginal-density}
\begin{figure}
  \centering
  \includegraphics[width=\columnwidth]{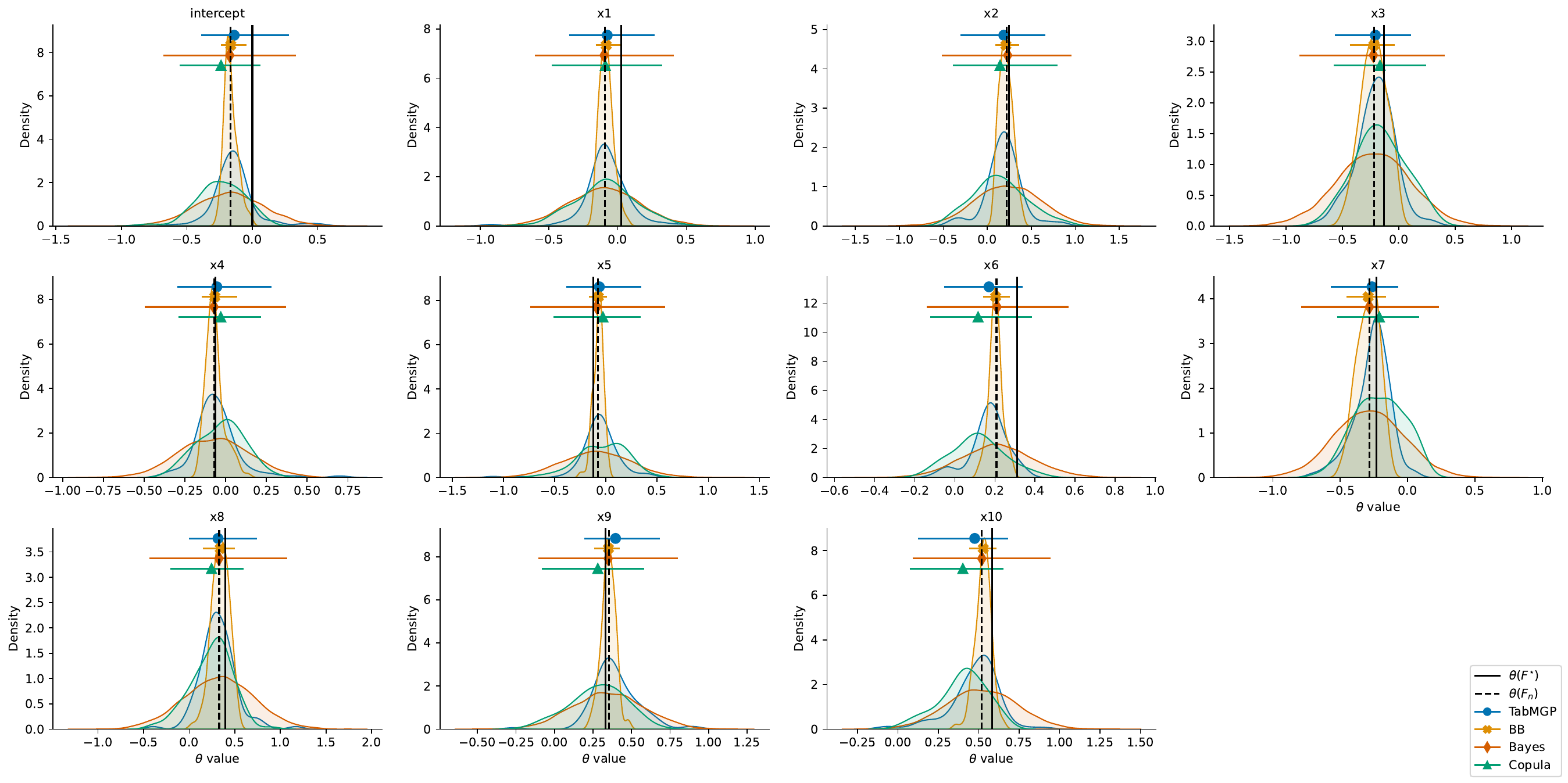}
  \caption{Marginal density plots for linear regression with synthetic
    i.i.d.~Gaussian data. For each posterior density, the 95\% marginal credible
    interval is shown as a horizontal bar, and the posterior mean is marked. The solid and dashed black vertical lines correspond to $\theta(F^{\star})$ and $\theta(F_{n})$, respectively.}
  \label{fig:density-regression}
\end{figure}

\begin{figure}
  \centering
  \includegraphics[width=\columnwidth]{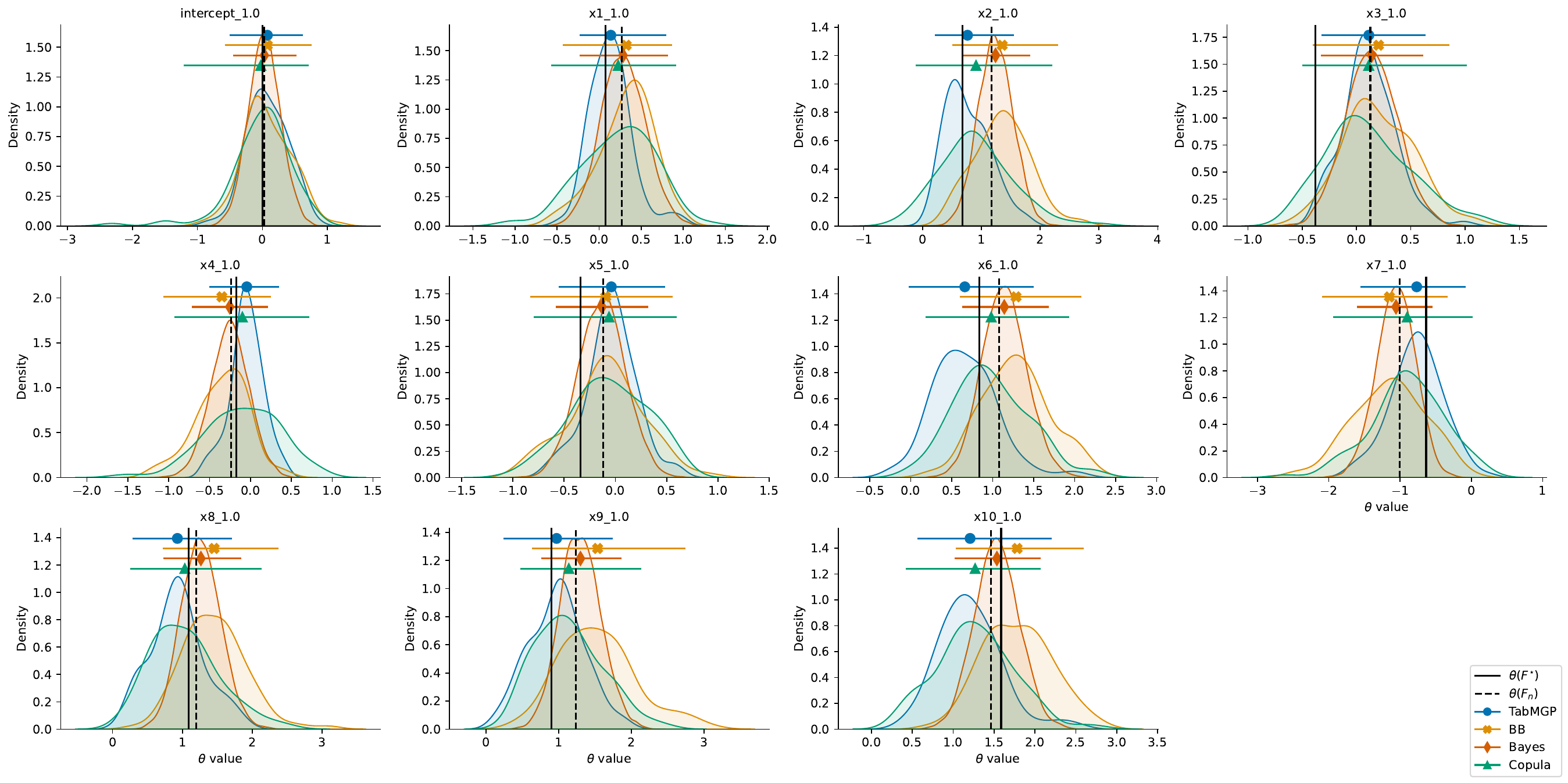}
  \caption{Marginal density plots for logistic regression with synthetic
    Bernoulli data and a logistic link. For each posterior density, the 95\%
    marginal credible interval is shown as a horizontal bar, and the posterior
    mean is marked. The solid and dashed black vertical lines correspond to $\theta(F^{\star})$ and $\theta(F_{n})$, respectively.}
  \label{fig:density-classification}
\end{figure}

\begin{figure}
  \centering
  \includegraphics[width=\columnwidth]{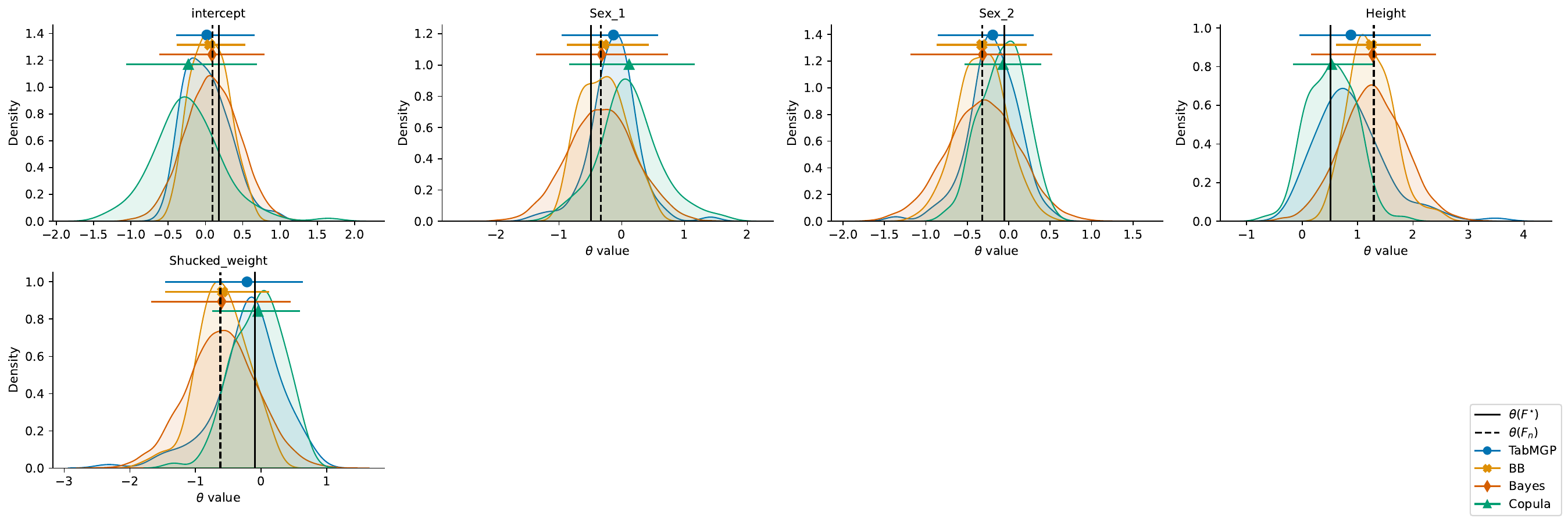}
  \caption{Marginal density plots for the \texttt{abalone} dataset. For each posterior
    density, the 95\% marginal credible interval is shown as a horizontal bar,
    and the posterior mean is marked. The solid and dashed black vertical lines correspond to $\theta(F^{\star})$ and $\theta(F_{n})$, respectively.}
  \label{fig:density-abalone}
\end{figure}

\begin{figure}
  \centering
  \includegraphics[width=\columnwidth]{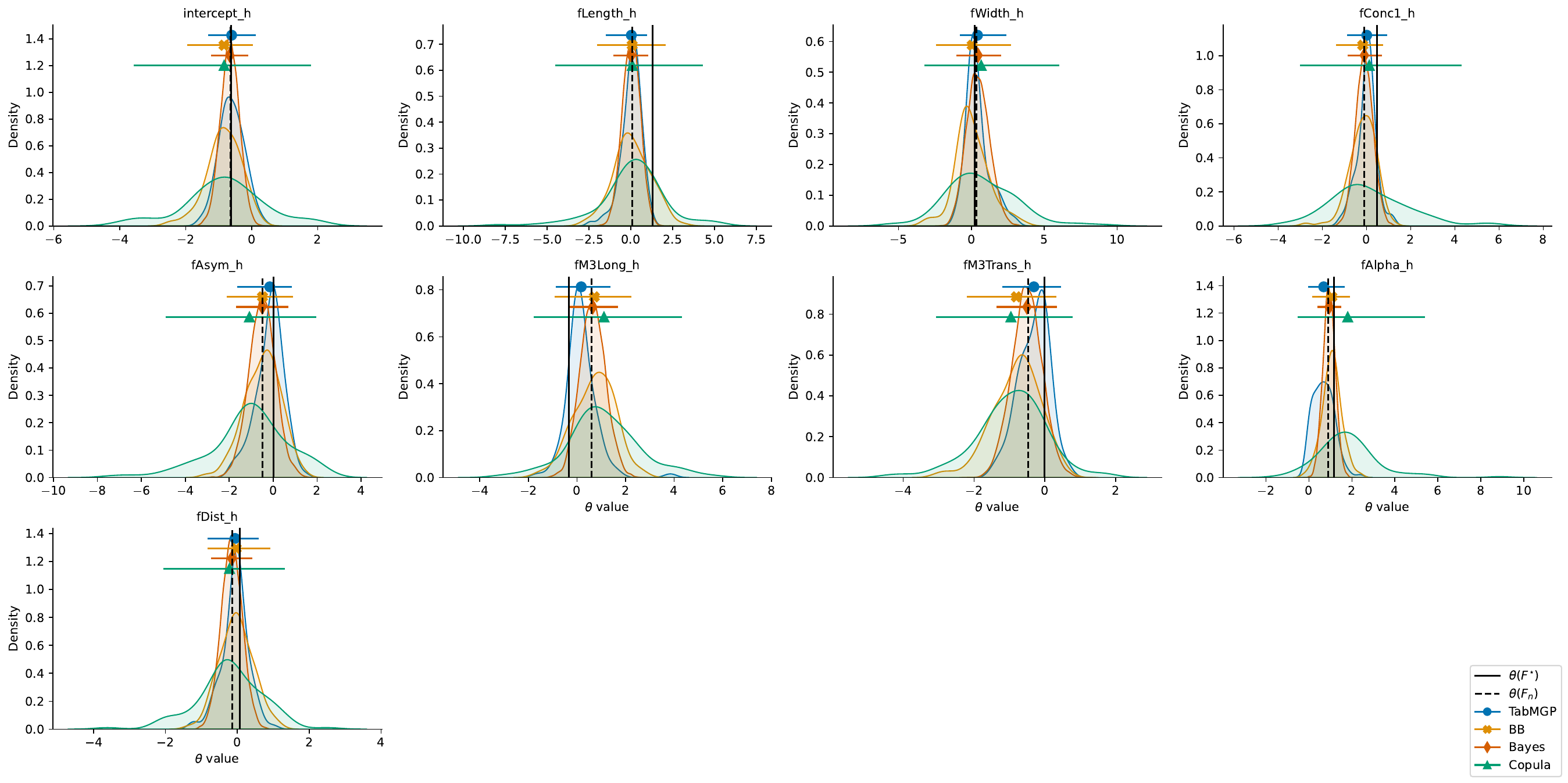}
  \caption{Marginal density plots for the \texttt{telescope} dataset. For each
    posterior density, the 95\% marginal credible interval is shown as a
    horizontal bar, and the posterior mean is marked. The solid and dashed black vertical lines correspond to $\theta(F^{\star})$ and $\theta(F_{n})$, respectively.
  }
  \label{fig:density-telescope}
\end{figure}

\begin{figure}
  \centering
  \includegraphics[width=\columnwidth]{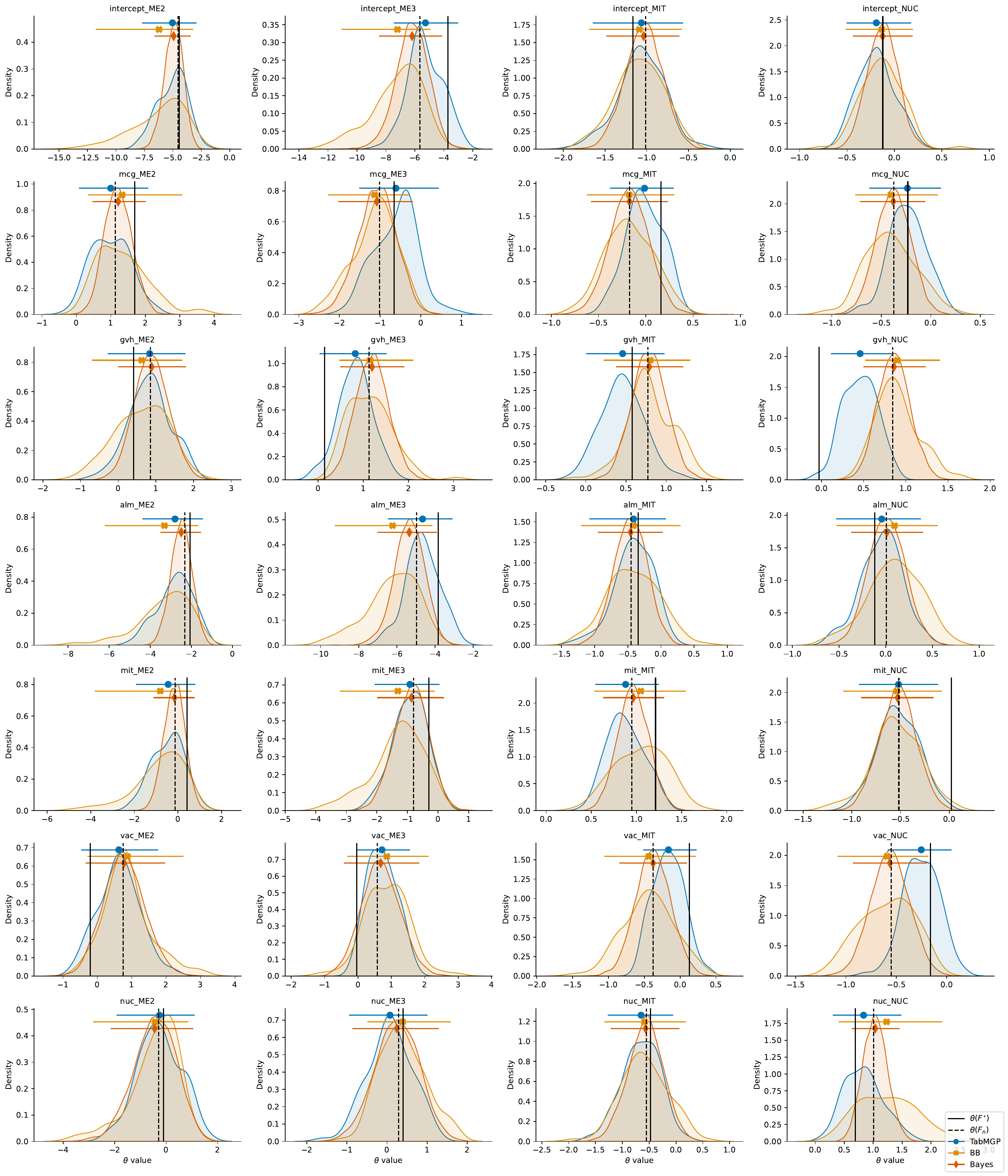}
  \caption{Marginal density plots for the \texttt{yeast} dataset. For each posterior
    density, the 95\% marginal credible interval is shown as a horizontal bar,
    and the posterior mean is marked. The solid and dashed black vertical lines correspond to $\theta(F^{\star})$ and $\theta(F_{n})$, respectively.}
  \label{fig:density-yeast}
\end{figure}

We present density plots of all posterior constructions for a single repetition
of $\vz_{1:n}$. $95\%$ marginal credible intervals are shown as coloured
horizontal bars in each plot. Due to space constraints, we display only the
following setups:
\begin{enumerate}
  \item Linear regression with synthetic i.i.d.~Gaussian data
        (Figure~\ref{fig:density-regression});
  \item Logistic regression with synthetic Bernoulli data with a logistic link
        (Figure~\ref{fig:density-classification});
    \item Linear regression with the \texttt{abalone} dataset
        (Figure~\ref{fig:density-abalone});
    \item Logistic regression with the \texttt{telescope} dataset
        (Figure~\ref{fig:density-telescope});
    \item Multinomial logistic regression with the \texttt{yeast} dataset
        (Figure~\ref{fig:density-yeast}).
\end{enumerate}


\newpage
\section{Coverage Rates of Marginal Credible Intervals}
\label{sec:coverage-marginal-ci}

\begin{figure}[ht!]
  \centering
  \begin{subfigure}[t]{\linewidth}
    \centering
    \includegraphics[width=\linewidth]{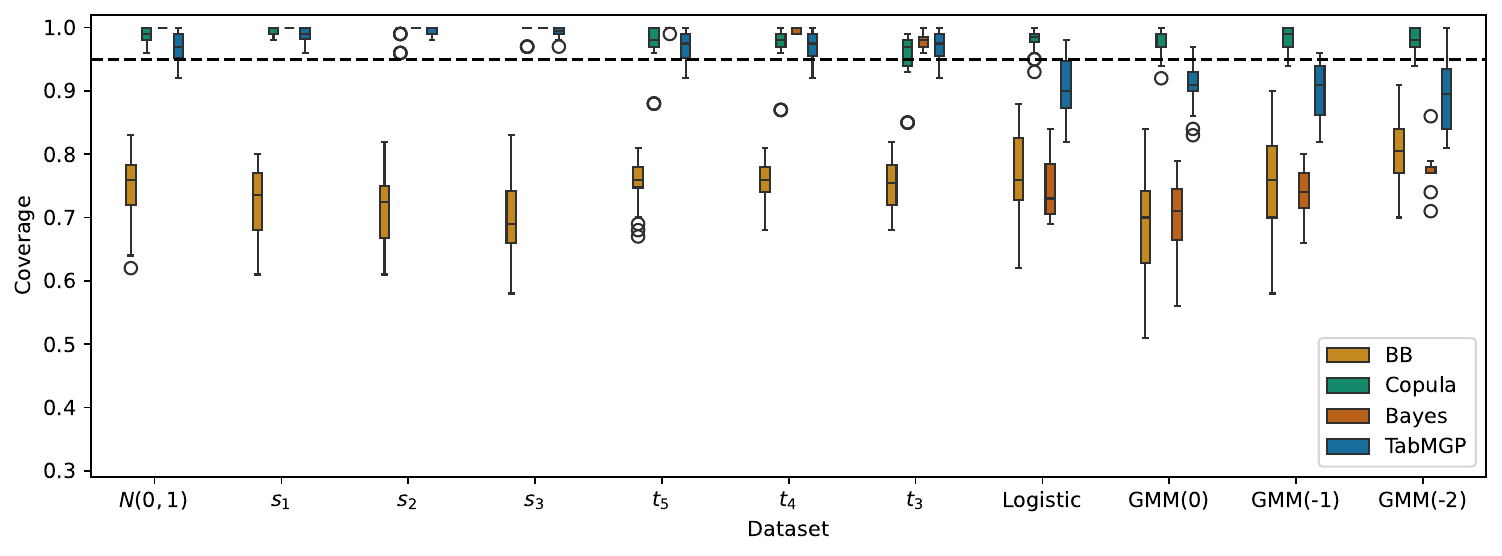}
    \caption{Synthetic data}
  \end{subfigure}%
  \\[1em]
  \begin{subfigure}[t]{\linewidth}
    \centering
    \includegraphics[width=\linewidth]{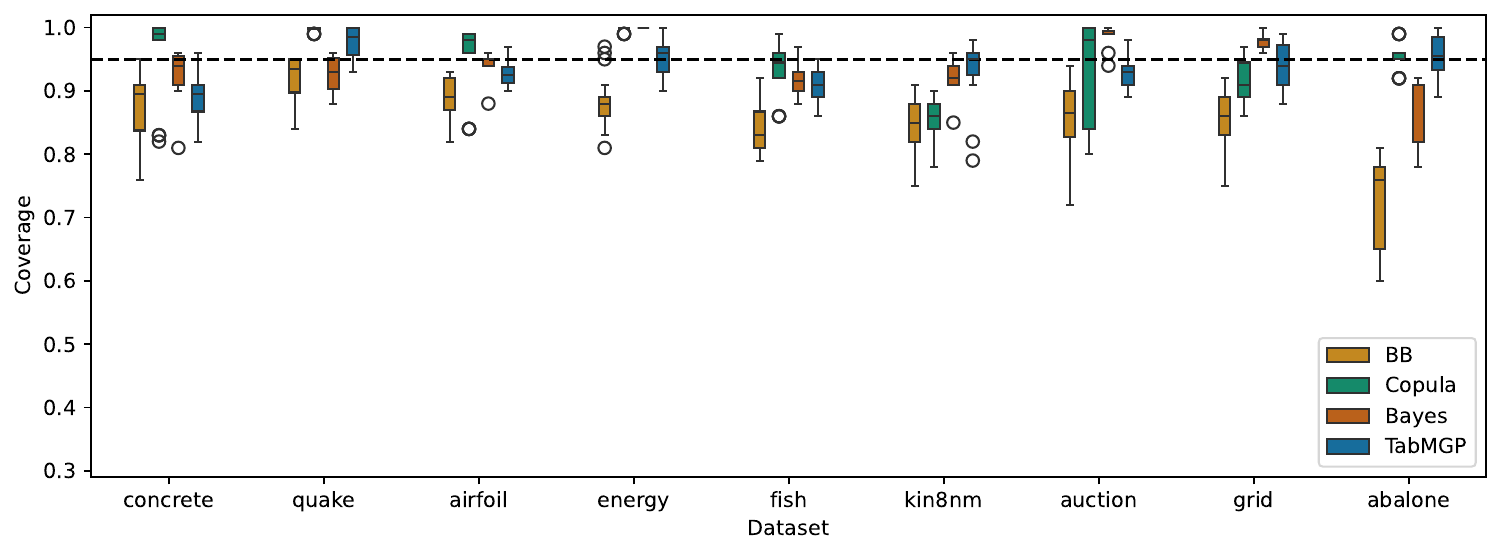}
    \caption{Real-data (linear regression)}
  \end{subfigure}
  \\[1em]
  \begin{subfigure}[t]{\linewidth}
    \centering
    \includegraphics[width=\linewidth]{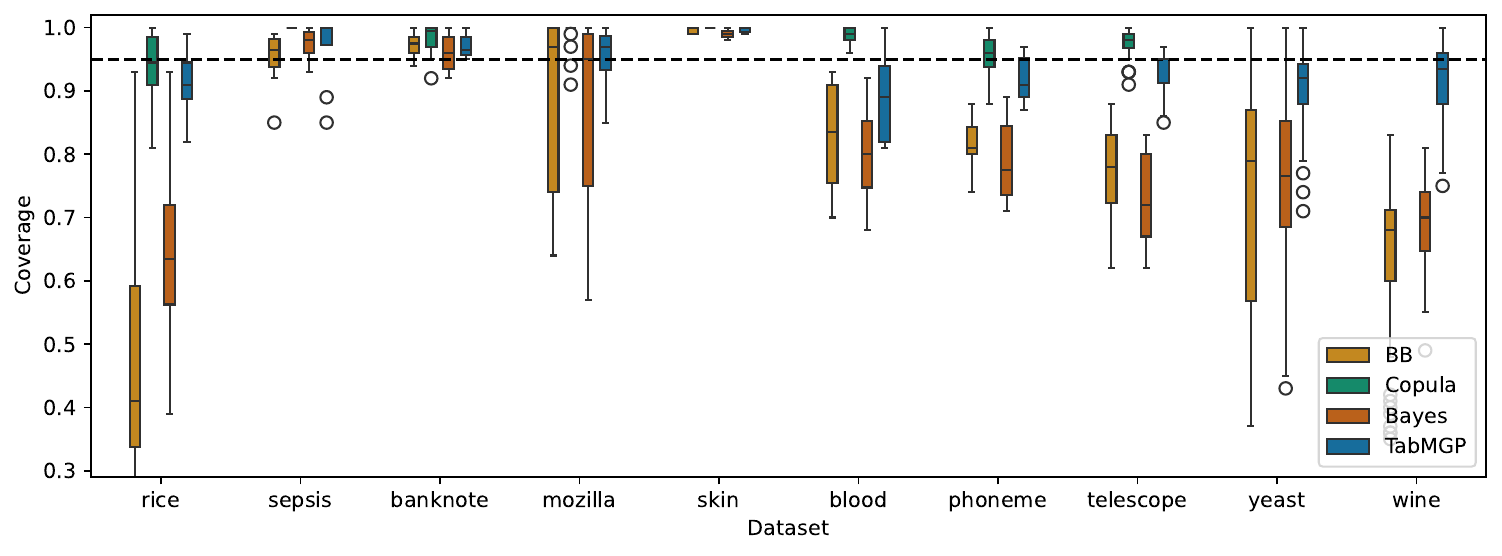}
    \caption{Real-data (logistic regression)}
  \end{subfigure}
  \caption{Coverage of the 95\% marginal credible intervals. Each data point in
    the boxplots represents the coverage for a single dimension of $\theta$ over
    100 realisations of $z_{1:n}$.}
  \label{fig:marginal-ci-coverage}
\end{figure}

In addition to the joint credible sets discussed in the main text, we evaluate
the performance of the $95\%$ marginal credible intervals \eqref{eq:marginal-ci}
for each parameter. Figure~\ref{fig:marginal-ci-coverage} shows boxplots of the
marginal coverage rates for each method across all parameters in our
experiments. While most methods' coverage is centred near the nominal 0.95
level, BB exhibits substantial undercoverage. Bayes also tends to exhibit
undercoverage in logistic regression.

%
%

%
%

The subsequent tables (Table~\ref{tab:marginal-ci-1} to
Table~\ref{tab:marginal-ci-8}) show the empirical coverage and the median width
for each parameter, computed over 100 data realisations. The method achieving at
least the target coverage with the smallest size is highlighted in bold. If none
of the methods achieve the target coverage, the method with the highest coverage
is highlighted in bold instead. Summarising across all 30 setups and 293
individual parameters, TabMGP is highlighted in bold 152 times, followed by Copula (100), the
classical Bayesian posterior (37), and Bayesian bootstrap (4). TabMGP performs
particularly well in multinomial logistic regression relative to the competing
methods.

\newpage

\begin{table}
\centering
\caption{Coverage (rate) and median width (size) over 100 repetitions of
  marginal credible intervals for linear regression with synthetic i.i.d.~data.
  The method achieving at least the target coverage with the smallest size is
  shown in bold. If none of the methods achieve the target coverage, the method
  with the highest coverage is shown in bold. The target coverage is 0.95.}
  \label{tab:marginal-ci-1}
\begin{tabular}{llcccccccc}
\toprule
\multicolumn{2}{c}{ } & \multicolumn{2}{c}{TabMGP} & \multicolumn{2}{c}{BB} & \multicolumn{2}{c}{Copula} & \multicolumn{2}{c}{Bayes} \\
\cmidrule(l{3pt}r{3pt}){3-4} \cmidrule(l{3pt}r{3pt}){5-6} \cmidrule(l{3pt}r{3pt}){7-8} \cmidrule(l{3pt}r{3pt}){9-10}
Setup & Feature Name & Rate & Size & Rate & Size & Rate & Size & Rate & Size\\
\midrule
 & intercept & 0.97 & 0.73 & 0.83 & 0.30 & \textbf{1.00} & \textbf{0.63} & 1.00 & 0.87\\

 & x1 & \textbf{1.00} & \textbf{0.64} & 0.77 & 0.31 & 1.00 & 0.67 & 1.00 & 0.94\\

 & x10 & 0.97 & 0.78 & 0.79 & 0.28 & \textbf{0.98} & \textbf{0.62} & 1.00 & 0.90\\

 & x2 & 0.94 & 0.65 & 0.75 & 0.30 & \textbf{0.99} & \textbf{0.65} & 1.00 & 0.88\\

 & x3 & \textbf{0.99} & \textbf{0.65} & 0.76 & 0.32 & 0.99 & 0.65 & 1.00 & 0.90\\

 & x4 & \textbf{0.99} & \textbf{0.63} & 0.81 & 0.29 & 0.99 & 0.64 & 1.00 & 0.90\\

 & x5 & \textbf{0.99} & \textbf{0.65} & 0.76 & 0.31 & 1.00 & 0.67 & 1.00 & 0.91\\

 & x6 & 0.95 & 0.68 & 0.72 & 0.31 & \textbf{0.97} & \textbf{0.65} & 1.00 & 0.92\\

 & x7 & 0.96 & 0.68 & 0.67 & 0.30 & \textbf{0.98} & \textbf{0.65} & 1.00 & 0.86\\

 & x8 & 0.94 & 0.69 & 0.78 & 0.30 & \textbf{0.96} & \textbf{0.66} & 1.00 & 0.90\\

\multirow{-11}{*}{\raggedright\arraybackslash $\sN(0, 1)$} & x9 & 0.97 & 0.70 & 0.79 & 0.33 & \textbf{0.98} & \textbf{0.67} & 1.00 & 0.90\\
\cmidrule{1-10}
 & intercept & 0.97 & 0.78 & 0.79 & 0.34 & \textbf{1.00} & \textbf{0.65} & 1.00 & 0.87\\

 & x1 & \textbf{1.00} & \textbf{0.67} & 0.78 & 0.36 & 0.99 & 0.67 & 1.00 & 0.95\\

 & x10 & 0.92 & 0.84 & 0.76 & 0.37 & 0.88 & 0.63 & \textbf{1.00} & \textbf{0.90}\\

 & x2 & 0.96 & 0.70 & 0.76 & 0.37 & \textbf{0.96} & \textbf{0.67} & 0.99 & 0.89\\

 & x3 & 0.99 & 0.66 & 0.76 & 0.36 & \textbf{1.00} & \textbf{0.65} & 1.00 & 0.91\\

 & x4 & 1.00 & 0.66 & 0.77 & 0.34 & \textbf{0.97} & \textbf{0.65} & 1.00 & 0.91\\

 & x5 & \textbf{0.99} & \textbf{0.66} & 0.75 & 0.35 & 0.99 & 0.68 & 1.00 & 0.91\\

 & x6 & 0.98 & 0.74 & 0.76 & 0.36 & \textbf{0.97} & \textbf{0.66} & 1.00 & 0.92\\

 & x7 & 0.99 & 0.69 & 0.78 & 0.36 & \textbf{1.00} & \textbf{0.66} & 1.00 & 0.87\\

 & x8 & 0.94 & 0.73 & 0.68 & 0.34 & \textbf{0.97} & \textbf{0.64} & 1.00 & 0.89\\

\multirow{-11}{*}{\raggedright\arraybackslash $t_5$} & x9 & 0.97 & 0.74 & 0.81 & 0.34 & \textbf{0.98} & \textbf{0.66} & 1.00 & 0.90\\
\cmidrule{1-10}
 & intercept & 0.97 & 0.80 & 0.81 & 0.35 & \textbf{1.00} & \textbf{0.64} & 1.00 & 0.87\\

 & x1 & 1.00 & 0.68 & 0.79 & 0.38 & \textbf{0.99} & \textbf{0.67} & 1.00 & 0.95\\

 & x10 & 0.93 & 0.83 & 0.75 & 0.40 & 0.87 & 0.63 & \textbf{0.99} & \textbf{0.90}\\

 & x2 & 0.94 & 0.68 & 0.76 & 0.39 & \textbf{0.96} & \textbf{0.66} & 0.99 & 0.88\\

 & x3 & 0.99 & 0.67 & 0.78 & 0.38 & \textbf{1.00} & \textbf{0.65} & 1.00 & 0.90\\

 & x4 & 1.00 & 0.68 & 0.78 & 0.36 & \textbf{0.98} & \textbf{0.65} & 1.00 & 0.92\\

 & x5 & 0.99 & 0.67 & 0.76 & 0.36 & \textbf{0.99} & \textbf{0.66} & 1.00 & 0.90\\

 & x6 & 0.99 & 0.75 & 0.75 & 0.38 & \textbf{0.97} & \textbf{0.67} & 1.00 & 0.91\\

 & x7 & 0.98 & 0.71 & 0.78 & 0.38 & \textbf{0.98} & \textbf{0.65} & 0.99 & 0.87\\

 & x8 & 0.94 & 0.71 & 0.70 & 0.34 & \textbf{0.97} & \textbf{0.64} & 0.99 & 0.89\\

\multirow{-11}{*}{\raggedright\arraybackslash $t_4$} & x9 & 0.97 & 0.72 & 0.81 & 0.36 & \textbf{0.98} & \textbf{0.67} & 1.00 & 0.89\\
\cmidrule{1-10}
 & intercept & 0.97 & 0.80 & 0.82 & 0.38 & \textbf{0.96} & \textbf{0.62} & 0.99 & 0.87\\

 & x1 & 1.00 & 0.68 & 0.79 & 0.40 & \textbf{0.99} & \textbf{0.66} & 0.99 & 0.95\\

 & x10 & 0.93 & 0.83 & 0.74 & 0.41 & 0.85 & 0.62 & \textbf{0.98} & \textbf{0.91}\\

 & x2 & 0.95 & 0.67 & 0.76 & 0.42 & \textbf{0.96} & \textbf{0.66} & 0.97 & 0.89\\

 & x3 & 0.99 & 0.67 & 0.78 & 0.40 & \textbf{0.98} & \textbf{0.65} & 0.98 & 0.90\\

 & x4 & 1.00 & 0.68 & 0.78 & 0.38 & \textbf{0.98} & \textbf{0.65} & 0.98 & 0.90\\

 & x5 & 0.99 & 0.66 & 0.73 & 0.39 & \textbf{0.98} & \textbf{0.66} & 1.00 & 0.91\\

 & x6 & \textbf{0.98} & \textbf{0.73} & 0.73 & 0.41 & 0.94 & 0.65 & 0.96 & 0.92\\

 & x7 & 0.98 & 0.69 & 0.77 & 0.40 & \textbf{0.98} & \textbf{0.64} & 0.98 & 0.87\\

 & x8 & 0.92 & 0.71 & 0.71 & 0.37 & 0.93 & 0.62 & \textbf{0.97} & \textbf{0.90}\\

\multirow{-11}{*}{\raggedright\arraybackslash $t_3$} & x9 & 0.98 & 0.73 & 0.79 & 0.38 & \textbf{0.97} & \textbf{0.66} & 0.97 & 0.90\\
\bottomrule
\end{tabular}
\end{table}

\begin{table}[ht]
  \centering
  \caption{Coverage (rate) and median width (size) over 100 repetitions of
    marginal credible intervals for linear regression with synthetic,
    dependent-error data. The method achieving at least the target coverage with
    the smallest size is shown in bold. If none of the methods achieve the
    target coverage, the method with the highest coverage is shown in bold. The
    target coverage is 0.95.}
  \label{tab:marginal-ci-2}
\begin{tabular}{llcccccccc}
\toprule
\multicolumn{2}{c}{ } & \multicolumn{2}{c}{TabMGP} & \multicolumn{2}{c}{BB} & \multicolumn{2}{c}{Copula} & \multicolumn{2}{c}{Bayes} \\
\cmidrule(l{3pt}r{3pt}){3-4} \cmidrule(l{3pt}r{3pt}){5-6} \cmidrule(l{3pt}r{3pt}){7-8} \cmidrule(l{3pt}r{3pt}){9-10}
Setup & Feature Name & Rate & Size & Rate & Size & Rate & Size & Rate & Size\\
\midrule
 & intercept & 0.99 & 0.66 & 0.79 & 0.18 & \textbf{1.00} & \textbf{0.63} & 1.00 & 0.86\\

 & x1 & \textbf{1.00} & \textbf{0.55} & 0.73 & 0.19 & 1.00 & 0.67 & 1.00 & 0.94\\

 & x10 & 0.99 & 0.71 & 0.78 & 0.19 & \textbf{0.98} & \textbf{0.64} & 1.00 & 0.91\\

 & x2 & \textbf{0.97} & \textbf{0.61} & 0.73 & 0.18 & 1.00 & 0.65 & 1.00 & 0.89\\

 & x3 & \textbf{1.00} & \textbf{0.59} & 0.80 & 0.20 & 1.00 & 0.66 & 1.00 & 0.90\\

 & x4 & \textbf{1.00} & \textbf{0.57} & 0.77 & 0.19 & 1.00 & 0.66 & 1.00 & 0.91\\

 & x5 & \textbf{1.00} & \textbf{0.59} & 0.77 & 0.20 & 1.00 & 0.66 & 1.00 & 0.91\\

 & x6 & \textbf{0.98} & \textbf{0.64} & 0.67 & 0.19 & 0.99 & 0.65 & 1.00 & 0.92\\

 & x7 & \textbf{0.99} & \textbf{0.59} & 0.65 & 0.19 & 1.00 & 0.64 & 1.00 & 0.87\\

 & x8 & \textbf{0.97} & \textbf{0.63} & 0.72 & 0.19 & 0.98 & 0.66 & 1.00 & 0.89\\

\multirow{-11}{*}{\raggedright\arraybackslash $s_1$} & x9 & \textbf{0.99} & \textbf{0.62} & 0.78 & 0.20 & 1.00 & 0.66 & 1.00 & 0.89\\
\cmidrule{1-10}
 & intercept & 1.00 & 0.65 & 0.82 & 0.14 & \textbf{1.00} & \textbf{0.63} & 1.00 & 0.86\\

 & x1 & \textbf{0.99} & \textbf{0.54} & 0.62 & 0.16 & 1.00 & 0.66 & 1.00 & 0.95\\

 & x10 & 0.99 & 0.70 & 0.75 & 0.15 & \textbf{0.96} & \textbf{0.64} & 1.00 & 0.91\\

 & x2 & \textbf{0.99} & \textbf{0.58} & 0.78 & 0.14 & 1.00 & 0.65 & 1.00 & 0.89\\

 & x3 & \textbf{1.00} & \textbf{0.55} & 0.72 & 0.15 & 1.00 & 0.65 & 1.00 & 0.90\\

 & x4 & \textbf{1.00} & \textbf{0.55} & 0.76 & 0.15 & 1.00 & 0.67 & 1.00 & 0.91\\

 & x5 & \textbf{0.99} & \textbf{0.58} & 0.76 & 0.16 & 1.00 & 0.67 & 1.00 & 0.91\\

 & x6 & \textbf{1.00} & \textbf{0.61} & 0.64 & 0.15 & 0.99 & 0.65 & 1.00 & 0.92\\

 & x7 & \textbf{0.99} & \textbf{0.57} & 0.75 & 0.15 & 1.00 & 0.63 & 1.00 & 0.87\\

 & x8 & \textbf{0.98} & \textbf{0.61} & 0.66 & 0.15 & 1.00 & 0.66 & 1.00 & 0.89\\

\multirow{-11}{*}{\raggedright\arraybackslash $s_2$} & x9 & \textbf{1.00} & \textbf{0.60} & 0.77 & 0.16 & 1.00 & 0.65 & 1.00 & 0.90\\
\cmidrule{1-10}
 & intercept & \textbf{1.00} & \textbf{0.63} & 0.82 & 0.13 & 1.00 & 0.64 & 1.00 & 0.87\\

 & x1 & \textbf{0.99} & \textbf{0.55} & 0.62 & 0.15 & 1.00 & 0.66 & 1.00 & 0.95\\

 & x10 & 0.99 & 0.70 & 0.74 & 0.14 & \textbf{0.97} & \textbf{0.64} & 1.00 & 0.90\\

 & x2 & \textbf{0.98} & \textbf{0.58} & 0.78 & 0.13 & 1.00 & 0.65 & 1.00 & 0.89\\

 & x3 & \textbf{0.99} & \textbf{0.56} & 0.68 & 0.15 & 1.00 & 0.66 & 1.00 & 0.91\\

 & x4 & \textbf{1.00} & \textbf{0.57} & 0.69 & 0.14 & 1.00 & 0.67 & 1.00 & 0.91\\

 & x5 & \textbf{1.00} & \textbf{0.58} & 0.75 & 0.15 & 1.00 & 0.67 & 1.00 & 0.90\\

 & x6 & \textbf{1.00} & \textbf{0.61} & 0.62 & 0.15 & 1.00 & 0.65 & 1.00 & 0.92\\

 & x7 & \textbf{1.00} & \textbf{0.56} & 0.76 & 0.14 & 1.00 & 0.63 & 1.00 & 0.87\\

 & x8 & \textbf{0.98} & \textbf{0.60} & 0.67 & 0.14 & 1.00 & 0.66 & 1.00 & 0.90\\

\multirow{-11}{*}{\raggedright\arraybackslash $s_3$} & x9 & \textbf{1.00} & \textbf{0.61} & 0.74 & 0.15 & 1.00 & 0.66 & 1.00 & 0.89\\
\bottomrule
\end{tabular}
\end{table}

\begin{table}[ht]
  \centering
  \caption{Coverage (rate) and median width (size) over 100 repetitions of
    marginal credible intervals for logistic regression with synthetic data.
    The method achieving at least the target coverage with the smallest size is
    shown in bold. If none of the methods achieve the target coverage, the method
    with the highest coverage is shown in bold. The target coverage is 0.95.}
  \label{tab:marginal-ci-3}
\begin{tabular}{llcccccccc}
\toprule
\multicolumn{2}{c}{ } & \multicolumn{2}{c}{TabMGP} & \multicolumn{2}{c}{BB} & \multicolumn{2}{c}{Copula} & \multicolumn{2}{c}{Bayes} \\
\cmidrule(l{3pt}r{3pt}){3-4} \cmidrule(l{3pt}r{3pt}){5-6} \cmidrule(l{3pt}r{3pt}){7-8} \cmidrule(l{3pt}r{3pt}){9-10}
Setup & Feature Name & Rate & Size & Rate & Size & Rate & Size & Rate & Size\\
\midrule
 & intercept\_1.0 & \textbf{0.96} & \textbf{1.11} & 0.88 & 1.36 & 0.98 & 1.67 & 0.84 & 0.93\\

 & x10\_1.0 & 0.86 & 1.76 & 0.66 & 2.07 & \textbf{0.99} & \textbf{2.24} & 0.69 & 1.25\\

 & x1\_1.0 & \textbf{0.97} & \textbf{0.97} & 0.86 & 1.44 & 0.99 & 1.69 & 0.77 & 0.95\\

 & x2\_1.0 & 0.85 & 1.19 & 0.76 & 1.56 & \textbf{0.98} & \textbf{1.82} & 0.71 & 1.01\\

 & x3\_1.0 & \textbf{0.95} & \textbf{1.04} & 0.81 & 1.41 & 1.00 & 1.69 & 0.80 & 0.97\\

 & x4\_1.0 & \textbf{0.98} & \textbf{0.99} & 0.85 & 1.41 & 0.99 & 1.71 & 0.73 & 0.94\\

 & x5\_1.0 & 0.94 & 1.04 & 0.86 & 1.44 & \textbf{0.99} & \textbf{1.73} & 0.81 & 0.95\\

 & x6\_1.0 & 0.90 & 1.34 & 0.77 & 1.56 & \textbf{1.00} & \textbf{1.88} & 0.75 & 1.04\\

 & x7\_1.0 & 0.89 & 1.21 & 0.79 & 1.54 & \textbf{1.00} & \textbf{1.85} & 0.72 & 1.00\\

 & x8\_1.0 & 0.85 & 1.48 & 0.75 & 1.74 & \textbf{0.99} & \textbf{2.01} & 0.70 & 1.11\\

\multirow{-11}{*}{\raggedright\arraybackslash Logistic} & x9\_1.0 & 0.89 & 1.44 & 0.73 & 1.67 & \textbf{0.99} & \textbf{1.94} & 0.69 & 1.08\\
\cmidrule{1-10}
 & intercept\_1.0 & 0.90 & 1.48 & 0.73 & 2.02 & \textbf{0.99} & \textbf{2.16} & 0.70 & 1.24\\

 & x10\_1.0 & 0.91 & 2.53 & 0.52 & 3.37 & \textbf{1.00} & \textbf{3.00} & 0.56 & 1.87\\

 & x1\_1.0 & \textbf{0.96} & \textbf{1.21} & 0.82 & 1.84 & 1.00 & 1.98 & 0.74 & 1.14\\

 & x2\_1.0 & 0.86 & 1.61 & 0.70 & 2.09 & \textbf{0.97} & \textbf{2.16} & 0.71 & 1.27\\

 & x3\_1.0 & 0.91 & 1.33 & 0.79 & 1.79 & \textbf{0.99} & \textbf{1.96} & 0.75 & 1.16\\

 & x4\_1.0 & \textbf{0.97} & \textbf{1.26} & 0.79 & 1.77 & 0.99 & 2.00 & 0.73 & 1.16\\

 & x5\_1.0 & 0.94 & 1.24 & 0.84 & 1.76 & \textbf{0.97} & \textbf{1.89} & 0.79 & 1.16\\

 & x6\_1.0 & 0.91 & 1.86 & 0.68 & 2.37 & \textbf{1.00} & \textbf{2.37} & 0.67 & 1.35\\

 & x7\_1.0 & 0.93 & 1.60 & 0.74 & 2.02 & \textbf{0.99} & \textbf{2.11} & 0.75 & 1.26\\

 & x8\_1.0 & 0.90 & 2.07 & 0.62 & 2.76 & \textbf{0.99} & \textbf{2.51} & 0.63 & 1.50\\

\multirow{-11}{*}{\raggedright\arraybackslash GMM$(0)$} & x9\_1.0 & \textbf{0.95} & \textbf{1.92} & 0.64 & 2.54 & 1.00 & 2.33 & 0.66 & 1.43\\
\cmidrule{1-10}
 & intercept\_1.0 & \textbf{0.96} & \textbf{1.10} & 0.89 & 1.36 & 0.98 & 1.67 & 0.80 & 0.94\\

 & x10\_1.0 & 0.90 & 1.78 & 0.66 & 1.96 & \textbf{1.00} & \textbf{2.25} & 0.71 & 1.26\\

 & x1\_1.0 & \textbf{0.96} & \textbf{0.96} & 0.82 & 1.46 & 1.00 & 1.68 & 0.76 & 0.96\\

 & x2\_1.0 & 0.85 & 1.24 & 0.77 & 1.54 & \textbf{0.99} & \textbf{1.80} & 0.76 & 1.02\\

 & x3\_1.0 & 0.91 & 1.06 & 0.83 & 1.43 & \textbf{0.99} & \textbf{1.68} & 0.78 & 0.97\\

 & x4\_1.0 & \textbf{0.96} & \textbf{0.99} & 0.82 & 1.39 & 0.99 & 1.69 & 0.74 & 0.95\\

 & x5\_1.0 & 0.93 & 1.02 & 0.88 & 1.42 & \textbf{1.00} & \textbf{1.71} & 0.79 & 0.95\\

 & x6\_1.0 & 0.88 & 1.36 & 0.75 & 1.59 & \textbf{1.00} & \textbf{1.86} & 0.72 & 1.04\\

 & x7\_1.0 & 0.86 & 1.18 & 0.78 & 1.55 & \textbf{1.00} & \textbf{1.82} & 0.74 & 0.98\\

 & x8\_1.0 & 0.86 & 1.51 & 0.68 & 1.81 & \textbf{0.99} & \textbf{2.02} & 0.66 & 1.11\\

\multirow{-11}{*}{\raggedright\arraybackslash GMM$(-1)$} & x9\_1.0 & 0.94 & 1.46 & 0.73 & 1.65 & \textbf{1.00} & \textbf{1.92} & 0.66 & 1.07\\
\cmidrule{1-10}
 & intercept\_1.0 & 0.92 & 1.05 & 0.84 & 1.35 & \textbf{0.98} & \textbf{1.67} & 0.79 & 0.89\\

 & x10\_1.0 & 0.85 & 1.42 & 0.73 & 1.48 & \textbf{1.00} & \textbf{1.82} & 0.77 & 1.03\\

 & x1\_1.0 & \textbf{0.97} & \textbf{0.81} & 0.89 & 1.22 & 0.97 & 1.56 & 0.86 & 0.85\\

 & x2\_1.0 & 0.87 & 0.94 & 0.78 & 1.28 & \textbf{0.99} & \textbf{1.60} & 0.77 & 0.89\\

 & x3\_1.0 & 0.94 & 0.85 & 0.88 & 1.24 & \textbf{0.97} & \textbf{1.58} & 0.78 & 0.86\\

 & x4\_1.0 & \textbf{0.99} & \textbf{0.84} & 0.91 & 1.21 & 1.00 & 1.54 & 0.78 & 0.85\\

 & x5\_1.0 & 0.90 & 0.88 & 0.82 & 1.26 & \textbf{1.00} & \textbf{1.53} & 0.78 & 0.86\\

 & x6\_1.0 & 0.91 & 1.05 & 0.84 & 1.33 & \textbf{1.00} & \textbf{1.67} & 0.77 & 0.91\\

 & x7\_1.0 & 0.83 & 0.97 & 0.84 & 1.33 & \textbf{1.00} & \textbf{1.61} & 0.77 & 0.87\\

 & x8\_1.0 & 0.81 & 1.18 & 0.78 & 1.37 & \textbf{1.00} & \textbf{1.74} & 0.74 & 0.92\\

\multirow{-11}{*}{\raggedright\arraybackslash GMM$(-2)$} & x9\_1.0 & 0.83 & 1.08 & 0.79 & 1.33 & \textbf{1.00} & \textbf{1.65} & 0.71 & 0.91\\
\bottomrule
\end{tabular}
\end{table}

\begin{table}[ht]
  \centering
  \caption{Coverage (rate) and median width (size) over 100 repetitions of
    marginal credible intervals for linear regression with real data.
    The method achieving at least the target coverage with the smallest size is
    shown in bold. If none of the methods achieve the target coverage, the method
    with the highest coverage is shown in bold. The target coverage is 0.95.}
  \label{tab:marginal-ci-4}
\begin{tabular}{llcccccccc}
\toprule
\multicolumn{2}{c}{ } & \multicolumn{2}{c}{TabMGP} & \multicolumn{2}{c}{BB} & \multicolumn{2}{c}{Copula} & \multicolumn{2}{c}{Bayes} \\
\cmidrule(l{3pt}r{3pt}){3-4} \cmidrule(l{3pt}r{3pt}){5-6} \cmidrule(l{3pt}r{3pt}){7-8} \cmidrule(l{3pt}r{3pt}){9-10}
Setup & Feature Name & Rate & Size & Rate & Size & Rate & Size & Rate & Size\\
\midrule
 & age & 0.83 & 0.36 & 0.83 & 0.35 & \textbf{0.83} & \textbf{0.32} & 0.81 & 0.29\\

 & cement & 0.90 & 0.32 & 0.91 & 0.30 & \textbf{0.98} & \textbf{0.50} & 0.92 & 0.34\\

 & coarse\_aggregate & 0.91 & 0.29 & 0.91 & 0.27 & \textbf{0.99} & \textbf{0.47} & 0.94 & 0.30\\

 & fine\_aggregate & 0.95 & 0.29 & \textbf{0.95} & \textbf{0.26} & 1.00 & 0.45 & 0.96 & 0.31\\

 & fly\_ash & \textbf{0.96} & \textbf{0.34} & 0.92 & 0.32 & 0.99 & 0.63 & 0.96 & 0.36\\

 & intercept & 0.91 & 0.29 & 0.92 & 0.26 & 1.00 & 0.55 & \textbf{0.95} & \textbf{0.28}\\

\multirow{-7}{*}{\raggedright\arraybackslash concrete} & superplasticizer & 0.91 & 0.37 & 0.85 & 0.35 & \textbf{0.99} & \textbf{0.49} & 0.90 & 0.36\\
\cmidrule{1-10}
 & col\_1 & \textbf{1.00} & \textbf{0.23} & 0.88 & 0.44 & 0.99 & 0.50 & 0.95 & 0.45\\

 & col\_2 & 0.93 & 0.25 & 0.94 & 0.51 & \textbf{1.00} & \textbf{0.80} & 0.88 & 0.41\\

 & col\_3 & \textbf{1.00} & \textbf{0.24} & 0.95 & 0.50 & 1.00 & 0.81 & 0.91 & 0.40\\

\multirow{-4}{*}{\raggedright\arraybackslash quake} & intercept & 0.96 & 0.47 & 0.95 & 0.50 & 1.00 & 1.02 & \textbf{0.96} & \textbf{0.39}\\
\cmidrule{1-10}
 & chord\_length & 0.96 & 0.44 & 0.91 & 0.38 & 0.98 & 0.55 & \textbf{0.96} & \textbf{0.42}\\

 & displacement\_thickness & 0.93 & 0.45 & 0.87 & 0.40 & \textbf{0.96} & \textbf{0.59} & 0.94 & 0.42\\

 & free\_stream\_velocity & 0.94 & 0.40 & 0.93 & 0.35 & 0.99 & 0.55 & \textbf{0.95} & \textbf{0.41}\\

 & frequency & \textbf{0.90} & \textbf{0.48} & 0.87 & 0.43 & 0.84 & 0.47 & 0.88 & 0.42\\

\multirow{-5}{*}{\raggedright\arraybackslash airfoil} & intercept & 0.93 & 0.42 & 0.93 & 0.36 & 0.99 & 0.63 & \textbf{0.95} & \textbf{0.40}\\
\cmidrule{1-10}
 & glazing\_area & 0.93 & 0.25 & 0.86 & 0.15 & \textbf{1.00} & \textbf{0.34} & 1.00 & 0.42\\

 & glazing\_area\_distribution & \textbf{0.96} & \textbf{0.25} & 0.89 & 0.15 & 0.99 & 0.33 & 1.00 & 0.43\\

 & intercept & 0.92 & 0.28 & 0.88 & 0.15 & 1.00 & 0.45 & \textbf{1.00} & \textbf{0.40}\\

 & orientation & \textbf{0.96} & \textbf{0.24} & 0.88 & 0.15 & 1.00 & 0.34 & 1.00 & 0.41\\

 & overall\_height & \textbf{0.97} & \textbf{0.26} & 0.89 & 0.15 & 1.00 & 0.49 & 1.00 & 0.42\\

\multirow{-6}{*}{\raggedright\arraybackslash energy} & wall\_area & 1.00 & 0.25 & \textbf{0.96} & \textbf{0.12} & 1.00 & 0.35 & 1.00 & 0.43\\
\cmidrule{1-10}
 & CIC0 & 0.89 & 0.46 & 0.86 & 0.43 & 0.86 & 0.49 & \textbf{0.90} & \textbf{0.43}\\

 & GATS1i & 0.92 & 0.41 & 0.92 & 0.34 & 0.96 & 0.46 & \textbf{0.97} & \textbf{0.42}\\

 & NdsCH & 0.95 & 0.45 & 0.81 & 0.36 & \textbf{0.95} & \textbf{0.45} & 0.93 & 0.45\\

 & NdssC & \textbf{0.94} & \textbf{0.46} & 0.89 & 0.44 & 0.94 & 0.47 & 0.93 & 0.48\\

 & SM1\_Dz & 0.89 & 0.48 & 0.83 & 0.38 & \textbf{0.92} & \textbf{0.48} & 0.88 & 0.45\\

\multirow{-6}{*}{\raggedright\arraybackslash fish} & intercept & 0.93 & 0.43 & 0.83 & 0.35 & \textbf{0.99} & \textbf{0.54} & 0.90 & 0.41\\
\cmidrule{1-10}
 & intercept & 0.94 & 0.47 & 0.91 & 0.38 & 0.90 & 0.35 & \textbf{0.94} & \textbf{0.42}\\

 & theta1 & \textbf{0.96} & \textbf{0.41} & 0.85 & 0.38 & 0.86 & 0.36 & 0.92 & 0.44\\

 & theta2 & \textbf{0.98} & \textbf{0.40} & 0.86 & 0.37 & 0.88 & 0.35 & 0.96 & 0.43\\

 & theta3 & \textbf{0.96} & \textbf{0.52} & 0.90 & 0.36 & 0.78 & 0.34 & 0.94 & 0.43\\

 & theta4 & \textbf{0.96} & \textbf{0.42} & 0.90 & 0.39 & 0.90 & 0.34 & 0.93 & 0.43\\

 & theta5 & 0.82 & 0.44 & 0.87 & 0.40 & 0.86 & 0.33 & \textbf{0.91} & \textbf{0.42}\\

 & theta6 & \textbf{0.92} & \textbf{0.43} & 0.79 & 0.39 & 0.78 & 0.34 & 0.85 & 0.43\\

 & theta7 & \textbf{0.94} & \textbf{0.42} & 0.84 & 0.38 & 0.84 & 0.35 & 0.91 & 0.43\\

\multirow{-9}{*}{\raggedright\arraybackslash kin8nm} & theta8 & \textbf{0.96} & \textbf{0.41} & 0.84 & 0.37 & 0.87 & 0.35 & 0.92 & 0.42\\
\bottomrule
\end{tabular}
\end{table}

\begin{table}[ht]
  \centering
  \caption{Coverage (rate) and median width (size) over 100 repetitions of
    marginal credible intervals for linear regression with real data.
    The method achieving at least the target coverage with the smallest size is
    shown in bold. If none of the methods achieve the target coverage, the method
    with the highest coverage is shown in bold. The target coverage is 0.95.}
  \label{tab:marginal-ci-5}
\begin{tabular}{llcccccccc}
\toprule
\multicolumn{2}{c}{ } & \multicolumn{2}{c}{TabMGP} & \multicolumn{2}{c}{BB} & \multicolumn{2}{c}{Copula} & \multicolumn{2}{c}{Bayes} \\
\cmidrule(l{3pt}r{3pt}){3-4} \cmidrule(l{3pt}r{3pt}){5-6} \cmidrule(l{3pt}r{3pt}){7-8} \cmidrule(l{3pt}r{3pt}){9-10}
Setup & Feature Name & Rate & Size & Rate & Size & Rate & Size & Rate & Size\\
\midrule
 & intercept & 0.93 & 0.74 & 0.89 & 0.53 & 1.00 & 1.70 & \textbf{0.99} & \textbf{0.87}\\

 & process.b1.capacity & \textbf{0.98} & \textbf{0.50} & 0.94 & 0.40 & 1.00 & 1.04 & 0.99 & 0.51\\

 & process.b2.capacity & 0.90 & 0.39 & 0.80 & 0.32 & 1.00 & 0.84 & \textbf{0.96} & \textbf{0.46}\\

 & process.b3.capacity & 0.92 & 0.33 & 0.94 & 0.26 & 0.98 & 0.66 & \textbf{1.00} & \textbf{0.48}\\

 & process.b4.capacity & 0.93 & 0.40 & 0.86 & 0.32 & 1.00 & 1.20 & \textbf{0.99} & \textbf{0.45}\\

 & property.price & 0.91 & 0.45 & 0.81 & 0.37 & 1.00 & 1.03 & \textbf{0.99} & \textbf{0.52}\\

 & property.product\_2 & 0.91 & 1.39 & 0.87 & 1.08 & 0.84 & 1.61 & \textbf{0.94} & \textbf{1.26}\\

 & property.product\_3 & 0.93 & 1.06 & 0.89 & 0.85 & 0.80 & 1.61 & \textbf{0.99} & \textbf{1.48}\\

 & property.product\_4 & \textbf{0.96} & \textbf{1.04} & 0.89 & 0.84 & 0.88 & 1.39 & 1.00 & 1.44\\

 & property.product\_5 & 0.94 & 1.14 & 0.89 & 0.89 & 0.97 & 1.67 & \textbf{1.00} & \textbf{1.64}\\

\multirow{-11}{*}{\raggedright\arraybackslash auction} & property.product\_6 & 0.94 & 0.99 & 0.93 & 0.77 & 0.82 & 1.50 & \textbf{0.99} & \textbf{1.34}\\
\cmidrule{1-10}
 & g1 & 0.92 & 0.39 & 0.82 & 0.30 & 0.86 & 0.34 & \textbf{0.97} & \textbf{0.45}\\

 & g2 & \textbf{0.98} & \textbf{0.40} & 0.88 & 0.30 & 0.94 & 0.33 & 1.00 & 0.45\\

 & g3 & \textbf{0.96} & \textbf{0.39} & 0.86 & 0.30 & 0.89 & 0.33 & 0.97 & 0.45\\

 & g4 & \textbf{0.98} & \textbf{0.40} & 0.92 & 0.31 & 0.93 & 0.33 & 0.97 & 0.45\\

 & intercept & 0.92 & 0.40 & 0.91 & 0.30 & 0.86 & 0.33 & \textbf{0.98} & \textbf{0.44}\\

 & p2 & 0.97 & 0.36 & 0.90 & 0.29 & \textbf{0.96} & \textbf{0.34} & 0.98 & 0.45\\

 & p3 & 0.99 & 0.36 & 0.91 & 0.29 & \textbf{0.97} & \textbf{0.34} & 0.99 & 0.44\\

 & p4 & 0.99 & 0.38 & 0.87 & 0.29 & \textbf{0.97} & \textbf{0.34} & 0.98 & 0.44\\

 & tau1 & 0.91 & 0.39 & 0.81 & 0.29 & 0.91 & 0.32 & \textbf{0.96} & \textbf{0.44}\\

 & tau2 & 0.94 & 0.39 & 0.89 & 0.29 & 0.89 & 0.33 & \textbf{0.99} & \textbf{0.44}\\

 & tau3 & 0.91 & 0.38 & 0.86 & 0.29 & 0.90 & 0.33 & \textbf{0.96} & \textbf{0.44}\\

\multirow{-12}{*}{\raggedright\arraybackslash grid} & tau4 & 0.94 & 0.40 & 0.89 & 0.30 & 0.91 & 0.33 & \textbf{0.98} & \textbf{0.44}\\
\cmidrule{1-10}
 & Height & 0.99 & 1.53 & 0.65 & 1.23 & \textbf{0.96} & \textbf{1.36} & 0.78 & 1.49\\

 & Sex\_1 & 0.92 & 1.74 & 0.78 & 1.53 & \textbf{0.95} & \textbf{1.78} & 0.92 & 1.92\\

 & Sex\_2 & \textbf{0.99} & \textbf{1.48} & 0.81 & 1.47 & 0.95 & 1.55 & 0.91 & 1.63\\

 & Shucked\_weight & \textbf{0.96} & \textbf{1.40} & 0.68 & 1.23 & 0.92 & 1.48 & 0.82 & 1.36\\

\multirow{-5}{*}{\raggedright\arraybackslash abalone} & intercept & 0.94 & 1.23 & 0.76 & 1.01 & \textbf{0.99} & \textbf{1.76} & 0.91 & 1.23\\
\bottomrule
\end{tabular}
\end{table}

\begin{table}[ht]
  \centering
  \caption{Coverage (rate) and median width (size) over 100 repetitions of
    marginal credible intervals for logistic regression with real data.
    The method achieving at least the target coverage with the smallest size is
    shown in bold. If none of the methods achieve the target coverage, the method
    with the highest coverage is shown in bold. The target coverage is 0.95.}
  \label{tab:marginal-ci-6}
\begin{tabular}{llcccccccc}
\toprule
\multicolumn{2}{c}{ } & \multicolumn{2}{c}{TabMGP} & \multicolumn{2}{c}{BB} & \multicolumn{2}{c}{Copula} & \multicolumn{2}{c}{Bayes} \\
\cmidrule(l{3pt}r{3pt}){3-4} \cmidrule(l{3pt}r{3pt}){5-6} \cmidrule(l{3pt}r{3pt}){7-8} \cmidrule(l{3pt}r{3pt}){9-10}
Setup & Feature Name & Rate & Size & Rate & Size & Rate & Size & Rate & Size\\
\midrule
 & Eccentricity\_Osmancik & 0.91 & 5.36 & 0.37 & 7.52 & \textbf{0.98} & \textbf{14.46} & 0.62 & 4.81\\

 & Extent\_Osmancik & \textbf{0.99} & \textbf{1.19} & 0.93 & 2.11 & 1.00 & 4.14 & 0.93 & 1.40\\

 & Minor\_Axis\_Length\_Osmancik & 0.93 & 3.91 & 0.49 & 5.55 & \textbf{0.97} & \textbf{11.15} & 0.65 & 3.38\\

\multirow{-4}{*}{\raggedright\arraybackslash rice} & intercept\_Osmancik & 0.89 & 2.65 & 0.30 & 3.42 & \textbf{0.97} & \textbf{6.20} & 0.39 & 2.28\\
\cmidrule{1-10}
 & age\_years\_1 & 0.85 & 1.72 & \textbf{0.96} & \textbf{2.67} & 1.00 & 4.88 & 0.93 & 2.10\\

 & episode\_number\_1 & \textbf{1.00} & \textbf{1.19} & 0.93 & 2.54 & 1.00 & 3.28 & 0.99 & 1.83\\

 & intercept\_1 & \textbf{1.00} & \textbf{1.77} & 0.98 & 2.91 & 1.00 & 6.42 & 0.97 & 1.95\\

\multirow{-4}{*}{\raggedright\arraybackslash sepsis} & sex\_0male\_1female\_1\_1 & \textbf{1.00} & \textbf{1.73} & 0.99 & 3.59 & 1.00 & 7.84 & 1.00 & 2.52\\
\cmidrule{1-10}
 & V1\_2 & 0.96 & 2.24 & 0.96 & 2.62 & 1.00 & 2.85 & \textbf{0.98} & \textbf{1.96}\\

 & V2\_2 & \textbf{0.97} & \textbf{2.23} & 0.98 & 2.49 & 0.97 & 2.44 & 0.94 & 1.71\\

 & V4\_2 & \textbf{0.98} & \textbf{1.99} & 0.98 & 2.25 & 0.99 & 2.30 & 0.92 & 1.38\\

\multirow{-4}{*}{\raggedright\arraybackslash banknote} & intercept\_2 & 1.00 & 1.33 & 1.00 & 1.44 & 1.00 & 1.45 & \textbf{1.00} & \textbf{1.01}\\
\cmidrule{1-10}
 & end\_1 & 0.98 & 0.99 & 0.97 & 1.10 & 1.00 & 2.08 & \textbf{0.99} & \textbf{0.74}\\

 & event\_1 & 1.00 & 0.78 & 1.00 & 0.92 & 1.00 & 2.05 & \textbf{1.00} & \textbf{0.68}\\

 & id\_1 & 0.99 & 0.77 & 1.00 & 0.91 & 1.00 & 1.83 & \textbf{0.95} & \textbf{0.66}\\

 & intercept\_1 & 0.94 & 1.44 & 0.77 & 1.76 & \textbf{1.00} & \textbf{2.40} & 0.75 & 1.04\\

\multirow{-5}{*}{\raggedright\arraybackslash mozilla} & size\_1 & 0.93 & 3.96 & 0.74 & 4.69 & \textbf{0.99} & \textbf{4.41} & 0.57 & 2.72\\
\cmidrule{1-10}
 & V1\_2 & 1.00 & 3.00 & 1.00 & 3.29 & 1.00 & 3.23 & \textbf{0.99} & \textbf{1.59}\\

 & V3\_2 & 0.99 & 2.91 & 0.99 & 3.18 & 1.00 & 3.37 & \textbf{0.98} & \textbf{1.76}\\

\multirow{-3}{*}{\raggedright\arraybackslash skin} & intercept\_2 & \textbf{1.00} & \textbf{1.70} & 1.00 & 1.85 & 1.00 & 2.11 & 1.00 & 1.76\\
\cmidrule{1-10}
 & V1\_2 & 0.82 & 1.80 & 0.91 & 2.64 & \textbf{0.98} & \textbf{4.18} & 0.83 & 1.63\\

 & V2\_2 & 0.91 & 2.30 & 0.74 & 3.51 & \textbf{0.99} & \textbf{4.98} & 0.68 & 2.19\\

 & V4\_2 & 0.81 & 1.76 & 0.79 & 2.74 & \textbf{1.00} & \textbf{4.01} & 0.77 & 1.91\\

\multirow{-4}{*}{\raggedright\arraybackslash blood} & intercept\_2 & \textbf{1.00} & \textbf{1.48} & 0.93 & 2.14 & 1.00 & 3.68 & 0.92 & 1.28\\
\cmidrule{1-10}
 & V1\_2 & \textbf{0.96} & \textbf{1.93} & 0.85 & 2.05 & 1.00 & 3.58 & 0.86 & 1.72\\

 & V2\_2 & \textbf{0.95} & \textbf{1.84} & 0.87 & 2.15 & 0.98 & 3.44 & 0.80 & 1.46\\

 & V3\_2 & 0.90 & 1.87 & 0.80 & 2.34 & \textbf{0.96} & \textbf{3.70} & 0.71 & 1.43\\

 & V4\_2 & 0.91 & 1.68 & 0.80 & 1.97 & \textbf{0.95} & \textbf{3.68} & 0.75 & 1.24\\

 & V5\_2 & 0.89 & 1.42 & 0.83 & 1.90 & \textbf{0.94} & \textbf{3.30} & 0.73 & 1.11\\

\multirow{-6}{*}{\raggedright\arraybackslash phoneme} & intercept\_2 & \textbf{0.97} & \textbf{1.73} & 0.83 & 1.94 & 1.00 & 3.37 & 0.89 & 1.39\\
\cmidrule{1-10}
 & fAlpha\_h & 0.87 & 2.48 & 0.79 & 2.96 & \textbf{0.93} & \textbf{8.26} & 0.71 & 1.54\\

 & fAsym\_h & \textbf{0.96} & \textbf{2.55} & 0.78 & 3.38 & 1.00 & 7.77 & 0.72 & 2.26\\

 & fConc1\_h & \textbf{0.95} & \textbf{2.59} & 0.76 & 3.47 & 0.99 & 9.79 & 0.65 & 1.97\\

 & fDist\_h & 0.93 & 1.99 & 0.81 & 2.66 & \textbf{0.98} & \textbf{6.56} & 0.72 & 1.60\\

 & fLength\_h & 0.87 & 4.56 & 0.69 & 5.94 & \textbf{0.98} & \textbf{13.39} & 0.62 & 3.57\\

 & fM3Long\_h & \textbf{0.95} & \textbf{2.44} & 0.87 & 3.45 & 0.99 & 7.24 & 0.80 & 2.22\\

 & fM3Trans\_h & \textbf{0.97} & \textbf{2.21} & 0.88 & 3.12 & 1.00 & 6.63 & 0.81 & 1.92\\

 & fWidth\_h & \textbf{0.95} & \textbf{ 4.38} & 0.70 & 5.57 & 0.98 & 11.94 & 0.67 & 3.62\\

\multirow{-9}{*}{\raggedright\arraybackslash telescope} & intercept\_h & \textbf{0.97} & \textbf{1.93} & 0.88 & 2.54 & 1.00 & 7.44 & 0.83 & 1.56\\
\bottomrule
\end{tabular}
\end{table}

\begin{table}[ht]
  \centering
  \caption{Coverage (rate) and median width (size) over 100 repetitions of
    marginal credible intervals for multinomial logistic regression with the \texttt{yeast} dataset.
    The method achieving at least the target coverage with the smallest size is
    shown in bold. If none of the methods achieve the target coverage, the method
    with the highest coverage is shown in bold. The target coverage is 0.95.}
  \label{tab:marginal-ci-7}
\begin{tabular}{llcccccc}
\toprule
\multicolumn{2}{c}{ } & \multicolumn{2}{c}{TabMGP} & \multicolumn{2}{c}{BB} & \multicolumn{2}{c}{Bayes} \\
\cmidrule(l{3pt}r{3pt}){3-4} \cmidrule(l{3pt}r{3pt}){5-6} \cmidrule(l{3pt}r{3pt}){7-8}
Setup & Feature Name & Rate & Size & Rate & Size & Rate & Size\\
\midrule
 & alm\_ME2 & \textbf{0.93} & \textbf{3.43} & 0.47 & 3.86 & 0.45 & 2.58\\

 & alm\_ME3 & \textbf{0.95} & \textbf{3.16} & 0.56 & 4.22 & 0.70 & 2.81\\

 & alm\_MIT & 0.85 & 1.20 & \textbf{0.89} & \textbf{1.49} & 0.71 & 1.03\\

 & alm\_NUC & \textbf{0.93} & \textbf{0.79} & 0.90 & 0.97 & 0.90 & 0.79\\

 & gvh\_ME2 & \textbf{0.94} & \textbf{2.39} & 0.56 & 3.27 & 0.57 & 2.19\\

 & gvh\_ME3 & \textbf{0.94} & \textbf{1.41} & 0.82 & 1.95 & 0.73 & 1.49\\

 & gvh\_MIT & \textbf{0.92} & \textbf{0.94} & 0.86 & 1.20 & 0.77 & 0.87\\

 & gvh\_NUC & \textbf{0.94} & \textbf{0.61} & 0.93 & 0.86 & 0.86 & 0.64\\

 & intercept\_ME2 & \textbf{0.89} & \textbf{ 6.35} & 0.39 & 10.10 & 0.43 & 4.42\\

 & intercept\_ME3 & \textbf{0.95} & \textbf{3.53} & 0.54 & 4.87 & 0.67 & 3.01\\

 & intercept\_MIT & \textbf{0.97} & \textbf{1.08} & 0.87 & 1.30 & 0.82 & 0.94\\

 & intercept\_NUC & 1.00 & 0.70 & 1.00 & 0.81 & \textbf{1.00} & \textbf{0.62}\\

 & mcg\_ME2 & \textbf{0.90} & \textbf{2.74} & 0.61 & 4.14 & 0.59 & 2.47\\

 & mcg\_ME3 & \textbf{0.92} & \textbf{1.63} & 0.82 & 2.11 & 0.81 & 1.64\\

 & mcg\_MIT & \textbf{0.90} & \textbf{0.90} & 0.90 & 1.12 & 0.78 & 0.85\\

 & mcg\_NUC & \textbf{0.96} & \textbf{0.66} & 0.94 & 0.89 & 0.86 & 0.64\\

 & mit\_ME2 & \textbf{0.93} & \textbf{2.48} & 0.57 & 2.93 & 0.62 & 2.62\\

 & mit\_ME3 & \textbf{0.97} & \textbf{1.99} & 0.84 & 2.65 & 0.77 & 2.07\\

 & mit\_MIT & \textbf{0.87} & \textbf{0.92} & 0.84 & 1.07 & 0.80 & 0.75\\

 & mit\_NUC & \textbf{0.93} & \textbf{0.76} & 0.90 & 0.93 & 0.87 & 0.70\\

 & nuc\_ME2 & \textbf{0.98} & \textbf{3.30} & 0.55 & 3.92 & 0.65 & 3.26\\

 & nuc\_ME3 & \textbf{0.74} & \textbf{1.67} & 0.74 & 2.23 & 0.69 & 1.77\\

 & nuc\_MIT & \textbf{0.94} & \textbf{1.34} & 0.83 & 1.61 & 0.76 & 1.28\\

 & nuc\_NUC & \textbf{0.89} & \textbf{0.81} & 0.81 & 1.04 & 0.72 & 0.70\\

 & vac\_ME2 & \textbf{0.99} & \textbf{2.06} & 0.63 & 2.62 & 0.71 & 2.31\\

 & vac\_ME3 & \textbf{0.97} & \textbf{1.60} & 0.87 & 2.22 & 0.91 & 1.98\\

 & vac\_MIT & \textbf{0.98} & \textbf{0.76} & 0.90 & 1.04 & 0.85 & 0.78\\

\multirow{-28}{*}{\raggedright\arraybackslash yeast} & vac\_NUC & \textbf{0.95} & \textbf{0.53} & 0.89 & 0.75 & 0.86 & 0.57\\
\bottomrule
\end{tabular}
\end{table}

\begin{table}[ht]
  \centering
  \caption{Coverage (rate) and median width (size) over 100 repetitions of
    marginal credible intervals for multinomial logistic regression with the \texttt{wine} dataset.
    The method achieving at least the target coverage with the smallest size is
    shown in bold. If none of the methods achieve the target coverage, the method
    with the highest coverage is shown in bold. The target coverage is 0.95.}
  \label{tab:marginal-ci-8}
\begin{tabular}{llcccccc}
\toprule
\multicolumn{2}{c}{ } & \multicolumn{2}{c}{TabMGP} & \multicolumn{2}{c}{BB} & \multicolumn{2}{c}{Bayes} \\
\cmidrule(l{3pt}r{3pt}){3-4} \cmidrule(l{3pt}r{3pt}){5-6} \cmidrule(l{3pt}r{3pt}){7-8}
Setup & Feature Name & Rate & Size & Rate & Size & Rate & Size\\
\midrule
 & V10\_3 & \textbf{0.97} & \textbf{1.25} & 0.75 & 2.30 & 0.73 & 1.65\\

 & V10\_4 & \textbf{0.96} & \textbf{1.27} & 0.71 & 2.18 & 0.71 & 1.61\\

 & V10\_5 & \textbf{0.96} & \textbf{1.34} & 0.74 & 2.25 & 0.70 & 1.69\\

 & V10\_6 & \textbf{0.96} & \textbf{1.64} & 0.79 & 3.03 & 0.73 & 2.09\\

 & V11\_3 & \textbf{0.93} & \textbf{1.94} & 0.77 & 2.72 & 0.76 & 2.09\\

 & V11\_4 & \textbf{0.97} & \textbf{1.98} & 0.73 & 2.78 & 0.78 & 2.09\\

 & V11\_5 & \textbf{0.93} & \textbf{2.16} & 0.71 & 2.93 & 0.78 & 2.21\\

 & V11\_6 & \textbf{0.93} & \textbf{2.83} & 0.64 & 4.22 & 0.67 & 2.85\\

 & V1\_3 & \textbf{0.97} & \textbf{1.21} & 0.72 & 2.22 & 0.74 & 1.50\\

 & V1\_4 & \textbf{0.90} & \textbf{1.26} & 0.74 & 2.20 & 0.72 & 1.48\\

 & V1\_5 & \textbf{0.88} & \textbf{1.35} & 0.74 & 2.33 & 0.68 & 1.59\\

 & V1\_6 & \textbf{0.91} & \textbf{1.73} & 0.74 & 3.17 & 0.74 & 2.19\\

 & V2\_3 & \textbf{0.90} & \textbf{1.12} & 0.79 & 1.97 & 0.74 & 1.29\\

 & V2\_4 & \textbf{0.81} & \textbf{1.30} & 0.73 & 2.06 & 0.68 & 1.36\\

 & V2\_5 & \textbf{0.82} & \textbf{1.45} & 0.69 & 2.23 & 0.70 & 1.51\\

 & V2\_6 & \textbf{0.84} & \textbf{1.99} & 0.67 & 3.12 & 0.66 & 2.15\\

 & V3\_3 & \textbf{0.96} & \textbf{1.14} & 0.76 & 2.16 & 0.65 & 1.49\\

 & V3\_4 & \textbf{0.97} & \textbf{1.18} & 0.72 & 2.21 & 0.63 & 1.49\\

 & V3\_5 & \textbf{0.99} & \textbf{1.37} & 0.80 & 2.33 & 0.76 & 1.66\\

 & V3\_6 & \textbf{1.00} & \textbf{1.85} & 0.74 & 3.19 & 0.71 & 2.37\\

 & V4\_3 & \textbf{0.98} & \textbf{1.49} & 0.70 & 2.60 & 0.74 & 1.94\\

 & V4\_4 & \textbf{0.96} & \textbf{1.50} & 0.74 & 2.66 & 0.75 & 1.93\\

 & V4\_5 & \textbf{0.93} & \textbf{1.65} & 0.72 & 2.69 & 0.75 & 2.04\\

 & V4\_6 & \textbf{0.90} & \textbf{2.17} & 0.71 & 3.78 & 0.64 & 2.66\\

 & V5\_3 & \textbf{0.98} & \textbf{1.33} & 0.68 & 2.46 & 0.74 & 1.74\\

 & V5\_4 & \textbf{0.96} & \textbf{1.37} & 0.77 & 2.55 & 0.77 & 1.77\\

 & V5\_5 & \textbf{0.94} & \textbf{1.98} & 0.82 & 2.90 & 0.81 & 2.22\\

 & V5\_6 & \textbf{0.97} & \textbf{2.47} & 0.74 & 4.78 & 0.75 & 3.36\\

 & V6\_3 & \textbf{0.79} & \textbf{1.77} & 0.69 & 3.10 & 0.56 & 1.75\\

 & V6\_4 & \textbf{0.84} & \textbf{1.78} & 0.70 & 3.06 & 0.57 & 1.77\\

 & V6\_5 & \textbf{0.84} & \textbf{1.92} & 0.72 & 3.19 & 0.60 & 1.85\\

 & V6\_6 & \textbf{0.83} & \textbf{2.15} & 0.70 & 4.07 & 0.59 & 2.37\\

 & V9\_3 & \textbf{0.97} & \textbf{1.29} & 0.75 & 2.50 & 0.64 & 1.72\\

 & V9\_4 & \textbf{0.98} & \textbf{1.34} & 0.72 & 2.50 & 0.65 & 1.74\\

 & V9\_5 & \textbf{0.95} & \textbf{1.46} & 0.71 & 2.57 & 0.65 & 1.82\\

 & V9\_6 & \textbf{0.94} & \textbf{1.81} & 0.73 & 3.51 & 0.70 & 2.33\\

 & intercept\_3 & \textbf{0.97} & \textbf{2.87} & 0.42 & 5.76 & 0.55 & 2.76\\

 & intercept\_4 & \textbf{0.99} & \textbf{2.90} & 0.37 & 5.72 & 0.49 & 2.75\\

 & intercept\_5 & \textbf{0.98} & \textbf{3.00} & 0.53 & 5.91 & 0.56 & 2.82\\

\multirow{-40}{*}{\raggedright\arraybackslash wine} & intercept\_6 & \textbf{0.99} & \textbf{3.94} & 0.81 & 9.06 & 0.69 & 4.35\\
\bottomrule
\end{tabular}
\end{table}

\end{document}